\documentclass{emulateapj}
\usepackage{natbib,graphicx,psfig}

\newcommand{\kms}{\mbox{km s$^{-1}~$}} 
\newcommand{\kmse}{\mbox{km s$^{-1}$}} 
 
\newcommand{\msun}{M$_{\odot}~$} 
\newcommand{\msune}{M$_{\odot}$} 
\newcommand{\vlsr}{$V_{\rm LSR}~$}
\newcommand{\vlsre}{$V_{\rm LSR}$}
\newcommand{\lms}{$L_{\rm MS}~$}
\newcommand{\lmse}{$L_{\rm MS}$}
\newcommand{\bms}{$B_{\rm MS}~$}
\newcommand{\bmse}{$B_{\rm MS}$}
\newcommand{\dgr}{$^{\circ}~$}

\newcommand{\h}{$^{\rm h}$}
\newcommand{\m}{$^{\rm m}$}
\newcommand{\s}{$^{\rm s}~$}
\newcommand{\se}{$^{\rm s}$}
\newcommand{\nhi}{$N_{\rm H \small{I}}$ }
\newcommand{\nhie}{$N_{\rm H \small{I}}$}
\newcommand{\pcm}{cm$^{-2}$ }
\newcommand{\pcme}{cm$^{-2}$}
\newcommand{\hi}{\ion{H}{1} }
\newcommand{\hie}{\ion{H}{1}}

\lefthead{Nidever et al.}
\righthead{Origin of the Magellanic Stream}

\begin{document}

\title{The Origin of the Magellanic Stream and Its Leading Arm}

\author{David L. Nidever\altaffilmark{1},
Steven R. Majewski\altaffilmark{1}, 
and W. Butler Burton\altaffilmark{2,3}
}

\altaffiltext{1}{Dept. of Astronomy, University of Virginia,
Charlottesville, VA, 22904-4325 (dnidever, srm4n@virginia.edu)}

\altaffiltext{2}{Sterrewacht Leiden, PO Box 9513, 2300 RA Leiden,
The Netherlands}

\altaffiltext{3}{National Radio Astronomy Observatory, 520 Edgemont Road,
Charlottesville, Virginia 22903, USA (bburton@nrao.edu)}

\begin{abstract}
We explore the Magellanic Stream (MS) using a Gaussian decomposition of
the \hi velocity profiles in the Leiden-Argentine-Bonn (LAB) all-sky \hi survey.
This decomposition exposes the MS to be composed of two filaments distinct both spatially
(as first pointed out by Putman et al.) and in velocity.
Using the velocity coherence of the filaments, one can be traced back
to its origin in what we identify as the SouthEast \hi Overdensity (SEHO) of the Large Magellanic Cloud (LMC),
which includes 30 Doradus.
Parts of the Leading Arm (LA) can also be traced back to the SEHO
in velocity and position.  Therefore, at least one-half of the trailing Stream and most of the LA originates
in the LMC, contrary to previous assertions that both the MS and the LA originate
in the Small Magellanic Cloud (SMC) and/or in the Magellanic Bridge.
The two MS filaments show strong periodic, undulating spatial and velocity patterns that we
speculate are an imprint of the LMC rotation curve.
If true, then the drift rate of the Stream gas away from the Magellanic Clouds 
is $\sim$49 \kms and the age of the MS is $\sim$1.74 Gyr.
The Staveley-Smith et al.~high-resolution \hi data of the LMC show gas outflows from supergiant shells
in the SEHO that seem to be creating the LA and LMC filament of the MS.
Blowout of LMC gas is an effect not previously accounted for but one that
probably plays an important role in creating the MS and LA.
\end{abstract}

\keywords{Galaxies: interactions -- Galaxies: kinematics and dynamics -- Galaxies: Local Group
-- Galaxy: halo -- Intergalactic Medium -- Magellanic Clouds -- Radio Lines: general}

\section{Introduction}
\label{sec:intro}

Under the prevailing concordance cold dark matter cosmology, large structures 
like galaxies form through hierarchical accretion and merging of dark matter subhalos
(e.g., White \& Rees 1978; Davis et al.~1985; Navarro, Frenk \& White 1996, 1997; Moore et al.~1999).
While much of the merging took place at early times, the process of accretion onto 
large spiral galaxies, such as our Milky Way (MW), continues at a reduced rate until late times 
(Bullock \& Johnston 2005).  Disruption and accretion of small galaxies 
gives rise to gaseous and stellar tidal streams that continue to orbit the accreting
galaxy as fossil relics of the cannibalistic activity.
Many striking examples of disruption around our Milky Way have been discovered in recent
years:
the Pal 5 stream (Odenkirchen et al.~2001; Grillmair \& Dionatos 2006), 
the Sagittarius stream (e.g., Ibata et al.~2001; Newberg et al.~2002; Majewski et al.~2003);
the Monoceros stream (e.g., Yanny et al.~2003);
the orphan stream (Belokurov et al.~2006; Grillmair 2006a); 
and the anticenter stream \citep{Grill06b}.
There are likely numerous more such streams of stripped debris remaining to be discovered.

But the most prominent and earliest discovered stream is the Magellanic Stream (MS),
which stretches over 100\dgr across the southern sky behind the Large and Small Magellanic Clouds
(LMC and SMC).  From the mid-1960s onwards there were many efforts to detect high-velocity \hi clouds,
but \citet{WW72} were the first to recognize the large extent of what was to become known as the Magellanic Stream.
\citet{Math74} associated the Stream with the Magellanic Clouds and more fully demonstrated its $\sim$100\dgr span.
A more thorough history and general review of the MS  is given by \citet[][hereafter P03]{Put03} and \citet[][hereafter B05]{Br05}.
Whereas, except for the MS, all the above mentioned streams are {\em stellar} and attributed
to tidal forces for their origin, the MS is still only recognized as a {\em gaseous}
feature\footnote{There is a sparse stellar population at the location of the Magellanic Bridge feature
\citep{IDK90}, but their connection to the \hi Bridge is unclear,
as is the connection of the Bridge to the classical MS.}  and is
one such structure whose origin is still debated.

Due to their proximity to each other and as the most massive of the
MW satellites, the Magellanic Clouds (MCs) have long been considered to have influenced each other
as well as  the growth and evolution of their host galaxy.
Thus, the MW-LMC-SMC system is regarded as an important laboratory with which to study the
formation, evolution, and interaction of galaxies and their stellar populations.  To this end, extensive
mappings of the MCs have recently been conducted at a number of wavelengths.  Surveys have been made
in \hi emission at 21-cm (HIPASS: Barnes et al.~2001; Br\"uns et al.~2005;
Staveley-Smith et al.~1997, 2003; Stanimirovi\'c et al.~1999; and Muller et al.~2003), 
in molecular spectral CO lines (NANTEN: Fukui et al.~1999; Mizuno et al.~2001), as well as in the
radio continuum (Haynes et al.~1991; Dickel et al.~2005),
in the thermal infrared (IRAS: Beichman et al.~1988; MSX: Mill et al.~1994; SAGE: Meixner et al.~2006;
S$^3$MC: Bolatto et al.~2007), in the near-infrared (DENIS: Epchtein et al.~1997; 2MASS: Skrutskie et al.~2006),
in broadband optical colors (MCPS: Zaritsky et al.~2002, 2004) as well as in optical
emission lines (MCELS: Smith et al.~1998),
and at ultraviolet (Smith, Cornett \& Hill 1987)
and X-ray wavelengths (ROSAT: Snowden \& Petre 1994; Chandra and ACIS: Townsley et al.~2006).
While these surveys have deepened our understanding 
of the Clouds themselves --- e.g., their star formation history, their stellar
content, and their overall structure (van der Marel 2001) --- 
the \hi investigations are central to understanding the most obvious
product of the MW-MC interaction --- the Magellanic Stream.

Obviously, the overall appearance of the MS is shaped by the dynamics of the MCs which are
moving almost entirely tangentially in the sky \citep{kalli06a,kalli06b,piatek07} and have recently just passed
perigalacticon in their polar orbit around the MW \citep{besla}.
The earliest dynamical studies of the MCs used the MS to constrain the orbit of the LMC and
thereby obtained an LMC space velocity of $\sim$350 \kms \citep{MF80,LLB82,HR94}.
However, most proper motion measurements of the LMC 
\citep[summarized in][hereafter vdM02]{vandermarel02}
have favored a lower space velocity of around 250 \kmse.  
\citet{MF80}, and similar subsequent MS modeling papers, used these space velocities
and an isothermal sphere MW potential to derive orbits for the MCs with 
an orbital period $\sim$1.5 Gyr and a last apogalacticon distance of $\sim$120 kpc.  All of these
models agree that the MCs had a close encounter $\sim$200 Myr ago.
However, the new HST proper motions of the MCs \citep{kalli06a,kalli06b} give MC space
velocities $\sim$100 \kms higher than those produced by the earlier proper motion surveys
and that increase the orbital period to $\sim$2--3 Gyr and increase the orbital ellipticity
\citep[i.e.~the last apogalacticon distance at $\sim$150--200 kpc;][]{kalli06b,besla}.
But the addition of a more
realistic NFW potential for the MW produces the startling result of {\em hyperbolic} orbits for
the MCs \citep{besla}.  Even with a high MW mass model, while the MCs ``become bound'' again, the
LMC orbital period is $\sim$7 Gyr and the last apogalacticon distance is $\sim$400 kpc. Clearly,
whether the MCs are bound or not, and the shape of their orbits, has a direct influence on the
interactions that produce and shape the MS.

Large-area 21-cm radio surveys have produced most of the information now available about
the detailed structure of the MS.
Since its discovery as a long stream of \hi gas trailing the MCs 
a number of models have attempted to explain the dynamics and origin of the MS.
Early N-body simulations with hundreds of particles by Lin \&
Lynden-Bell (1977, 1982) and by Murai \& Fujimoto (1980) were able to reproduce the general 
features of the Stream (such as its length and velocity distribution) through tidal 
stripping by the MW.
Later on, it was proposed that the MS could have been created by ram pressure forces
(Meurer, Bicknell, \& Gingold 1985; Moore \& Davis 1994) as the MCs move through the hot
gaseous halo of the MW.
Ram pressure strips some gas from the Clouds and creates a {\em trailing} gaseous stream.
A persistent problem with the tidal models is that they predict a stellar MS component 
that, to date, has not been observed despite numerous efforts (e.g., Philip 1976a,b; 
Recillas-Cruz 1982; Br\"uck \& Hawkins 1983; Kunkel et al.~1997; Guhathakurta \& Reitzel 1998).  
Because ram pressure only affects gas and not
stars, these models seemed initially to be more consistent with a gas-only structure.

However, more recent large scale \hi surveys have revealed new complexities in the
MS that are difficult to account for in a ram pressure model.  For example,
\citet{Put98} used the HIPASS data \citep{Barnes01} to discover a gaseous
{\em leading} arm of the MS.  This is a feature readily accounted for by the tidal models 
but creating a leading arm by ram pressure forces remains a formidable problem.  
The HIPASS data have also shown that the trailing MS is spatially bifurcated 
(P03), although Cohen (1982) and Morras (1983) previously pointed out that the MS
splits into two branches.
\citet{Mastro05} performed a large ram pressure+tidal force simulation of the
Magellanic Stream (with only the LMC as a progenitor) and were able to reproduce 
the general features of the Stream, including its extent, shape, column density 
gradient,  and velocity gradient; however, Mastropietro et al. could not reproduce 
the spatial bifurcation of the Stream nor the Leading Arm Feature.
The recent N-body tidal simulations by Connors et al.~(2004, 2006),
in which most particles are stripped from the SMC during a close encounter with the 
LMC and MW $\sim$1.5 Gyr ago, give the closest reproduction of the Stream to date, 
including the spatial bifurcation of the Stream, the Leading Arm (and
its bent shape), and the MS velocity distribution.  However, a problem with most
tidal models, including those by Connors et al., is that they have trouble reproducing 
the column density gradient along the Stream, whereas ram pressure models match this 
particular feature of the observations better, and, of course, account for
an entirely gaseous Stream.  
While there is some evidence that stellar tails may be offset (or completely 
missing) from the gaseous tails in tidally interacting galaxies (Mihos 2001; 
Hibbard, Vacca, \& Yun 2000), the study of \citet{johnston98} indicates that
stellar debris from the LMC should have already been found.

Most of the current literature has supported either the SMC or Magellanic Bridge as the source of the 
MS gas (e.g., P03, B05) since it appears to emanate from these
regions in maps of \hi column density on the sky.
Moreover, the mass of the SMC is much less than that of the LMC (by an order of magnitude)
and it is therefore, presumably, much easier for SMC gas to be stripped than LMC gas.  Many of the
tidal models have used the SMC-origin assumption in their N-body simulations
(Gardiner \& Noguchi 1996; Yoshizawa \& Noguchi 2003;  Connors et al.~2004, 2006).  
On the other hand, Mastropietro et al.~were able to reproduce the general characteristics
of the MS by ram pressure stripping from the LMC alone.

In this paper we follow the tradition of using large-area 21-cm data to investigate the relation of
the MS and  the MCs.  We take advantage of the high velocity resolution of the Leiden-Argentine-Bonn
(LAB) \hi datacube \citep{LAB} to investigate the detailed structure of both the leading and  trailing arms of the
MS across their entire known length.  Particular attention is paid to disentangling MS features from
other overlapping structures in the datacube, including the MW disk, other Intermediate- and
High-Velocity Clouds, and the MCs themselves.  This has
allowed us to uncover several key aspects of the MS that differ from the earlier interpretations and
models of the MS and that lead us to a new mechanism to explain the origin of the Stream.

First, we show that one of the two trailing MS filaments as well as the Leading Arm
originate in the LMC, {\it not} in the SMC or in the Bridge as previously suggested.  
Moreover, the specific site that is the source of this gas we identify as
the SouthEast \hi Overdensity (SEHO), a region of dense \hi and intense star-formation in
the southeast of the LMC.
Analysis of the high spatial resolution \hi Parkes data of the LMC \citep{ss03} indicates
that supergiant shells in the SEHO are probably responsible for blowing out much
of the gas from the LMC, creating (at least one of) the filaments of the MS
and Leading Arm.  Once blown out and free of the gravitational grip
of the LMC, the gas experiences tidal stretching from the MW potential and separates 
into   the leading and trailing components.  This blowout mechanism for releasing 
Magellanic gas represents an alternative to the tidal and ram
pressure models.  Finally, we propose that the periodic sinusoidal weaving of the trailing
Stream filament may be a result of the off-center position of the SEHO in 
the rotating LMC disk.  Coupled with knowledge of the rotation period of the
LMC at the SEHO radius, we can use the sinusoidal patterns
to estimate the total age of the MS as $\sim$1.74 Gyr under the proposed scenario.

This paper is organized as follows: Section 2 is a description of the 21-cm LAB data. 
In Sections 3 and 4 we describe the Gaussian Decomposition of the LAB data and the 
separation of the MS from the MW features; the casual reader uninterested in these
details may wish to proceed to Section 5.
In Section 5 we present the results of our investigation of the structure of the Magellanic Stream
using our database of Gaussian Decomposition centers.  Section 6 details our analysis of the SE \hi
overdensity in the LMC.  A discussion of our findings and conclusions are given in Section 7,
and a summary of the primary conclusions is given in Section 8.

\section{Brief Description of Leiden-Argentine-Bonn (LAB) data}
\label{sec:lab}

The Leiden-Argentine-Bonn all-sky \hi survey \citep{LAB} is a combination of the Leiden/Dwingeloo Survey
(LDS: Hartmann \& Burton 1997), covering the sky north of $\delta= -30$\degr, and the Instituto Argentino
de Radioastronom\'{\i}a Survey (IAR: Arnal et al.~2000; Bajaja et al.~2005) at more southern declinations.
The combined material has a velocity resolution of 1.3 \kmse, a spatial resolution of 36\arcmin~on a grid
spacing of $0\fdg5$ in Galactic latitude ($b$) and $0\fdg5/{\rm cos}(b)$ in Galactic longitude ($l$), and
is corrected for stray radiation.  The
velocity range of $-450$ \kms to $+400$ \kms is adequate for (almost) all Galactic work.  The root
mean square noise is 0.09 K.
Here, we exploit the extensive sky coverage of the LAB data and its relatively high velocity resolution
to follow the MS in detail over its full length, and to unravel its filaments.

\section{Description of Automated Gaussian Decomposition}
\label{sec:gaussdecomp}

To improve our ability to trace structures of the MS and disentangle them from MW gas
we wrote an automated Gaussian analysis program in the Interactive
Data Language (IDL)\footnote{A product of ITT Visual Information Systems, formerly Research Systems, Inc.}
using an algorithm similar to that used by \citet{Haud00}.
Kalberla \& Haud (2006) have also performed a Gaussian decomposition of the LAB database but
with different goals, namely the physical interpretation of the structure of High Velocity Clouds.
Our use of Gaussian decomposition is predicated upon the expected continuity of the filamentary structures
in terms of velocity, position, velocity-dispersion, and integrated column density, which allows us
to track features even through complex, crowded \hi environments.

The general algorithm to decompose an \hi velocity profile into Gaussians proceeds in two stages.
In the first stage new Gaussians are added to the model fit until the root mean square
of the residuals (observed $-$ fitted profile; $rms$) drops below the noise level, or any new
Gaussians do not improve the model fit significantly ($\delta rms<$2\%).  In the second stage
an attempt is made to reduce the number of Gaussians in the fit without
increasing $rms$ appreciably.  The details of these stages are discussed in the following sections.

One of the benefits of the Gaussian analysis performed here is that faint \hi structures can be
enhanced by plotting the integrated column density of each Gaussian at its central velocity.
This avoids spreading the flux of the Gaussian over a range of velocities and instead
concentrates it to one point.   Figures \ref{allsky}, \ref{zerobeforeafter},
\ref{msprofile}, \ref{ms_vmlon_zoom}, \ref{ms_onskycut}, \ref{la_vmlon} and \ref{lmcfil_fit}
in this paper use this technique.

\subsection{Gaussian Fitting}
\label{subsec:gaussfit}

In order to find the best-fitting Gaussian decomposition to a velocity distribution
along a given line-of-sight
we use the general purpose IDL curve-fitting package MPFIT written by Craig
Markwardt.\footnote{Available at http://cow.physics.wisc.edu/$\sim$craigm/idl/idl.html}
MPFIT is a set of routines for robust least-squares minimization (curve fitting)
based on the MINPACK-1 FORTRAN package.

In order to adopt MPFIT to our purposes we wrote an IDL function that returns the output
of the Gaussian function, $T_{\rm B} = T_{\rm{B,0}}~ {\rm exp(-({\it v-v_0})^2
/ 2\sigma_{\it v}^2 )}$, given an array of velocities and the Gaussian parameters
($T_{\rm B}$, $v_0$, and $\sigma_v$).  The necessary inputs for MPFIT
are the set of values for the independent and dependent variables (the observed data:
velocity -- $v$, and brightness temperature -- $T_{\rm{B}}$) and
a first guess for the Gaussian parameters.  One can also give upper and lower limits for each
parameter that is allowed to vary in the fit; this is a useful feature
for constraining the Gaussians to meaningful solutions.  MPFIT returns
the parameters for the best fit, the formal 1-$\sigma$ uncertainties for those
parameters, and $\chi^2$ of the fit.  We also used $rms$ (root mean square of the residuals)
as another means to ascertain the ``goodness'' of the fit.

Our Gaussian decomposition program uses MPFIT to fit Gaussians to peaks in the
\hi profiles or to peaks in the residuals (profile $-$ previous best fit).
For the initial guess of the Gaussian parameters for a given single peak we use
the height of the peak for $T_{\rm B,0}$, and the velocity of the peak
for $v_0$.  To get an initial guess for $\sigma_v$ we used the fact that the
ratio of the derivative of the Gaussian to itself is a line with a slope
of $-1/\sigma_v^2$; $T_{\rm B}'/T_{\rm B} = (-1/\sigma_v^2)\times(v-v_0)$.
A line was fit to this ratio near the central part of the peak, and $\sigma_v$
computed from its slope.  In MPFIT the parameters were constrained to lie between
the limits: 0.01 K $ < T_{\rm B,0} < 2\times T_{\rm B,max}$, $v_{\rm min} < v_0 < v_{\rm max}$,
0.5 \kms $ < \sigma_v < \frac{1}{2}\times(v_{\rm max}-v_{\rm min})$; where $T_{\rm B,max}$ is
the maximum $T_{\rm B}$ of the whole profile, and $v_{\rm min}$ and $v_{\rm max}$ are the minimum
and maximum velocities of the whole profile ($-450$ and $+400$ \kmse).  $T_{\rm B,0}$ 
was also constrained so that only peaks higher than a certain threshhold
above the noise level were chosen for Gaussian fitting (see \S \ref{subsec:GaussAdd} below).

To obtain the noise level we smoothed the \hi profile with a [16,16,2] Savitzky-Golay filter
\citep{SavGol} (where the numbers in brackets refer respectively to the number of data points to the left
and right of each point to include in the filter, and the order of the derivative desired)
and then subtracted this from the original \hi profile to remove any real features.  Then
we found the standard deviation of points with $|v|>250$ \kms after $5\sigma$ outliers were
rejected.  This was used as the noise level for the given \hi profile.

\subsection{Adding New Gaussians}
\label{subsec:GaussAdd}

Gaussians were added to the velocity profile fit one at a time.  The current best fit was
subtracted from the observed \hi profile to find the residuals.  These residuals
were then searched for peaks higher than
5 times the noise level (although this was lowered to 2 times the noise level
if no peaks were found) and MPFIT was used to find the best-fitting Gaussian for
each peak along with its $rms$ using the first guesses described in the previous section.
To smooth over the noise and search for features on various scales five
smoothed versions of the residuals were created using Savitzky-Golay filters of [4,4,2],
[16,16,2], [30,30,2], [50,50,2] and [100,100,2] and Gaussians were fit to all the peaks in
these smoothed profiles.

The best-fit Gaussians to peaks in the six versions of the residuals (the original and the five
smoothed) that had parameters within our acceptble limits (see above) were kept for further
fitting.
Each one of these candidate Gaussians is taken in turn and
added to the current best-fit Gaussian decomposition of the entire \hi profile. MPFIT is then rerun
with this new Gaussian decomposition as a first guess, fitting the multiple Gaussians
at once, and the change of $rms$ compared to the previous best-fit was computed.
The candidate Gaussian that gives the greatest decrease in $rms$ is then added to the
overall decomposition of the \hi profile.

This procedure is repeated and Gaussians added to the decomposition until the $rms$
drops to or below the noise level, or the decrease in $rms$ is less than 2\%.

\subsection{Removing Gaussians}
\label{subsec:remgauss}

The best-fitting Gaussian decomposition is that which minimizes {\it both} the $rms$
as well as the number of Gaussians.  To achieve this goal, we attempt to remove Gaussians that do
not significantly improve the fits to the velocity profiles.  The Gaussians in the best-fit decomposition are
sorted in order of area ($A = T_{\rm B,0} \times \sigma_v \sqrt{2\pi}$ K \kmse), and the smallest half
of the Gaussians were picked for possible removal.  Each one in turn was temporarily removed
from the decomposition and the new best-fit and $rms$ for the whole profile are found with MPFIT.
If the increase in $rms$ is less than 2\% the Gaussian is permanently removed from the
decomposition.  Even though it might appear at first glance that this is repeating work
done in the ``Adding Gaussians'' stage, 
this particular decomposition might not have been looked at before because
of the order in which Gaussians were added.  Gaussians are also removed from the decomposition
if two quite similar Gaussians were found at the same velocity.  In that case they are replaced
by a single Gaussian with parameters given by Equations~11--13 in \citet{Haud00}.

\begin{figure}
\includegraphics[scale=0.50]{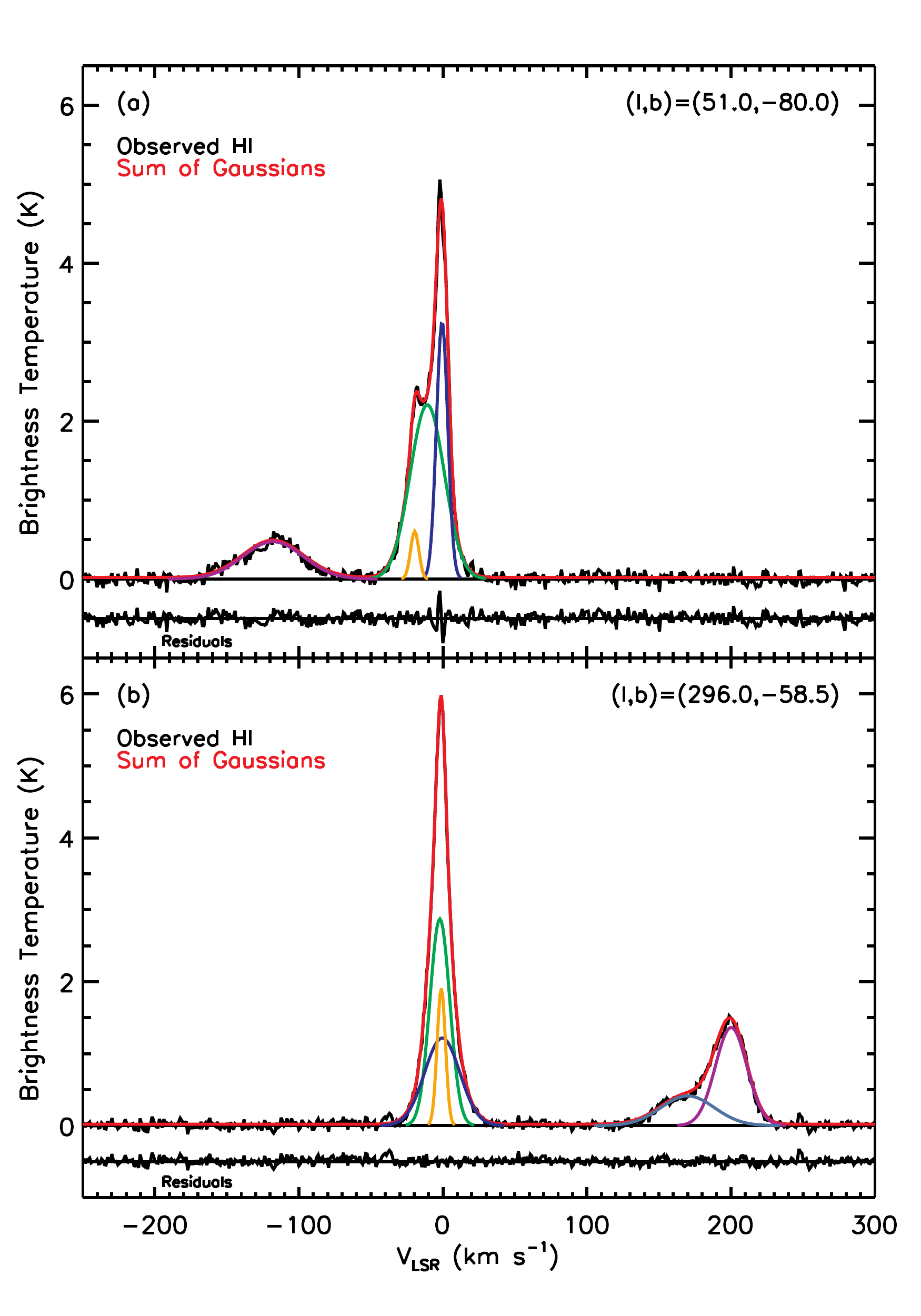}
\caption{Examples of Gaussian decomposition for LAB \hi profiles (top plot in each panel) and their residuals
(bottom plot in each panel).
({\em a}) The Gaussian decomposition at $(l,b) = $ (51.0\degr,$-80.0$\degr), with four Gaussian components
(each shown by a different color, their sum by red), showing the Magellanic Stream at negative velocities.
({\em b}) The Gaussian decomposition at $(l,b) = $ (296.0\degr,$-58.5$\degr), with five Gaussian components,
showing two separated Magellanic Stream components at positive velocities.}
\label{hidecomp}
\end{figure}

\subsection{Selecting the Next Position}
\label{subsec:nextpos}

We initially used the same procedure as \citet{Haud00} (see his section 3.3) to select the next
position on the sky to decompose.  If the profile at a neighboring position has already been decomposed
but has a worse decomposition (either larger $rms$ or more Gaussians) than the decomposition at the
current position, then the program tries to re-decompose the neighboring profile using the best-fit
solution of the current position as the initial guess.
The scheme also allows the program to wander around re-decomposing \hi profiles until the
re-decomposition is not an improvement.  We found
that this ``wandering'' scheme was too CPU intensive and did not improve the solutions substantially.  Therefore,
we used a modified scheme that forced the program to return to the 
previous position after re-decomposing a neighboring position.
For new positions, that hadn't been decomposed yet, the best-fit solution at the previous position was always used
for the initial guess.

\begin{figure}
\includegraphics[scale=0.45]{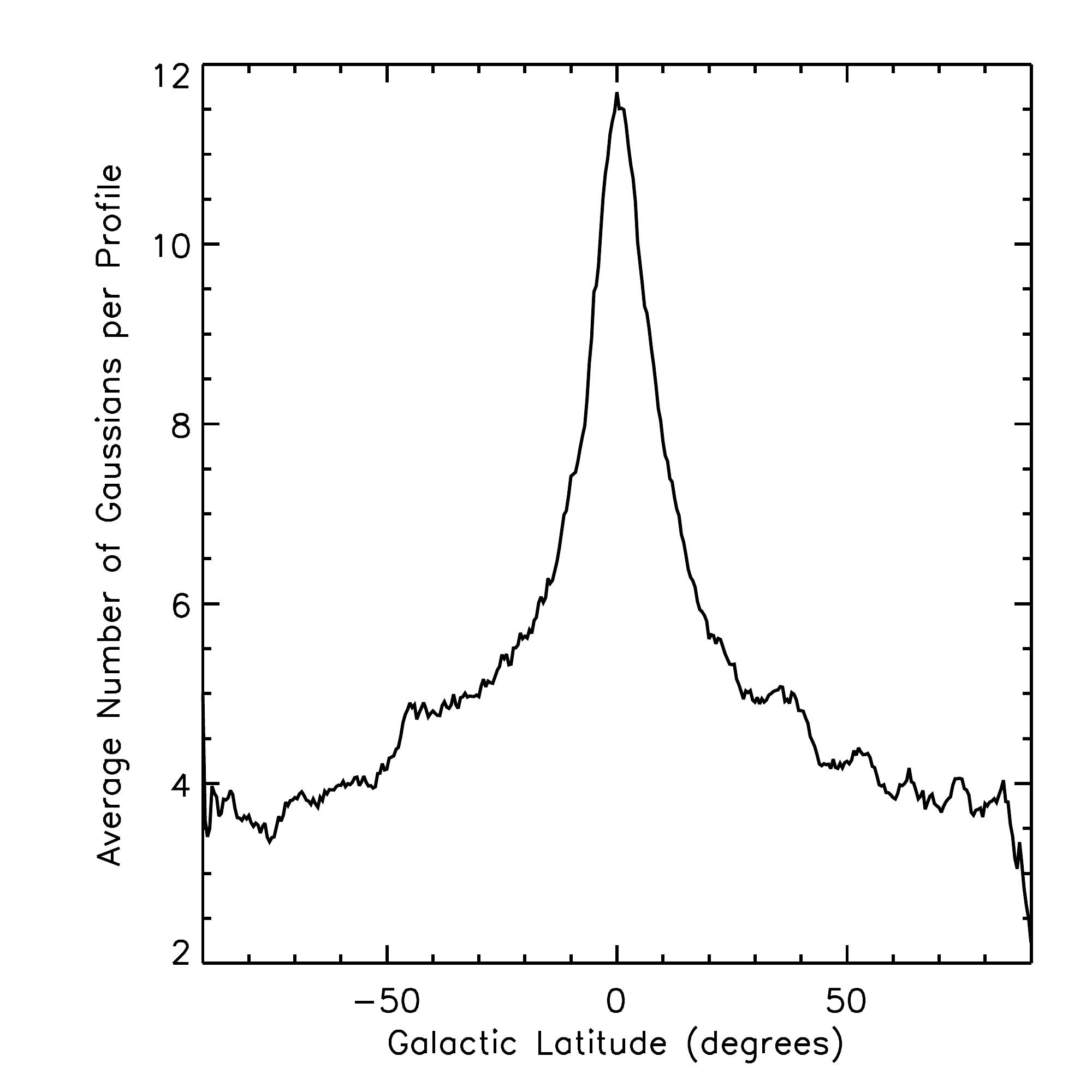}
\caption{Average number of Gaussians in the \hi decomposition per position as
a function of Galactic latitude $b$.}
\label{statgauss}
\end{figure}

For low latitudes the program was not allowed to re-decompose profiles since it took much
longer in these regions.  Our program did not take into account any self-absorption, so some low
latitude profiles near the Galactic center will not be correctly represented by the Gaussian
decomposition.  This does not affect our study of the MS here because it should
not have significant self-absorption, if any.

\subsection{Statistics of the Gaussians}
\label{subsec:statgauss}

\begin{figure}
\includegraphics[scale=0.43]{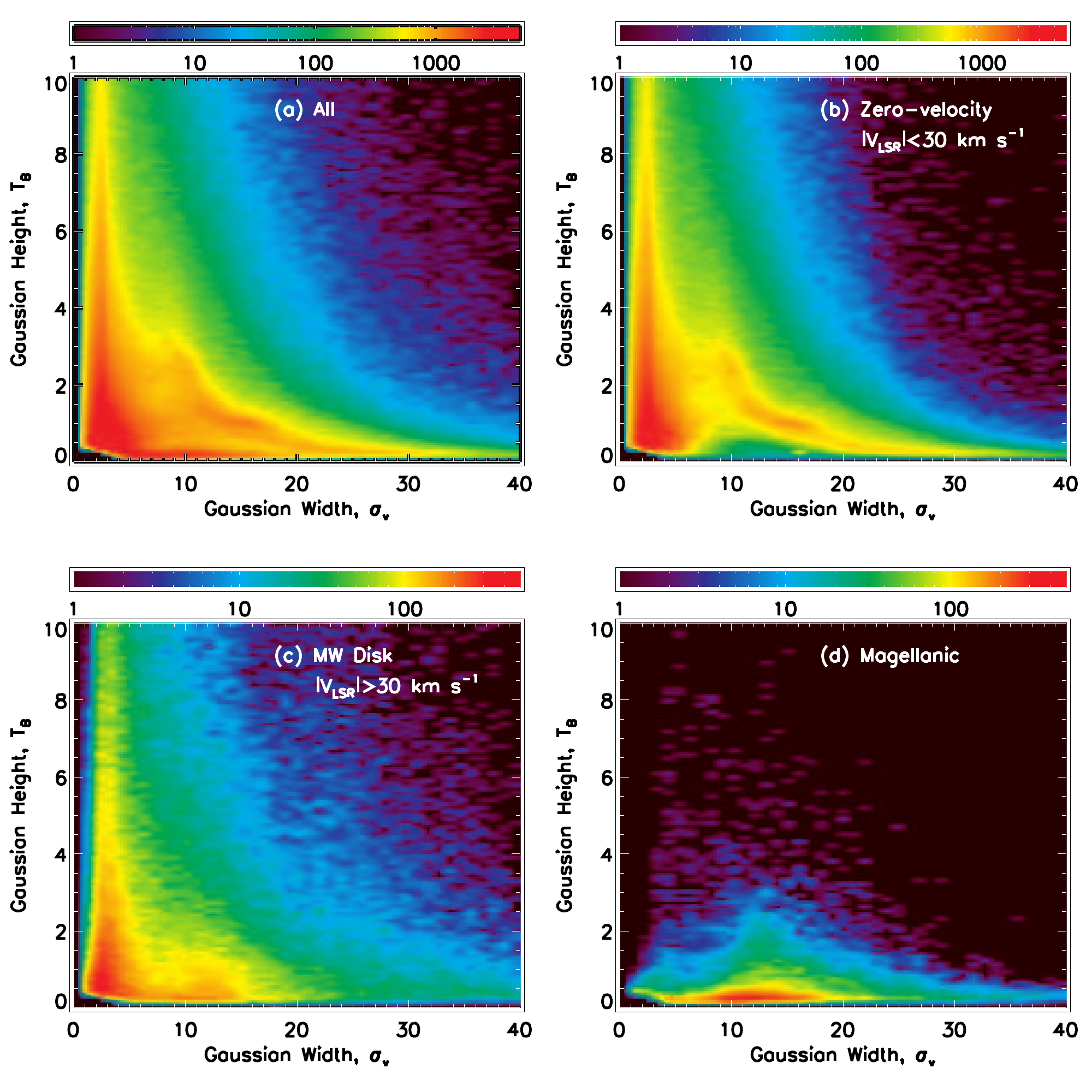}
\caption{The distribution of Gaussian parameters ($T_{\rm B}$ and $\sigma_v$)
for various \hi populations. ({\rm a}) All Gaussians;  ({\rm b}) local zero-velocity
Gaussians ($|$\vlsre$| < 30$ \kmse); ({\rm c}) Milky Way disk Gaussians with $|$\vlsre$| > 30$ \kms
(see \S 4.2 for how these are defined);  ({\rm d}) Magellanic Clouds
and Stream Gaussians.  Note the different color scalings for a+b and c+d.}
\label{gausshtsig}
\end{figure}

\begin{figure*}
\includegraphics[scale=0.72]{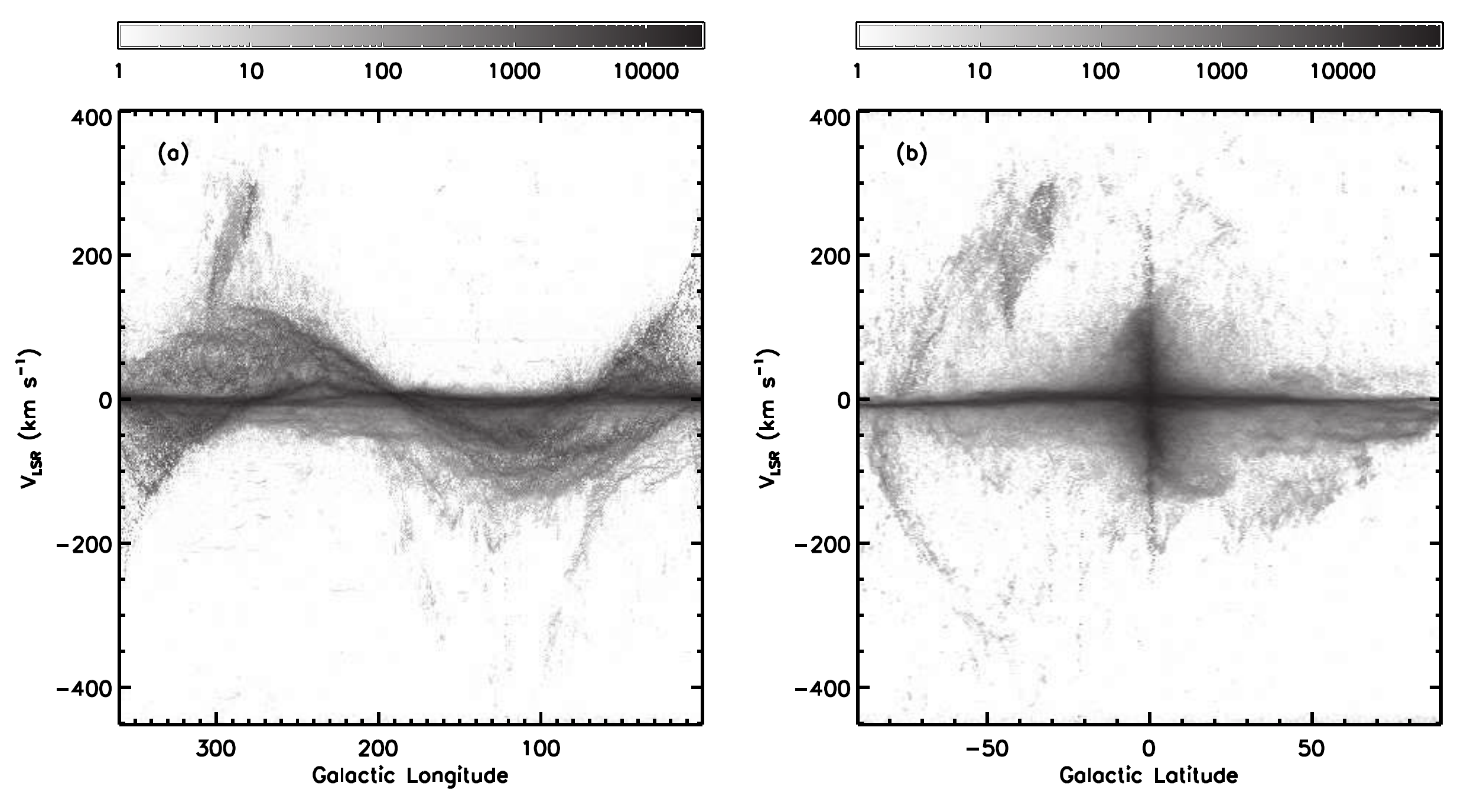}
\caption{Position-velocity distribution of detected \hi Gaussians in the LAB datacube where each is represented
by a single point with weight equal to the Gaussian area and the results summed over all Gaussians.
({\rm a}) \vlsr vs.~$l$, summed along $b$, showing the characteristic velocity curve of the Milky Way
as well as the Magellanic Clouds and Stream at larger velocities;  ({\rm b}) \vlsr vs.~$b$, summed along $l$,
showing the zero-velocity gas (even at high latitudes), Milky Way disk ($l\sim$0), Magellanic Clouds (to upper left)
and Stream (arcing on left side).  The greyscales are in units of K \kmse.}
\label{allsky}
\end{figure*}

Our automated Gaussian decomposition program was run on all 259,920 \hi profiles (720 values of $l$
$\times$ 361 values of $b$, in steps of $0\fdg5$) of the
LAB all-sky survey.  In the end, the entire sky was decomposed into 1,370,801 Gaussians.
Several examples of Gaussian decompositions at various sky positions are shown in Figure \ref{hidecomp}.
We find that the majority of \hi profiles (at latitudes above the disk, $|b|\gtrsim15$\degr) are well fit with
four to six Gaussians.  The average number of Gaussians per profile as a function of $b$ is
shown in Figure \ref{statgauss}; the number peaks at 12 at the Galactic center and levels off to four near
the poles.  The majority of Gaussians at higher latitudes are from local MW, zero-velocity gas.
The distribution of fitted Gaussian parameters for various populations are shown in
Figure \ref{gausshtsig}, a ``2D histogram'' indicating the number of Gaussians with a particular height
($T_{\rm B,0}$) and Gaussian width ($\sigma_v$).  An intriguing structure is apparent in the distribution
of zero-velocity gas Gaussians in Figure \ref{gausshtsig}b following a 1/$\sigma_v$ trend (for 8 $\gtrsim \sigma_v \gtrsim $ 35 \kms
and 1 $\gtrsim T_{\rm B,0} \gtrsim$ 3 K) and nearly conserving its area at $\sim$35 K \kmse.  It is not clear what
this structure corresponds to.
The distributions of all Gaussians as a function of $l$, $b$, and $v_{\rm LSR}$
are shown in Figure \ref{allsky}.  The disk of the MW, the local zero-velocity MW gas,
and the Magellanic Clouds and Stream are readily apparent.

\subsection{Validity of the Gaussian Decomposition}
\label{subsec:validity}

It might be asked whether Gaussian decomposition is the correct way to analyze these data.
Though Gaussian decomposition has been widely used as a tool to analyze all forms of \hi data
going back to the 1960s (e.g., Kaper et al.~1966; Takakubo \& van Woerden 1966; Burton 1970; Schwarz \& van Woerden 1974;
see review by Haud 2000), it is only physically well-motivated for isolated, internally
virialized clouds, and here it finds its most common application (e.g., Br\"uns, Kerp \& Pagels 2001; Wakker, Oosterloo, \& Putman 2002;
Kalberla \& Haud 2006).
In contrast, here we are using Gaussian decomposition primarily as a tool to disentangling overlapping
\hi structures under the presumption that \hi structures along the line-of-sight only slowly vary those
properties encapsulated by a Gaussian description (velocity, position, velocity-dispersion, and integrated column density).

We have looked at various isolated \hi clouds in the LAB data and
found that they are well-fit by Gaussians.
Moreover, by using Gaussians we are able to disentangle different \hi filaments even
when they are overlapping in velocity.  In those situations it is clear that the Gaussian decomposition
traces structures that are real and they may even hold physical information about the structures.
We are still successful in tracking tenuous structures through even more complicated environments
even though the decompositions of those environments (local MW zero-velocity gas, the MW disk, and the Magellanic Clouds)
likely holds no physical meaning.
Whether or not it holds physical meaning, we are primarily interested in how the Gaussian decomposition
provides a representation of the data that enables us to track large features of the MS.  In addition
the decomposition allows us to reduce our data to a manageable size (from a datacube to a database of Gaussians).

There are a high number of Gaussians with low $T_{\rm B,0}$ and $\sigma_v$ which are most
likely due to fitting of noise features.  We make a simple parameter space boundary $T_{\rm B,0}  > -0.08\sigma_v + 0.45$
to remove these ``noise'' Gaussians.  The presence of possibly extraneous narrow or wide Gaussians
does not influence our interpretation of the LAB data, which is based on significant features in the database.

\section{Removal of Milky Way gas}
\label{sec:removemwgas}

In order to study the MS, we need to separate the Gaussians of MS gas from those of MW gas, which is
a particulary difficult problem at low $|b|$, and at any latitude when the velocity of the MS differs
little from the velocities expected for gas in the conventional MW disk.  The velocities
of the conventional MW disk gas roughly follow the expectations for material moving in
circular orbits around the Galaxy, i.e.

\begin{equation}\label{rotcurve}
V_{\rm LSR}(l,b,R) = \left[{R_0 \over R} V(R) - V_0 \right] \rm sin({\it l}) \rm cos({\it b})
\end{equation}

\noindent where $V_0$ and $R_0$ are the solar velocity and Galactocentric distance, respectively,
and $V(R)$ is the rotation curve at $R$.  At high latitudes the gas is (mostly) concentrated to
velocities of $V_{\rm LSR} \approx 0$ (zero-velocity gas) because the gas is (mostly) nearby, since we are looking
out of the plane of the disk, and therefore $R_0/R\sim$1.
Essentially, this gas is local ISM gas moving in nearly the same orbit as the Sun around the MW.
At lower latitudes the MW disk gas has a larger range of velocities of approximately,
$-V_0 {\rm cos}(b) \lesssim V_{\rm LSR} \lesssim +V_0 {\rm cos}(b)$, but of course there is a strong
$\rm sin({\it l})$ dependence.
Due to the different characteristics of high-latitude versus low-latitude MW gas 
different methods were employed to separate them from the MS gas.  It is worth emphasizing that
ultimately no conclusions in this paper depend on the details or ultimate accuracy of these population
decompositions; our care in pursuing these strategies is to make improved maps of the MS.

\subsection{Zero-Velocity Gas at High Latitudes}
\label{subsec:remzerovel}

At most positions in the sky, the zero-velocity MW disk gas is easily distinguishable from
the MS gas because they have very different velocities.  However, since the
MS stretches from $V_{\rm LSR}\approx300$ \kms to $V_{\rm LSR}\approx-400$ \kms
it must cross $V_{\rm LSR}=0$ at some point.  This happens in the region $-84^{\circ} \lesssim b \lesssim -78^{\circ}$
and $288^{\circ} \lesssim l \lesssim 327^{\circ}$.  In this area it becomes
challenging to distinguish the zero-velocity MW gas from the MS gas.  It is difficult to
disentangle the two populations of gas without some decomposition
scheme as we have used here.  Most earlier column density maps of the MS show gaps in this
region (see Figs.\ 4--5 in P03, and Fig.\ 2 in B05) which we intend to
remedy in our maps.

\begin{figure}
\includegraphics[scale=0.43]{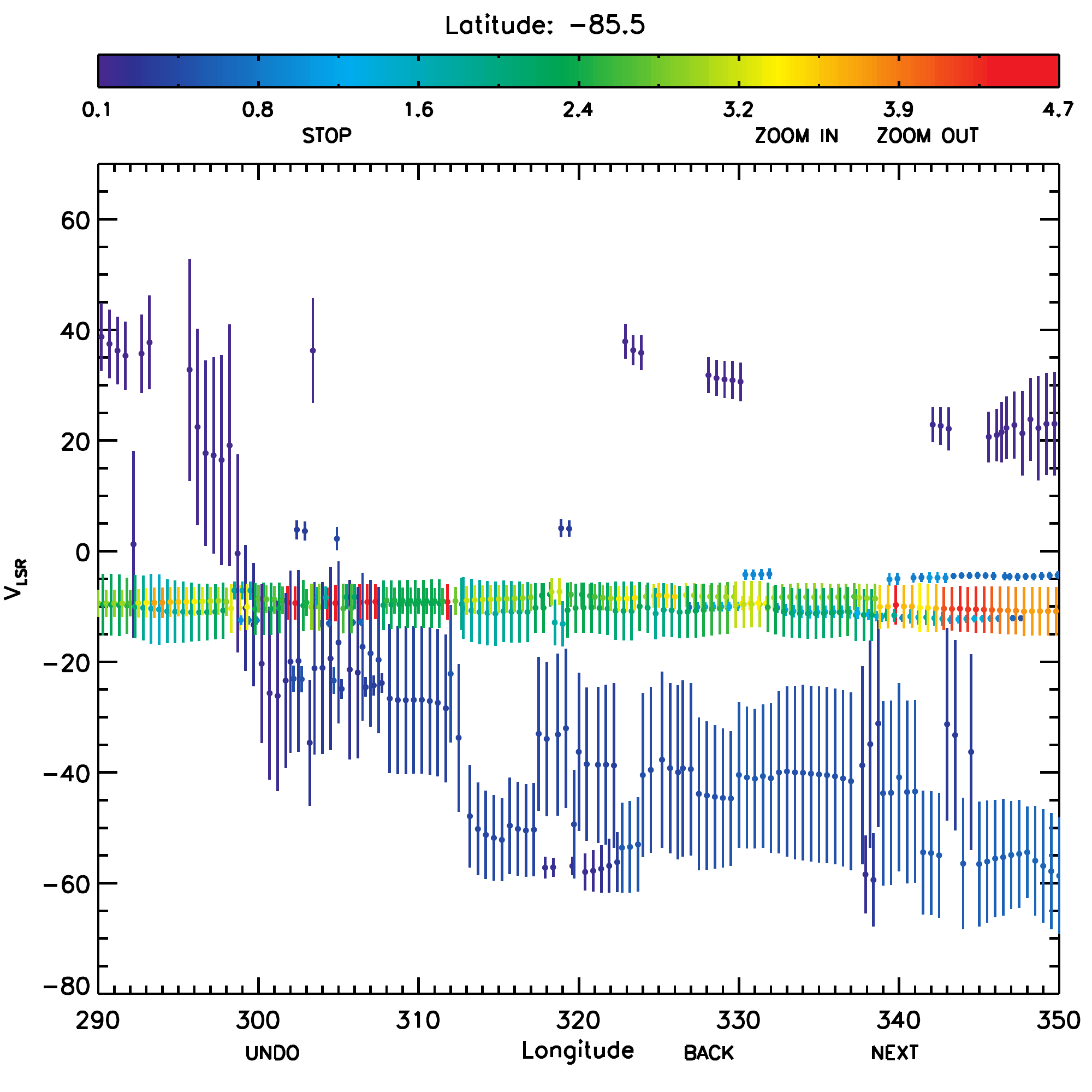}
\caption{A snapshot of the analysis used to remove the zero-velocity Milky Way Gaussians in
the region where they overlap the Magellanic Stream.  At a given Galactic latitude
the central velocity of each Gaussian is plotted against its Galactic
longitude. Each Gaussian is represented as a line, where its length
corresonds to $2\sigma_v$ and the color corresponds to $T_{\rm B}$ (the colorbar shows
the scale).  The continuity of features in position, velocity, $T_{\rm B}$, and
$\sigma_v$ is used to identify them as due either to local Milky-Way zero-velocity gas or to the Magellanic
Stream.  An interactive program was developed whereby any Gaussian can be removed from
the datacube by clicking on it on the display screen.  The labels ``STOP'', ``ZOOM IN'', etc.~are buttons
on the interactive display to perform actions during the Gaussian de-selection process.}
\label{zerorem}
\end{figure}

\begin{figure}
\includegraphics[scale=0.58]{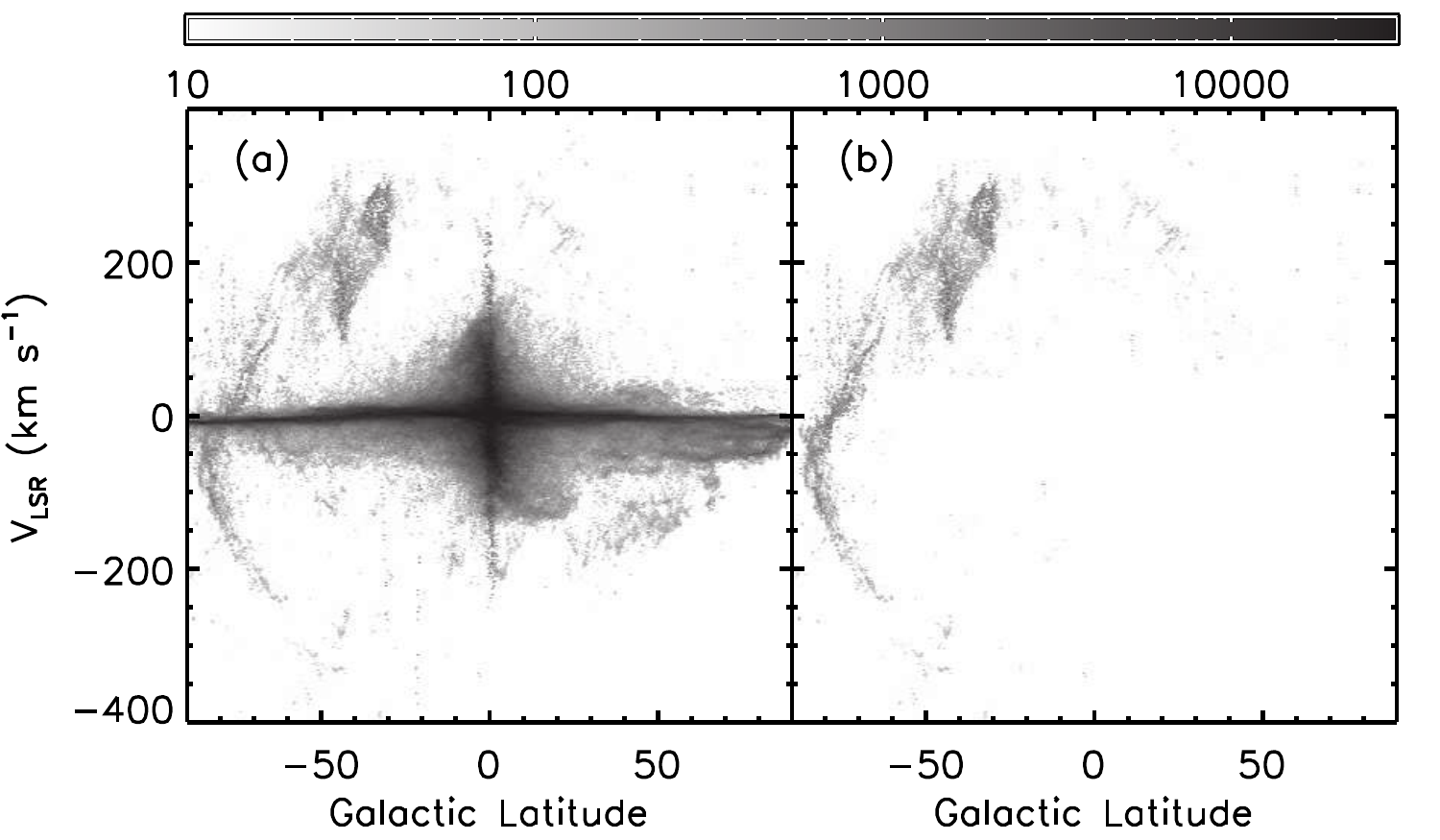}
\caption{\vlsr vs.~$b$ of \hi Gaussians (summed along $l$)
({\em a}) before the Milky Way and zero-velocity \hi gas has been removed, and ({\em b}) after it has been
removed. The greyscale is in units of K \kmse.}
\label{zerobeforeafter}
\end{figure}

We attempted to separate the zero-velocity MW Gaussians from the low-velocity MS Gaussians by
using the Gaussian parameters $\sigma_v$ and $T_{\rm B,0}$ alone, since velocity would not
be of much use in this case.
However the zero-velocity MW and MS Gaussians also overlap in $\sigma_v-T_{\rm B,0}$
space (see Fig.\ \ref{gausshtsig}) which makes this separation strategy untenable.
The zero-velocity gas was eventually removed interactively.  We made \vlsr vs.~$l$ plots
of all of the Gaussians at a single $b$ in the region where the MS crosses
$V_{\rm LSR}=0$.  Each Gaussian was represented by a vertical line, where
the length of the line corresponded to the Gaussian's 2$\sigma_v$ and the color of the line to 
its $T_{\rm B,0}$.  Figure \ref{zerorem} shows an example of one of these plots, and demonstrates
the relative ease with which such a representation makes it possible to distinguish the MS
Gaussians from MW Gaussians because of the nearly-constant, but different,  $T_{\rm B,0}$ and
$\sigma_v$ trends for each and the straight versus arcing trends that differentiate them.
An interactive program allows us to remove any Gaussian represented in this way by clicking on it.
All of the Gaussians consistent with being zero-velocity MW Gaussians were removed at a given
$b$, and the process was repeated for all $b$ where the MS and zero-velocity MW gas
overlapped.  This process was iterated a few times to ensure that no residual zero-velocity
patterns were left over.  The results of this zero-velocity MW Gaussian removal scheme can be seen in
Figure \ref{zerobeforeafter}.  For the regions where there is no overlap
between MS gas and Milky Way zero-velocity gas all Gaussians with $|$\vlsre$|<45$ \kms have been removed.
A separate scheme was used to remove the MW gas at somewhat higher velocities (i.e.~$|$\vlsre$|>45$ \kms
which is described in the next section.

\subsection{Milky Way Disk Gas at Low Latitudes}
\label{subsec:remlowlat}

\begin{figure*}
\includegraphics[scale=0.71]{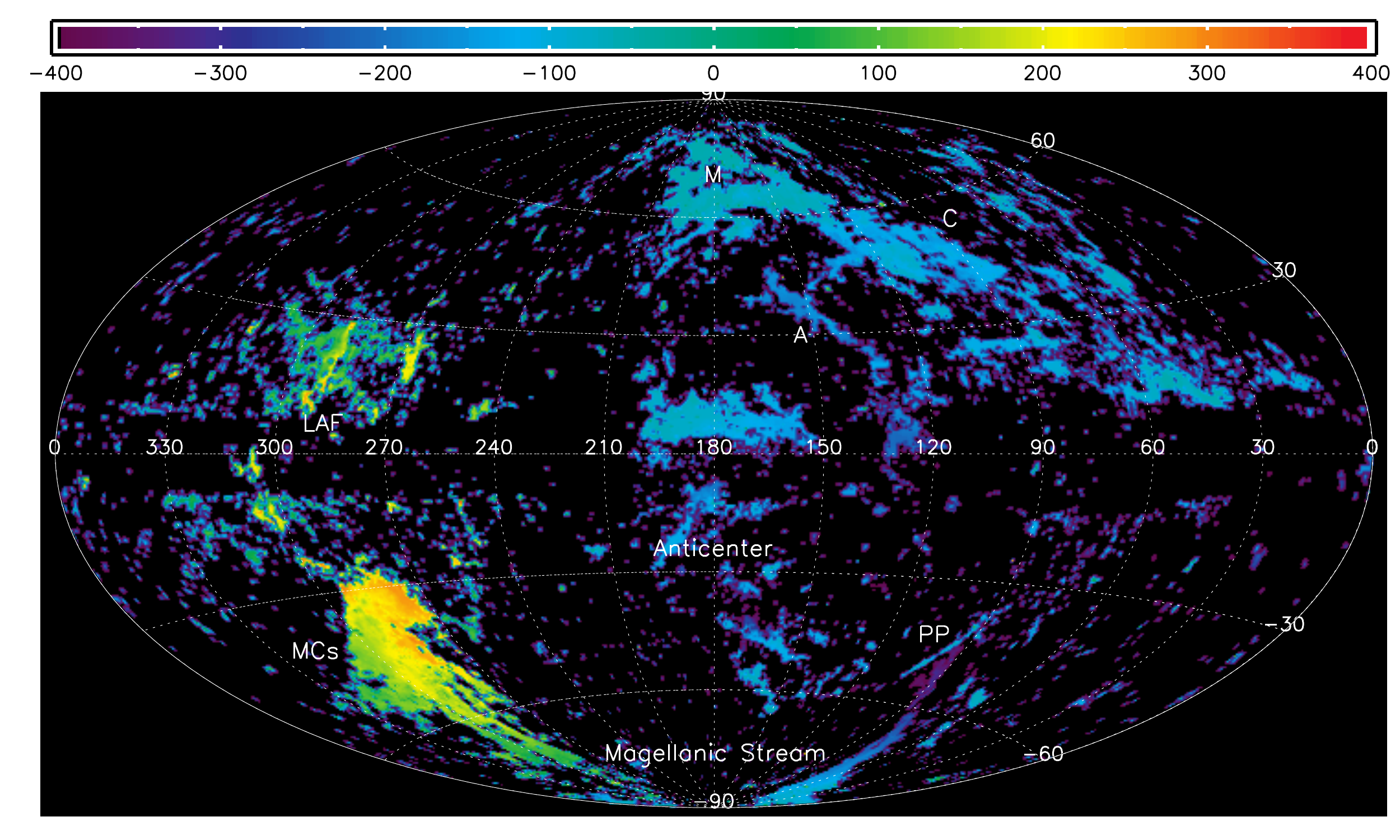}
\caption{Sky distribution of IVC and HVC \hi gas in Galactic coordinates, where the color represents \vlsr
in \kms and the individual Gaussian components are collapsed into single points (compare to Wakker 2004).
The Magellanic Clouds (lower left) and Stream (arcing across the bottom) as well as Complex C (upper right)
and other HVC complexes are clearly visible.}
\label{nonmwplots}
\end{figure*}

Since we were interested in investigating the Leading Arm, 
we also needed to remove the MW gas at low latitudes, because the Leading Arm passes through $b=0$\dgr.
We first attempted to 
use a simple, symmetric analytical model of the MW to remove the disk gas, but
this failed to remove a significant portion of MW gas because of the exaggerated simplicity of the model.
We instead adopted an empirical method to remove the disk gas.  At a given $l$ and
$V_{\rm LSR}$, the $T_{\rm B}$ due to the MW disk gas drops off  quickly with
$b$.  The basis for our empirical strategy is to find where this drop-off occurs and call that the end
of the disk.  We used the profiles from the original \hi datacube to accomplish this.  In
order to see the global trend and remove noise, each profile was smoothed
with a Savitzky-Golay [15,15,2] filter.  For each $l$ and $V_{\rm LSR}>$ 30 \kms
the ``edges'' of the disk (where it fell below 5\% of the central $T_{\rm B}$ value) in $b$ were found.  The
values of these ``edges'' were then used to remove all Gaussians that fell within this $b$ range at
that particular $l$ and $V_{\rm LSR}$.  In addition, all Gaussians with $V_{\rm LSR}<$ 30 \kms have been
automatically removed. 
The Gaussians left-over after the MW disk-gas removal (primarily HVCs, some IVCs, and MS gas)
can be seen in Figure \ref{nonmwplots}.
The distributions of the Gaussian parameters ($T_{\rm B,0}$, $\sigma_v$) of the MW disk
gas Gaussians and the left-over Gaussians are shown in Figure \ref{gausshtsig}c.

\subsection{General Features Observed in the HVC and IVC Distributions}

Though a thorough analysis of Figure \ref{nonmwplots} is beyond the scope of this paper,
several general characteristics of this figure are worth pointing out, especially in consideration of
similar all-sky maps produced earlier.  We specifically compare our figure to Figure 1a of \citet{Wakker04}.
Many of the same general features may be seen in both maps, but some structures are not consistently seen because 
different schemes were used to discriminate MW disk from other gas.
Wakker employed a symmetric, analytical model for the Milky Way disk, whereas we 
define and remove the Milky Way gas empirically.  
Morever, because of the two to four times higher net spatial resolution of the LAB data,
our Figure \ref{nonmwplots} map shows
some finer structures, and our colorscale levels reveal more detail because of the 
$>10$ times higher velocity resolution.
Most germane to the present discussion is that the tendril-like structure of the MS and Leading Arm Feature (LAF)
is more obvious in Figure \ref{nonmwplots} than in the Wakker maps.  The 
same kind of fine-structure is also seen in other HVC complexes in our maps (e.g., Complexes
A, C, and M).  In this regard, our data are more similar to those of P03 and B05,
with only slightly lower spatial resolution, equivalent or higher velocity resolution, 
but covering the entire sky. It is because of the higher velocity resolution data and the
analysis techniques we use to analyze them (\S \ref{sec:gaussdecomp}) that we are able
to build on P03's previous work on the MS.  The B05 data, which are higher in spatial resolution
than the LAB data, were unavailable to us when we began this analysis.

\section{Results of the Gaussian Decomposition}
\label{sec:results}

\subsection{Magellanic Stream Coordinate System}
\label{subsec:coordsystem}

Because the Magellanic Stream consistently follows such a long trace across the sky, it is useful to have
a coordinate system for which the equator lies along the great circle of the
MS.  A Magellanic coordinate system was defined by \citet{Wakker01},
in terms of the great circle along the $l=90^{\circ}$
and $l=270^{\circ}$ Galactic meridian.  Although this is close to the MS, the equator of
that coordinate system is not exactly along the Stream.
We define here a new coordinate system which we call
the ``Magellanic Stream'' coordinate system and whose equator more closely bisects the Magellanic
Stream.  The equator of the system was set by finding the great
circle best-fitting the MS.  The pole of this great circle is at 
$(l,b)=(188.5^{\circ},-7.5^{\circ})$; the longitude scale is defined in such a way that the
center of the LMC ($l,b=280.47^{\circ}, -32.75^{\circ}$: van der Marel et al.~2002)
has $L_{\rm MS}=0^{\circ}$.  Like the Magellanic coordinate system of Wakker, $L_{\rm MS}$ 
decreases along the MS (towards its tail).  Many subsequent figures
in this paper will use the Magellanic Stream coordinate system (\lmse, \bmse).

\subsection{Representations of the Magellanic Stream}

\begin{figure*}
\includegraphics[scale=0.95]{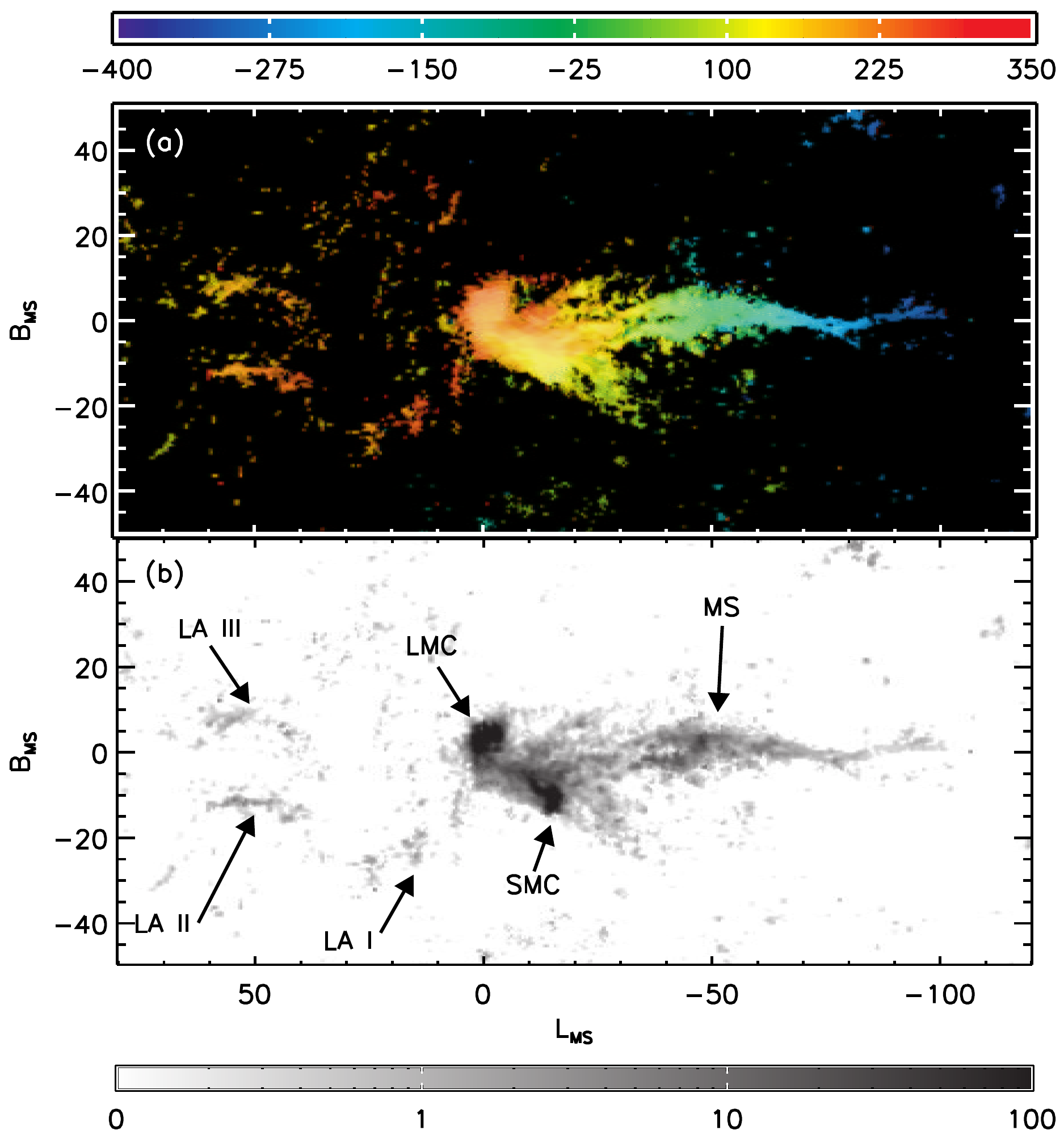}
\caption{The Magellanic Clouds and Stream \hi Gaussians as distributed on the sky. ({\em a})
Hue indicates $<$\vlsre$>$ and intensity indicates \nhi (on a logarithmic scale).
({\em b})  The three Leading Arm complexes I--III (including the three ``clumps'' of LA I clearly seen in
panel ({\em a})), the Large Magellanic Cloud (LMC), Small Magellanic Cloud (SMC), and the Magellanic
Stream (MS) are shown in this greyscale representation of \hi column density, \nhie, in units
of $10^{19}$ atoms cm$^{-2}$.}
\label{msskyplot}
\end{figure*}

\begin{figure*}
\includegraphics[scale=0.85]{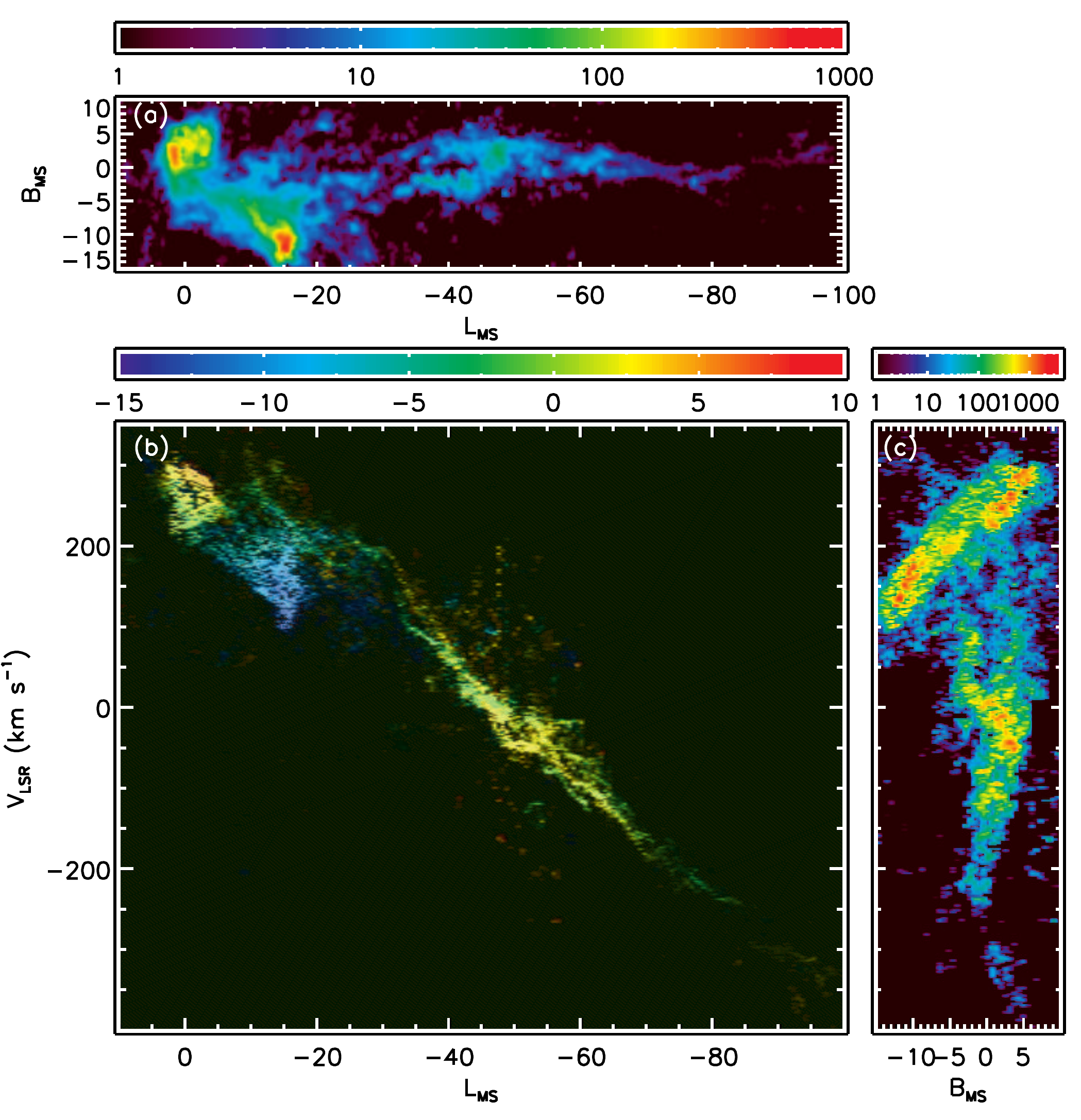}
\caption{Integrated intensity (sum of Gaussian areas) of the Magellanic Cloud and Stream \hi Gaussians (at their
central \vlsr velocity). ({\em a})
Column density, \nhie, in units of $10^{19}$ atoms cm$^{-2}$ (a blowup of Figure \ref{msskyplot}b).  A spatial
bifurcation of the Magellanic Stream into two filaments (first pointed out by P03) can be seen for
$-40$\dgr $\lesssim$ \lms $\lesssim-20$\degr.  Other bifurcations are seen farther down the Stream.  ({\em b})
\vlsr vs.~\lms (hue indicates \bmse, and brightness indicates integrated intensity along
\bmse).  This panel shows the two filaments of the Magellanic Stream also to be bifurcated in
velocity from $-40$\dgr $<$ \lms $< -20$\degr.  One of the filaments is discernible all the way
to (\lmse, \vlsre) $\approx$ ($-5$, $+247$ \kmse) where it connects to the LMC.  The second filament can be\
followed only to (\lmse, \vlsre) $\approx$ ($-16.5$, $+220$ \kmse).  More velocity bifurcations (or multiple
splits) are evident at more negative velocities.  The two filaments show strong periodic patterns
for $-40$\dgr $\lesssim$ \lms $\lesssim-5$\degr, after which they follow a fairly linear negative velocity gradient
($-95 \lesssim$ \lms $\lesssim-40$\degr).
({\em c}) Integrated intensity of \hi Gaussians in \vlsr vs.~\bms (colorscale in units of K \kmse).  The bifurcation
of the MS is discernible as well as some spiraling patterns of the filaments.}
\label{msprofile}
\end{figure*}

The results of the Gaussian decomposition, with the MW disk gas and local
zero-velocity gas removed, can be seen in Figures \ref{msskyplot} and \ref{msprofile}, which
show integrated intensity of the Gaussians in three
perspectives (\bms vs.~\lmse, \vlsr vs.~\lmse, \vlsr vs.~\bmse).
Figure \ref{msskyplot}a also shows the Gaussians of the Magellanic System in 
\bms vs.~\lms with color representing $<$\vlsre$>$.
The large velocity gradient is evident in this figure.
Figure \ref{msprofile}b similarly shows the sum of Gaussian centers of the Magellanic System in 
\vlsr vs.~\lms with color representing $<$\bmse$>$.
Some readily apparent, prominent features in these figures, such as the
LMC, SMC, the MS, and the LAF are labeled in Figure \ref{msskyplot}.

\subsection{The Two Filaments of the Magellanic Stream and their Source}
\label{subsec:twofil}

\begin{figure}
\includegraphics[scale=0.68]{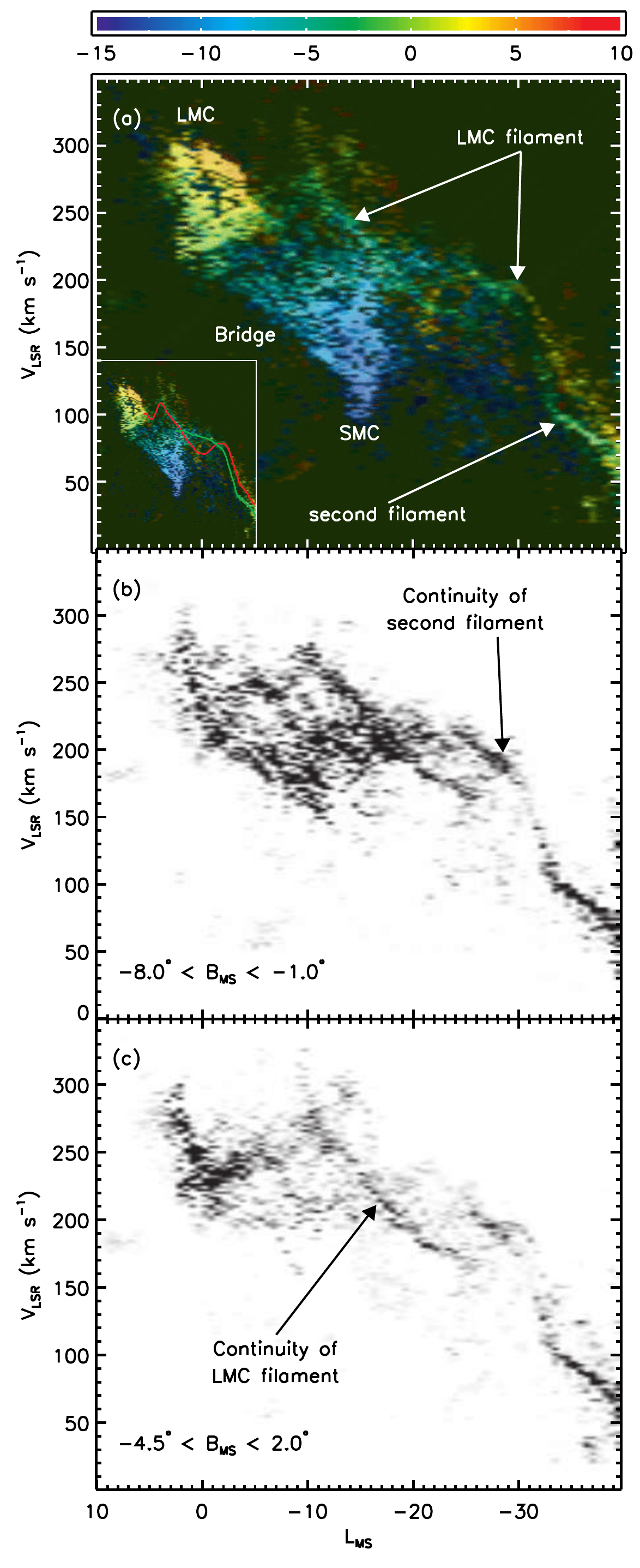}
\caption{({\em a}) Close up of the \vlsr vs.~\lms distribution of Magellanic Clouds and Stream \hi Gaussians from
Figure \ref{msprofile}b (hue indicates \bmse, and brightness indicates integrated intensity along \bmse).
The bifurcation of the two Magellanic Stream filaments is clearly discernable as are strong
periodic patterns in the radial velocities of the filaments.  The inset illustrates the shape of the
two filaments (red -- LMC filament, green -- second filament).  ({\em b}) A narrower range in \bms to show that
the second filament is continuous at the point (near \lms $\approx -28$\degr) where the two filaments cross
in this projection (on a linear scale from 0--100 K \kms [white to black]).
({\em c}) A slightly different range in \bms to indicate that the LMC filament is
continuous as it crosses the SMC/Bridge gas near \lms $\approx -16$\dgr
(on a linear scale from 0--200 K \kms [white to black])}
\label{ms_vmlon_zoom}
\end{figure}

The two-filament structure of the MS, previously pointed out by Cohen (1982) and by Morras (1983), and
studied recently by P03,
is clearly visible and separated in our datacube of Gaussian centers.
The MS filaments can be distinguished  between
$-40^{\circ}<L_{\rm MS}<0^{\circ}$ in the \vlsr vs.~\lms plot (Fig.\ \ref{msprofile}b).
For \lms$<-45$\dgr it becomes difficult to disentangle the two MS filaments in any projection of the data
and we therefore focus here on the \lms$>-45$\dgr region of the Magellanic System.
Figure \ref{msprofile} allows us to distinguish both filaments in the head of the Stream.
Until now the Stream was only known to be {\it spatially} bifurcated
and since the Magellanic Bridge gas overlaps the MS filaments near the Clouds it was
not possible to distinguish or separate the filaments for \lms$\gtrsim-20$\degr.
The {\it velocity} bifurcation at the head of the Stream allows us to trace the filaments
back further to their source than previously possible.  One striking characteristic of the filaments is
their oscillating pattern (Fig.\ \ref{msprofile}b and \ref{ms_vmlon_zoom}), which is discussed further in
\S \ref{subsec:periodic}.

Figure \ref{ms_vmlon_zoom}a is a close-up view of Figure \ref{msprofile}b and gives a clearer picture of
the two filaments (traced with red and green lines in the lower-left inset).  The filaments cross near
\lms $\approx -28$\dgr, but a narrower \bms range ($-8.0$\dgr$ < $ \bms $ < -1.0$\degr) shows the
continuity of the ``green'' filament at this point (Fig.\ \ref{ms_vmlon_zoom}b) and allows us to track the
filament across this longitude.  Near \lms $\approx -16$\dgr the ``red'' filament
crosses the SMC/Bridge gas.  Another \bms range ($-4.5$\dgr$ < $ \bms $ < 2.0$\degr) reveals the continuity
of the ``red'' filament through this region (Fig.\ \ref{ms_vmlon_zoom}c).  Beyond this point the ``red''
filament connects to the LMC (hereafter the ``red'' filament will be called the ``LMC'' filament and
the ``green'' filament the ``second'' filament).
\citet{Put98} pointed out an ``emission filament'' emanating from the LMC (seen in the sky channel maps),
but claimed it went into the Bridge; \citet{McGee86} earlier also saw possible indications of an LMC filament.
However, the P--V representation of Figure \ref{ms_vmlon_zoom} may be the
first conclusive evidence that any part of the MS comes from the LMC.
P03 claimed that the two MS filaments came from the SMC and Bridge, and most subsequent tidal simulations
(Connors et al.~2006, and references therein) have adopted these origins as
their starting point (e.g., by modeling the SMC as an N-body and the LMC as a rigid potential).  We argue here 
that the assumptions that are the foundation of these simulations need to be reconsidered.

\begin{figure}
\includegraphics[scale=0.49]{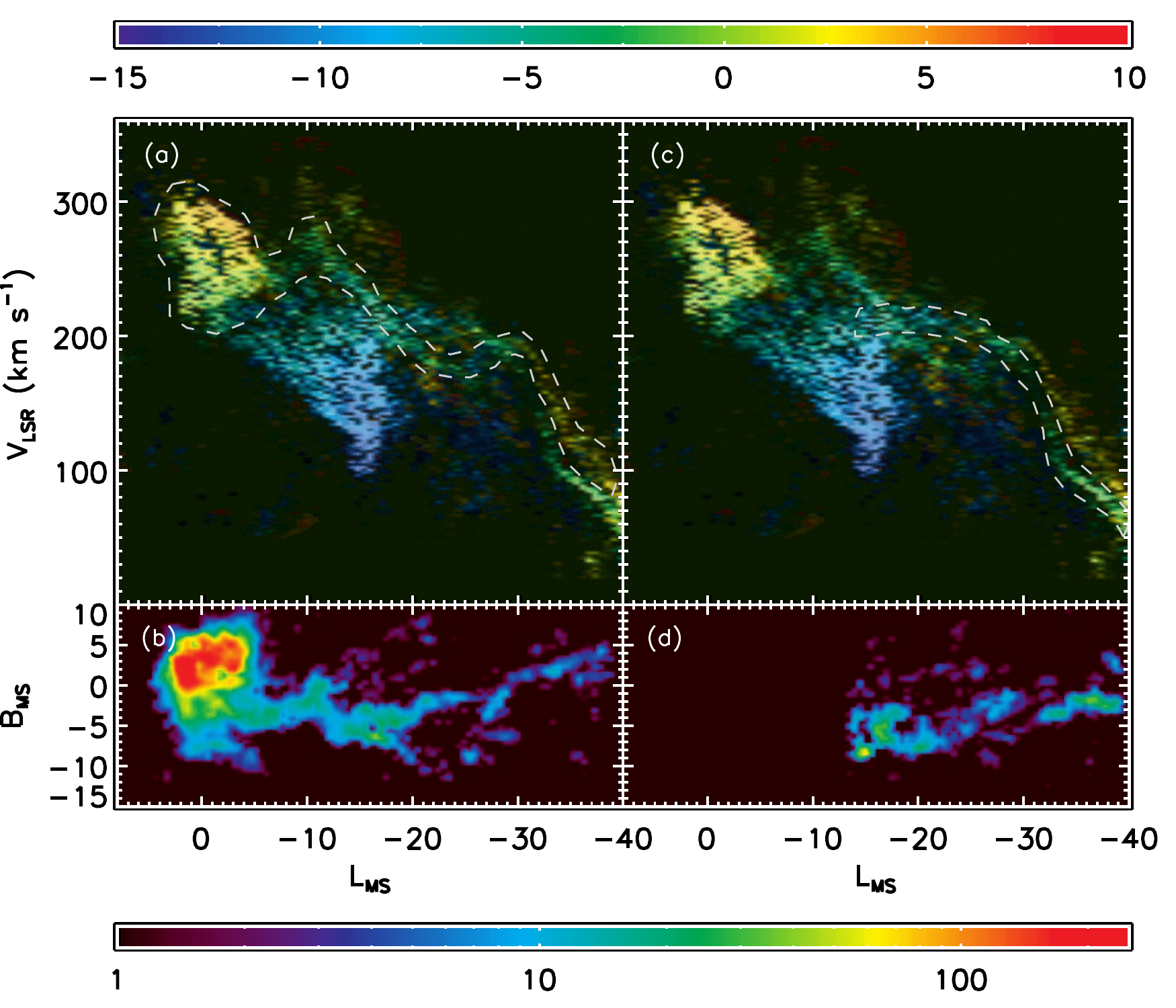}
\caption{The two Magellanic Stream filaments isolated by velocity.
{\bf The LMC filament:}
({\em a})  \vlsr vs~\lms distribution for the Magellanic Cloud and Stream \hi Gaussians showing two Magellanic Stream
filaments (same as Fig.\ \ref{msprofile}b).  The gray dashed lines show the velocity limits used to isolate the
LMC filament.
({\em b}) Sky distribution of the column density, \nhie, of the \hi Gaussians for the LMC filament
selected by the velocity limits shown in panel ({\em a}).
The association of the LMC filament with the LMC and the spatial periodic patterns are apparent.  
{\bf The second filament:}
({\em c})  \vlsr vs.~\lms for the Magellanic Cloud and Stream \hi Gaussians showing the two filaments (same
as Figure \ref{msprofile}b).  The gray dashed lines show the velocity limits used to isolate the second filament.
({\em d}) As in panel ({\em b}), but for the second filament selected by the velocity limits shown in panel ({\em c}).
The second filament can only be distinguished for \vlsr $\lesssim -17$\dgr and its source remains unclear.
The top color bar indicates \bms for panels ({\em a}) and ({\em c}), while the bottom color bar indicates
column density, \nhi (in units of $10^{19}$ atoms cm$^{-2}$), for panels ({\em b}) and ({\em d}).}
\label{ms_onskycut}
\end{figure}

\begin{figure}
\includegraphics[scale=0.62]{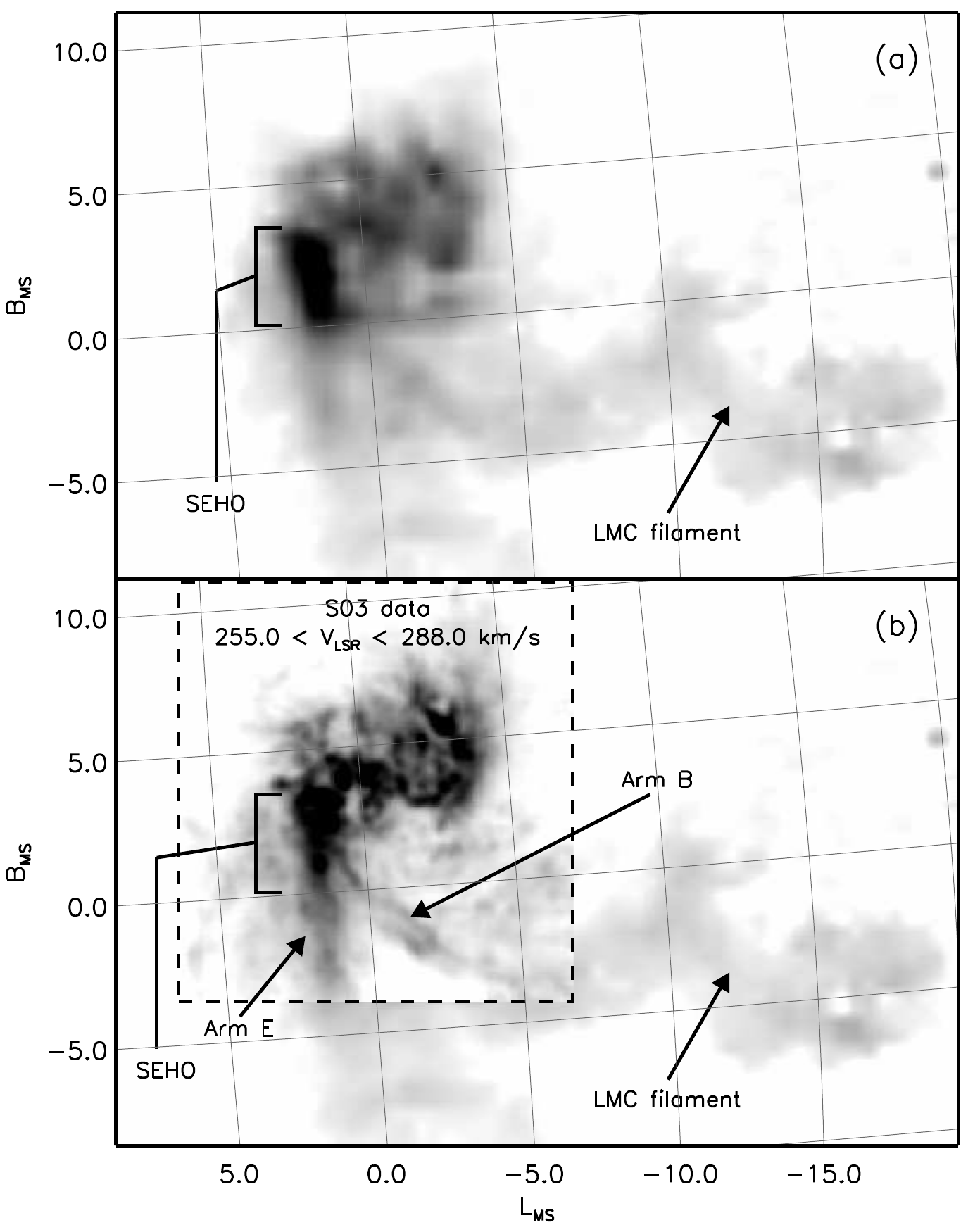}
\caption{Close-up of the Fig.\ \ref{ms_onskycut}b map of the integrated intensity (sum of Gaussian areas)
of the LMC and LMC filament \hi Gaussians on the sky after a velocity filter (see Fig.\ \ref{ms_onskycut}a)
is applied.  These maps show that the LMC filament is emanating from the SEHO in
the LMC when viewed with either ({\em a}) the LAB data only, or ({\em b}) the high-resolution
\hi data (255.0 $<$ \vlsr $<$ 288.0 \kmse) from S03 are substituted in the region outlined
by the dashed lined box.  The filament can be associated with S03 arm B. A square root transfer function is used
in these greyscale images.}
\label{lmcfil_source}
\end{figure}

In order to track the LMC filament back to the location of its origin in the LMC, we isolated the filament
in the \vlsr vs.~\lms plot by eye with the dashed lines shown in Figure \ref{ms_onskycut}a.
Figure \ref{ms_onskycut}b shows the distribution of the LMC filament on the sky.  The filament
appears to have a spatially periodic pattern.  The filament
emanates from a region of dense \hi on the southeastern, or leading, edge of the LMC, namely the SEHO
(throughout this paper we will use the term SE \hi overdensity to 
mean the entire region of high--density \hi in the southeast of
the LMC, 05\h 34\m $\lesssim$ $\alpha$ $\lesssim$ 05\h 52\m, $-68$\degr 28\arcmin $\lesssim$ $\delta$
$\lesssim$ $-71$\degr 53\arcmin).
The connection to the LMC can be even more clearly seen by overlaying the high spatial resolution \hi data of the LMC
from Staveley-Smith et al.~(2003; hereafter S03) on our map (Fig.\ \ref{lmcfil_source}b).  The LMC filament is
clearly arm ``B'' seen by S03.  

The SEHO is a natural place for an \hi stream to originate,
due to the high-density of \hi there (the highest concentration of \hi in the LMC).
Furthermore, the SEHO is near the end of the LMC bar,
is rich in CO \citep{yama01}, H$\alpha$ emission \citep{Kim99},
giant molecular clouds \citep{yama01}, and young star clusters \citep{Bica99}.
There are also several supergiant shells \citep{Kim99} and two CO filaments \citep{yama01} in this region.
Their relation to the MS is further explored in \S \ref{sec:outflow}.

The origin of the second filament is not as well defined.  In our maps (Figs.\ \ref{msprofile} and
\ref{ms_vmlon_zoom}) the second filament can only be clearly traced to higher longitude as far as the
Magellanic Bridge (near \lms$\approx -15$\degr, \vlsr$\approx +200$ \kmse).
It is not clear whether the second filament eventually connects to the SMC, to the Bridge, or to the LMC.
We used a by-eye selection in the \vlsr vs.~\lms plot (similar to the one used for
the LMC filament above) to extract the second filament gas from the database of Gaussians (the dashed lines in
Fig.\ \ref{ms_onskycut}c); its distribution on the sky is shown in Fig.\ \ref{ms_onskycut}d.

\begin{figure*}
\includegraphics[scale=1.04]{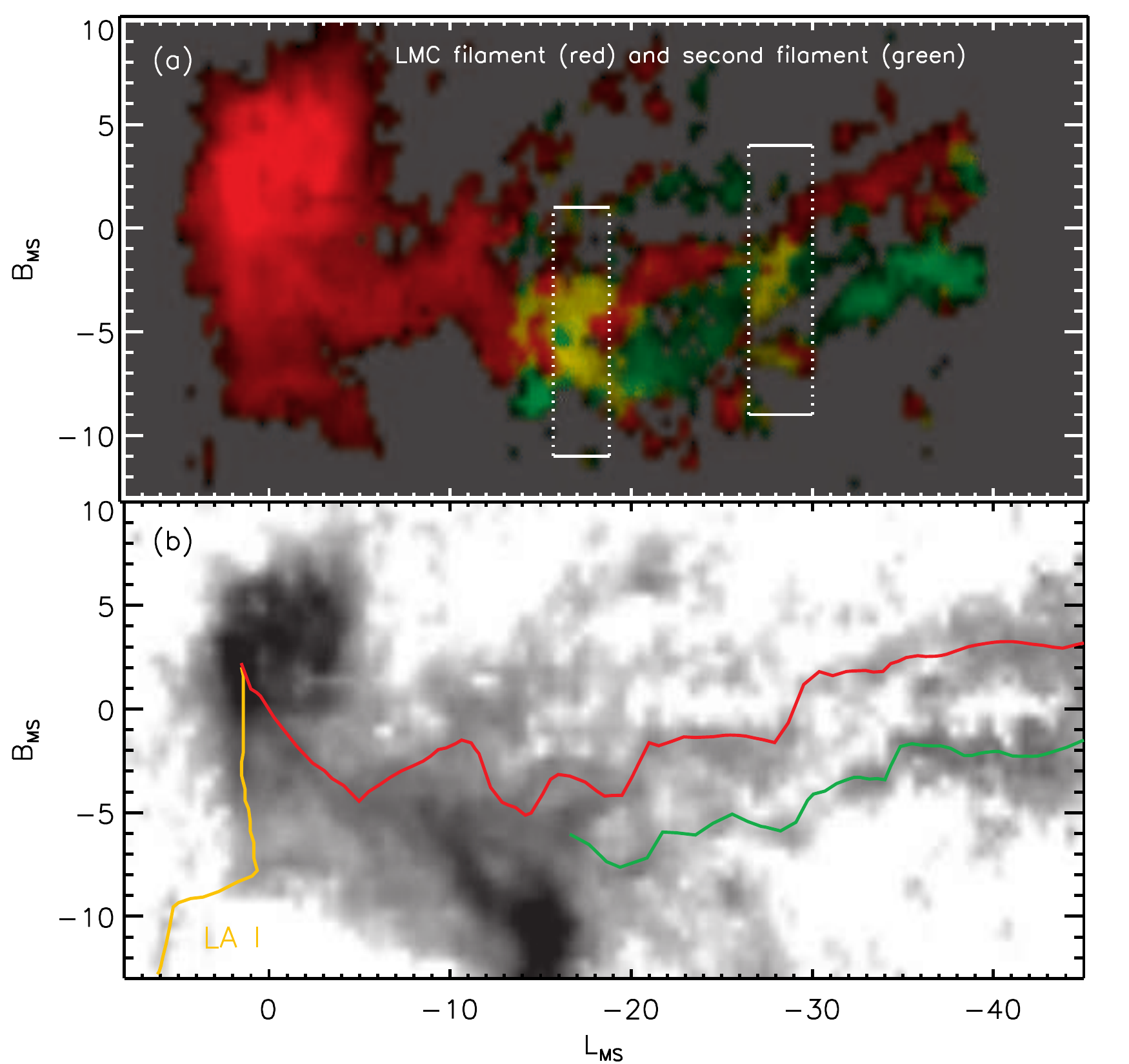}
\caption{({\em a}) The sky distribution of the two Magellanic Stream filaments as we have extracted them with velocity
filters shown in Figure \ref{ms_onskycut}a,c (using a logarithmic transfer function for the intensity).  The LMC
filament is shown in red and the second filament in green.  Most of the LMC is included in the velocity
filter of the LMC filament.  The two velocity filters (and the filaments) overlap in 
\vlsr and \lms near \lms$\sim-$17\dgr and \lms$\sim-$28\dgr (as indicated by the dotted lines).
Therefore, in these regions the separation of the filaments using the velocity selection {\em alone} 
is not as good.  However, in these regions the filaments are spatially well-separated
(the continuity of the two filaments in these overlap regions can also be seen in Figs.\ \ref{ms_vmlon_zoom}b and c).
({\em b}) Column density, \nhie, of the \hi Gaussians (on a logarithmic scale from 1--200$\times 10^{19}$
atoms cm$^{-2}$ [white to black]).  The ``guiding'' centers of the LMC filament (red), the
second filament (green), and LA I (orange, discussed in \S \ref{subsec:laorigins}) are overplotted
in order to indicate their positions relative to more familiar structures (e.g., SMC, Bridge, etc.).}
\label{sep_fils}
\end{figure*}

Figure \ref{sep_fils}a compares the positions of the two filaments on the sky (after the velocity selections
in Fig.\ \ref{ms_onskycut}a and c have been applied).  The filaments can be more
easily distinguished in this representation because the Bridge gas has been removed.  
The two velocity filters (and the filaments themselves) overlap in \vlsr and \lms near
\lms$\sim-$17\dgr and \lms$\sim-$28\dgr (as indicated by the dotted lines).
Therefore, in these regions the separation of the filaments using the velocity selection {\em alone} 
is not as good, but, fortunately, the filaments are spatially well-separated in these regions.
The continuity of the two filaments in these overlap regions can also be seen in Figure \ref{ms_vmlon_zoom}b and c.
The patterns of the two filaments on the sky are similar (also see Fig.\ \ref{msskyplot}b),
shifted by $\sim$5\dgr in \lms and $\sim$1\dgr in \bmse.  Possible physical
explanations for these patterns are discussed in \S \ref{subsec:periodic}.
Figure \ref{sep_fils}b shows the \hi column density on the sky of the Magellanic Gaussians with
the ``guiding'' centers of the LMC filament (red), the second filament (green), and the connection of LA I to
the SEHO (orange, discussed in \S \ref{subsec:laorigins}) overplotted in order to
indicate their positions relative to more familiar structures (e.g., SMC, Bridge, etc.).

\subsection{The Source of the Leading Arm Feature (LAF)}
\label{subsec:laorigins}

The LAF consists of three complexes of gas north of the MCs.
Pieces of these structures were first seen by van Kuilenburg (1972)
and \citet{WWW72}. Although there was no direct
connection, \citet{Math74} suggested a possible association of these features to the MCs and this 
hypothesis was further explored by \citet{Math79}, \citet{Morras82}, and \citet{Bajaja89}.
With reprocessed HIPASS data \citet{Put98} demonstrated that the
LAF gas is indeed associated with the Magellanic Clouds and is an extension 
of the Magellanic Stream.
The association of the LAF with the Magellanic Clouds and the MS lent support
to the tidal origin of the MS over the ram-pressure model, because a leading
feature of the MS would {\it not} be expected if ram-pressure were the dominant force.

The three primary complexes of the LAF can be  seen in Figure \ref{msskyplot} at positive \lmse:
LA I: ($3^{\circ} <$ \lms $< 29^{\circ}$, $-34^{\circ} <$ \bms $< -6^{\circ}$);
LA II: ($36^{\circ} <$ \lms $< 61^{\circ}$, $-17^{\circ} <$ \bms $< -10^{\circ}$);
and LA III: ($35^{\circ} <$ \lms $< 62^{\circ}$, $-2^{\circ} <$ \bms $< 11^{\circ}$)
(nomenclature by B05).  Our analysis here focuses on LA I, the closest LAF complex to the Magellanic Clouds.
LA I consists of three nearly rectangular $\sim$2.5$^{\circ} \times 8^{\circ}$ 
concentrations of gas that each lie almost parallel to lines of constant \lmse, and, combined, form a linear structure
making an $\sim$40$^{\circ}$ angle with the \bmse=0\dgr line (Fig.\ \ref{msskyplot}). \citet{Put98} showed that the two
concentrations nearest the LMC are nearly continuous (see their Fig.\ 3) and it is therefore likely
that the entire LA I feature is a physically connected structure.
The first concentration of LA I is close to the south--eastern edge of the LMC, both in
position and in radial velocity.
\citet{Put98} claim, however, that the Leading Arm material comes mainly from the SMC, based on a
filamentary feature that is nearly parallel with the \bmse=0\dgr line and that begins
near the SMC and stretches to the first concentration of LA I (see their Fig.\ 1).
However, S03 noted several \hi features of the LMC (arms B, E, S, and W) and remarked that arm E pointed
to the Leading Arm clouds, which lay beyond the coverage of their survey.  S03 go on to say that
deep, reprocessed HIPASS data (P03) shows a continuous connection between
arm E and the Leading Arm.  Nevertheless, S03 conclude that the Leading Arm gas
mainly arises from the SMC, and only some LMC gas ``leaks'' into the Leading Arm.
B05 showed that the first concentration of LA I is directly connected in position and
velocity to the \hi clouds close to the LMC, but claim that it is associated with the
Bridge.  We believe that there is a much firmer association of the Leading Arm with the LMC.

\begin{figure}
\includegraphics[scale=0.98]{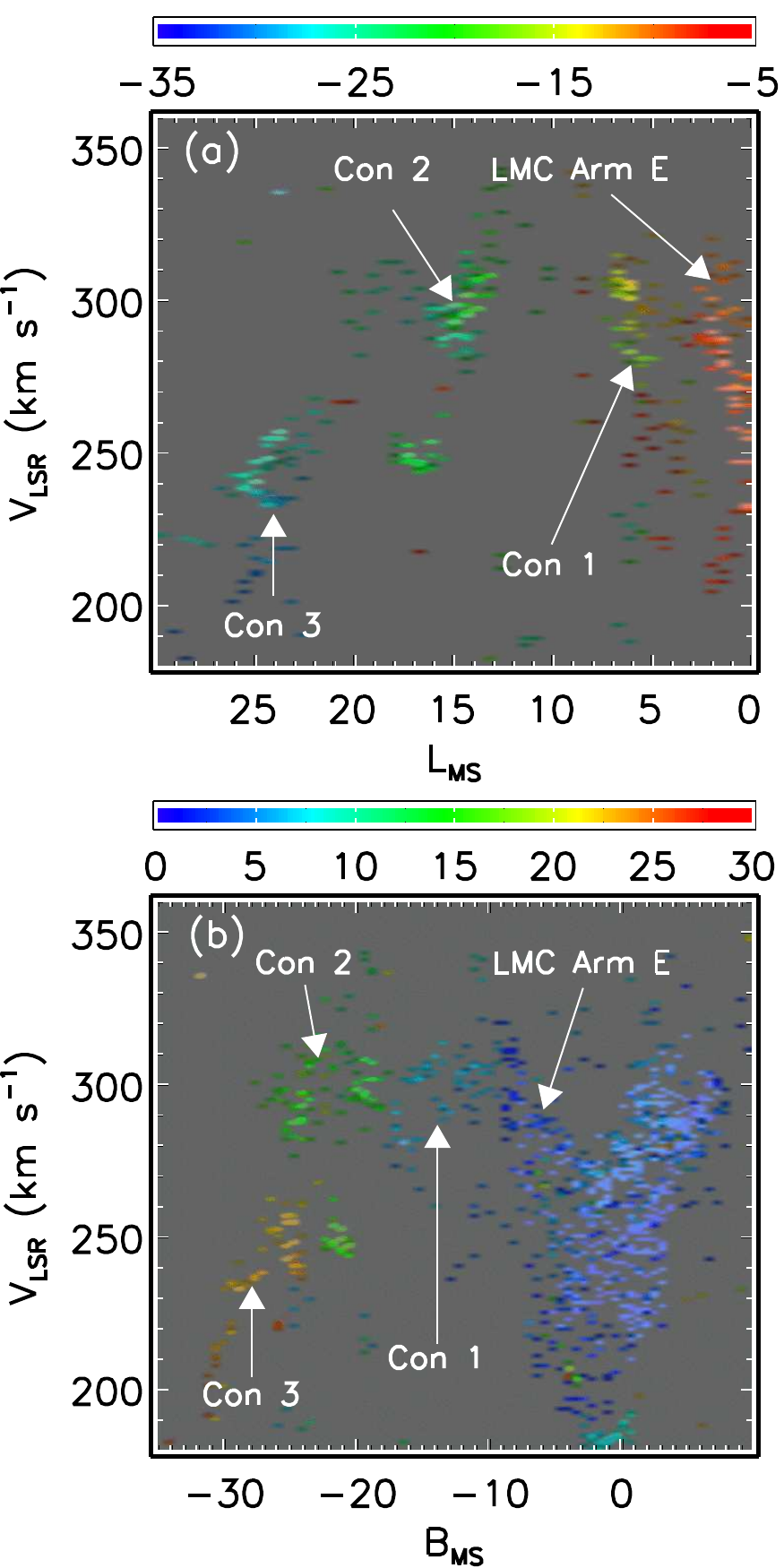}
\caption{Integrated intensity distributions of the Leading Arm Feature (LA I) and some LMC \hi Gaussians:
({\em a}) \vlsr vs.~\lms distribution (hue indicates $<$\bmse$>$, brightness indicates integrated intensity
along \bmse), and ({\em b}) \vlsr vs.~\bms (hue indicates $<$\lmse$>$, and brightness indicates integrated
intensity along \lmse).  The three concentrations and the LMC arm E are labeled (also see Fig.\ \ref{la_onsky_lacut}).
These two figures show that the first concentration of LA I ($-17$\dgr $\lesssim$ \bms $\lesssim -10$\degr) connects
in position and radial velocity to an extension of the eastern part of the LMC ($-10$\dgr $\lesssim$
\bms $\lesssim -5$\degr), which is arm E of \citet{ss03}.}
\label{la_vmlon}
\end{figure}

\begin{figure}
\includegraphics[scale=0.78]{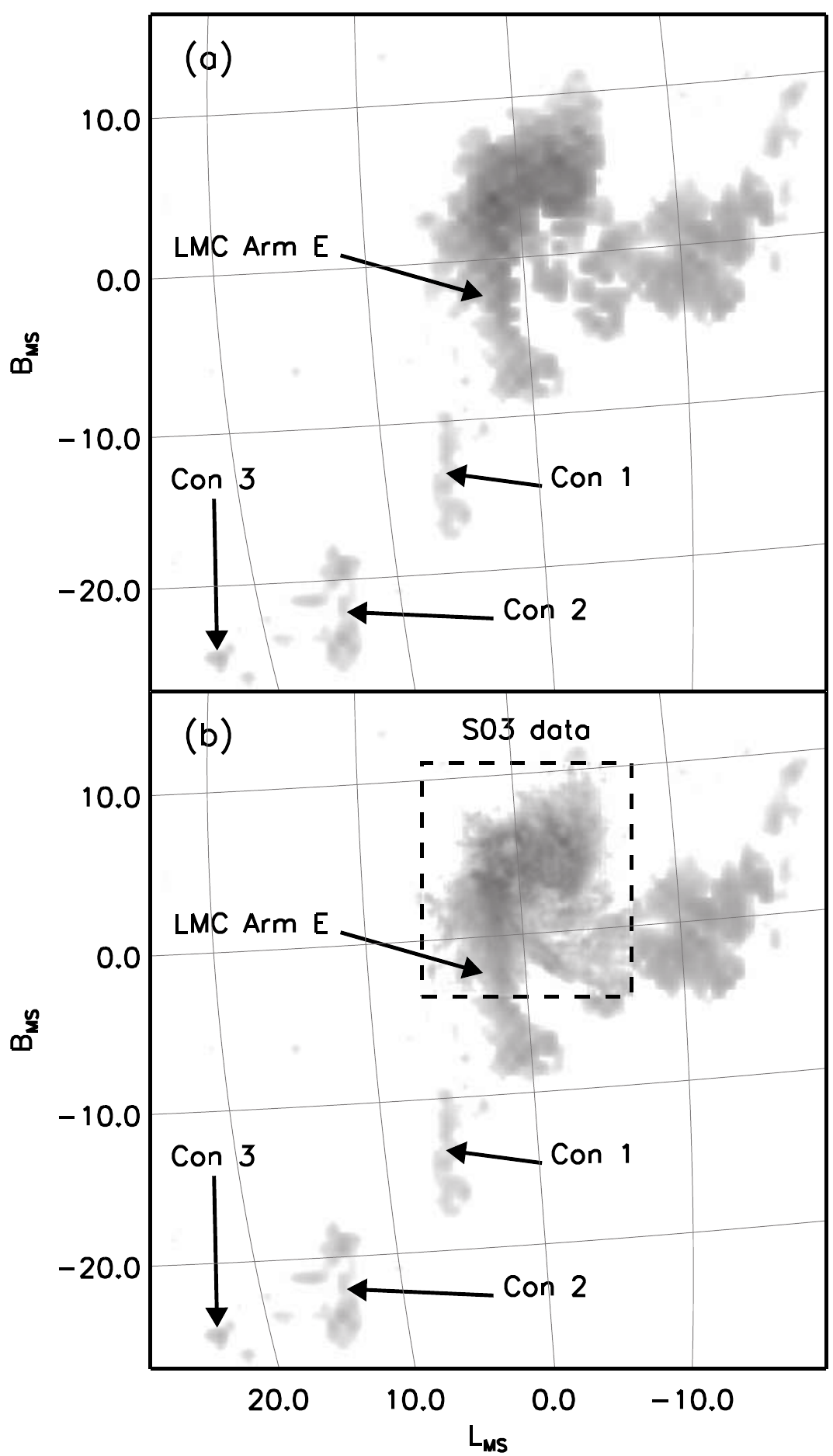}
\caption{Maps of integrated intensity (sum of Gaussian areas) of the Leading Arm Feature (LA I) and the LMC \hi
Gaussians on the sky for gas with $250 < $ \vlsr $ < 320$ \kms (on a logarithmic scale): ({\em a}) LAB data
only; ({\em b}) LAB data augmented with the high-resolution \hi data from \citet{ss03} in the dashed lined box.
The three concentrations of LA I and the LMC arm E are marked.  These maps show that the first concentration
of LA I is close to the southern end of LMC arm E.  Arm E (which originates in the SEHO) is similar in shape to the
three concentrations of LA I and continues the spatial progression.  This indicates that LA I originates in
the SEHO of the LMC.}
\label{la_onsky_lacut}
\end{figure}

The radial velocities of the LA I complexes are quite similar to those of the LMC (Fig.\ \ref{allsky}b),
but typically some $\sim$100 \kms higher than those of the SMC.  This
can be seen especially well in Figure \ref{la_vmlon}b, where the first concentration of LA I
($-17$\dgr $\lesssim$ \bms $\lesssim -10$\degr) is connected in position and velocity to an extension of 
the LMC ($-10$\dgr $\lesssim$ \bms $\lesssim -5$\degr), namely S03's arm E.  To better probe the association
of LA I to the LMC we show the distribution of gas in the velocity range 250 $<$ \vlsr $<$ 320 \kms
in Figure \ref{la_onsky_lacut}a.  In Figure \ref{la_onsky_lacut}b we also overlay the S03 high
resolution ATCA \hi data over our own (using the same velocity cut) to confirm the basic geometry seen
in the LAB data, namely S03's arm E extending out of the LMC and towards the first two concentrations of LA I
(which are not covered by the S03 survey).
Arm E has the same elongated shape (parallel to \bmse) and continues the spatial progression of
the three LA concentrations (more negative \bms towards higher \lmse).
Arm E also continues the velocity trend with \bms as seen in Figure \ref{la_vmlon}b.
There is a gap of a few degrees between the end of arm E and the beginning of the first concentration of
LA I (although there is a small clump of gas between them at [\lmse, \bmse]$\approx$[4\degr,$-$11\degr]).
However, our data also show gaps between the
three concentrations of LA I that the deeper HIPASS data show are contiguous.  Therefore, it is likely
that arm E and the first concentration of LA I are also connected.  For all of these reasons, we strongly suspect
that LA I is physically connected to arm E (which starts in the SEHO) and has its origins
in the LMC.  Therefore, we conclude that both the trailing LMC filament of the MS and LA I have their
origin in the SEHO of the LMC.  We discuss the implications of these findings further in \S \ref{sec:outflow}.

As previously mentioned, \citet{Put98} argue that the LAF comes from the SMC, based on a horizontal feature
south of the LMC which seems to connect the SMC to the Leading Arm (see their Figs.\ 1 and 2).  This
feature is extended nearly parallel to the \bmse=0\dgr line and the first of these features is
also visible in our maps (see Fig.\ \ref{msprofile}a; $-10$\dgr $\lesssim$ \lms $\lesssim$ 1\dgr and
$-8$\dgr $\lesssim$ \bms $\lesssim$ $-7$\degr)
and does connect to the end of arm E in position as well as in velocity.
It is this apparent connection, as well as the presence of the other ``horizontal'' 
cloud between the first feature and the SMC,
that is the basis for Putman et al.'s claim that the LAF originates in the SMC.
We cannot definitively refute this hypothesis but believe that the geometry of the gas
in LA I and arm E is more suggestive of the LMC as the originator of the LAF:
whereas arm E is already oriented in the same direction as the LA I concentrations and continues
their staggered vertical striping spatial pattern (Figs.\ \ref{msskyplot} and \ref{la_onsky_lacut}),
the Putman et al.~horizontal features are oriented orthogonally.

If this gas is moving towards the LMC, then it might be possible that 
the SMC is contributing some gas to the Leading Arm
or creating its own leading arm.
On the other hand, if the horizontal feature is moving from the LMC to the direction of the SMC,
then it may be originating from the end of arm E and is part of the trailing MS.
The structures and patterns of the horizontal features look much like the filaments of the MS,
and, when extrapolated, the horizontal features seem to connect to the second MS filament (Fig.\ \ref{msprofile}a).
If this is the case, then the entire Magellanic Stream and the Leading Arm might originate from LMC gas;
more work is needed to clarify this part of the Magellanic Bridge.  Based on the evidence that we
have presented in this section (summarized in Figs.\ \ref{la_vmlon} and \ref{la_onsky_lacut}), however, 
we conclude that most (if not all) of the Leading Arm gas originates in the SEHO of the LMC.

\subsection{The Cause of the Periodic Pattern}
\label{subsec:periodic}

\begin{figure}
\includegraphics[scale=0.49]{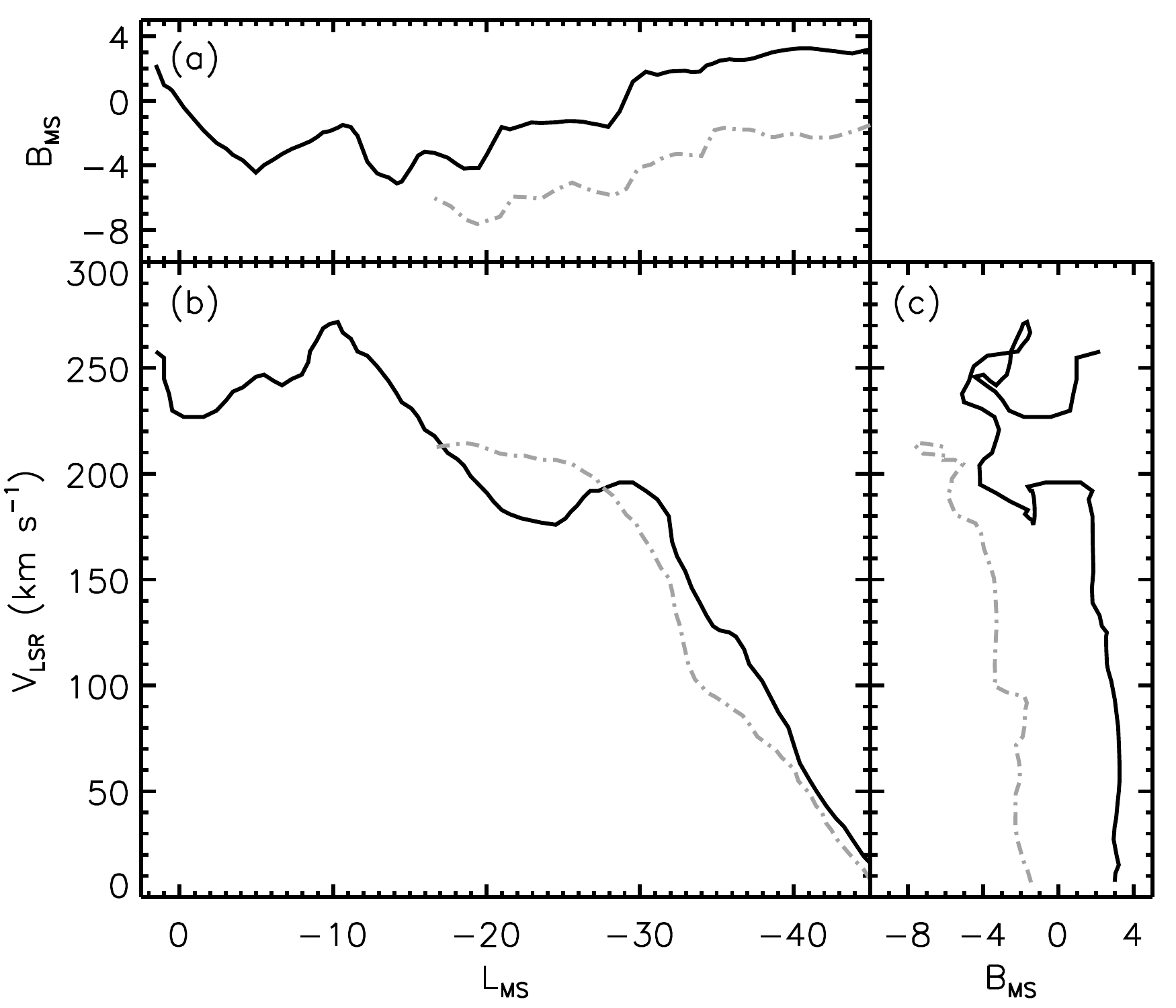}
\caption{This figure is analogous to Figure \ref{msprofile}.
The path of the central concentrations of the two MS filaments from three perspectives: ({\em a})
\bms vs.~\lmse, ({\em b}) \vlsr vs.~\lmse, and ({\em c}) \vlsr vs.~\lmse.
The LMC filament (solid) and the second filament (grey, dash--dotted)
do not wrap around each other, but move parallel to one another, in this \lms range.}
\label{msfilaments}
\end{figure}

As pointed out in \S \ref{subsec:twofil}, the two filaments of the MS have
pronounced sinusoidal patterns in velocity and in position, especially for \lms $>-40$\dgr
(Fig.\ \ref{msprofile}). P03 
noted that the two filaments give the impression of a double helix and postulated that
it might be due to the ``pseudobinary motion of the LMC and SMC''.  This tumbling motion thus
provides a first hypothesis as to the cause of these patterns.  However, there
are several problems with this hypothesis.  The path of two bodies orbiting each other
would create a double-helix pattern in our position--position--velocity (ppv) datacube
and in the projections of the datacube.  That is not seen for the two
filaments at the head of the stream (\lms $>-40$\degr), although they might cross farther downstream.
Figure \ref{msfilaments} shows the paths of the central concentrations of the
two filaments,  in the three projections (\bms vs.~\lmse, \vlsr vs.~\lmse, and \vlsr vs.~\bmse),
and reveals that the two filaments do not
cross or spiral around each other (at least in this \lms range), but rather create two independent
spirals parallel to each other.  Therefore, the spiraling motion that we see cannot be
explained by the tumbling of the LMC and SMC about each other (if they are indeed bound to one another).
It is also not clear where the second filament originates, whereas its origin {\em must} lie in the SMC
if the tumbling hypothesis is to work.
The two filaments do eventually cross each other at \lms $\sim -47$\dgr (Fig.\ \ref{msprofile}a) and other
places, beyond the large velocity oscillations seen at the head of the MS ($-40$\dgr$<$\lms$<-5$\degr).
The spatial crossing of MS strands we see at \lms $\sim-47$\dgr has become apparent in our maps
because the zero-velocity gas has been removed; this MS crossing is also apparent in Figure 2 of B05, but
was not noted by P03.
It is possible that these ``later'', more widely separated from the MCs, crossings could be due to the
binary motion of the Magellanic Clouds (if the second filament actually originates in the SMC), but the
spirals at the head of the Stream cannot.  There must be another explanation for these spiraling patterns
close to the MCs.

\begin{figure}
\includegraphics[scale=0.50]{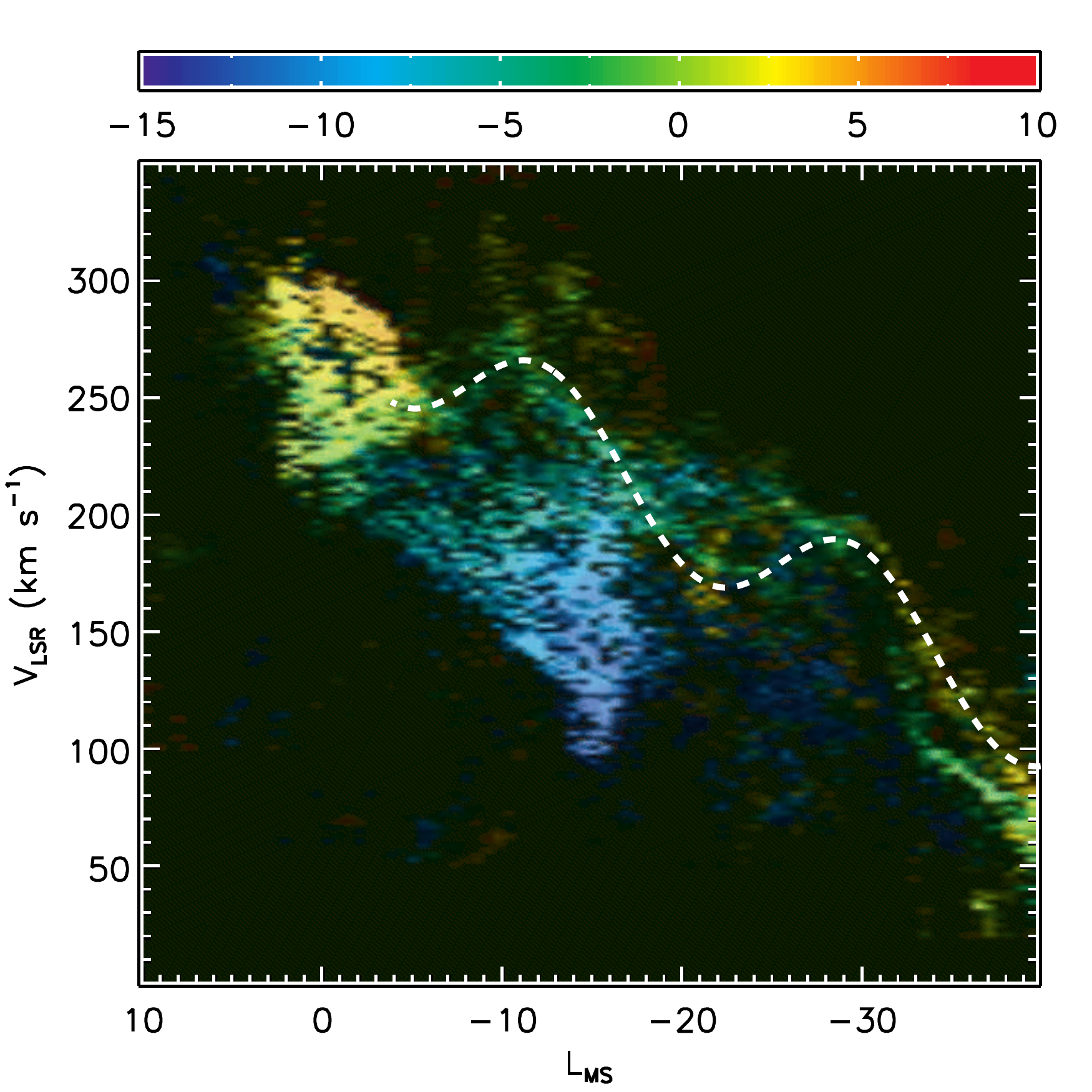}
\caption{\vlsr vs.~\lms of the Magellanic Stream Gaussians (same as Fig.\ \ref{ms_vmlon_zoom}) with the sine+linear
fit (Equation 2) to the LMC filament.}
\label{lmcfil_fit}
\end{figure}

We postulate instead that the spiraling motion is an imprint of the rotation curve of the LMC (and possibly
the SMC).  This scenario fits the data better 
than the previous hypothesis, especially for the LMC filament where the sinusoidal variation is
pronounced and the filament can be reliably traced back to its origin in the LMC,
which has a well-defined disk \citep[e.g.,][]{weinberg01,vandermarel01} and rotation curve
\citep[e.g.,][]{Kim98,vandermarel02}.
The amplitude of the spatial variations is $\sim$2\dgr (Fig.\ \ref{ms_onskycut}b) which is 
close to the radius of the SEHO from the LMC center.  A sinusoid$+$line fit to the LMC
filament in \vlsr vs.~\lms gives


\begin{eqnarray}
V_{\rm LSR} = &(26.4~\mbox{km s$^{-1}$})~ {\rm sin}\left(\frac{360^{\circ}}{20.9^{\circ}} L_{\rm MS} -5.9^{\circ} \right) + \\
  & (4.40~\mbox{km s$^{-1}$ deg$^{-1}$})~ L_{\rm MS} + 291.75~\mbox{km s$^{-1}$}
\nonumber
\end{eqnarray}

\noindent
and can be seen in Figure \ref{lmcfil_fit}.
Assuming that the linear portion is due to the orbital motion of the LMC about the
MW,  the sinusoidal amplitude of 26.4 \kms is not that different from the projected LMC rotation
velocity at the radius of the SEHO ($\sim$2.5$^{\circ}$ from the center of the LMC), i.e.
$V_{\rm rot}$ sin $i \approx 39.4$ \kms sin $34.7^{\circ} = 
22.4$ \kms \citep[vdM02;][]{kalli06a}.
It therefore seems that the data are reasonably consistent with the hypothesis that the spiral 
pattern in the LMC filament was created by the rotation of the SEHO (the filament's
apparent birthplace) about the center of the LMC as the entire system orbited the MW.  Our hypothesis
is that some force --- tidal, ram pressure or otherwise --- must have pulled/pushed the \hi gas out of the
SEHO (see \S \ref{sec:outflow}) and the trailing gas bears the imprint of the SEHO's
particular velocity and position.

Assuming this hypothesis is true, we can estimate the drift rate of the \hi gas in the LMC filament by
using the angular period of the sinusoidal pattern.
The orbital period of the SEHO around the LMC is 
 $T=(2\pi \times 2.5$\degr$\times 0.875~{\rm kpc~deg^{-1}})/39.4~{\rm km~ s^{-1}} \approx 340~ {\rm Myr}$.
In this time the SEHO's radial velocity undergoes one full cycle,
which we can identify with one full cycle of the LMC filament's velocity pattern.
Even though we have a spatial period from the sinusoidal fit, it is
apparent from Figure \ref{lmcfil_fit} that the sinusoid+line deviates slightly from the 
periodic pattern.
We can get a more accurate estimate of the spatial period by looking at
the maxima at (\lmse,\vlsre) = ($-10.3$\degr, $271.4$ \kmse) and at
(\lmse,\vlsre) = ($-29.8$\degr, $194.6$ \kmse), which gives 19.5\dgr for the
spatial period.  The LMC filament
gas is probably still close to the LMC distance from the Sun of $\sim$50 kpc which we can
use to calculate the distance between the maxima as $D = 19.5$\dgr $\times$ 0.875 kpc~deg$^{-1}$ = 17.06 kpc.
If the time difference between when these two points of gas were in the SEHO
is just 340 Myr, the orbital period of the SEHO,
and if the drift rate of the LMC filament gas away from the LMC is also roughly constant, then we
can estimate it as $V_{\rm drift} = 17.06$ kpc$/340$ Myr = 49.1 \kmse.  This is a lower limit
since it does not include any radial component of motion.
The drift velocity is $\sim$1/9 the tangential velocity of the LMC, $v_{\rm LMC,tan}=367$ \kms \citep{kalli06a}.

Lastly, we can use the drift rate of the LMC filament gas to roughly
estimate the age of the Magellanic Stream.  If we assume that the drift rate
has been approximately constant for the entire MS as well as an average distance of
50 kpc to the Stream, then the age of the Stream
is $(100$\dgr $\times$ 0.875 kpc deg$^{-1}$) / 49.1 \kms = 1.74 Gyr.
A more direct way of calculating this would be to count the number of oscillations
along the MS and multiply by 340 Myr (the orbital period at the SEHO radius).  However,
the amplitude of the modulations decreases quite dramatically after about $\sim$40\dgr
from the LMC and the two filaments
are not as well separated in velocity so it is not yet possible to do this in practice.
If the distance to the tip of the MS is larger than 50 kpc \citep[as predicted by
the models of][see their Fig.\ 6]{C06}, then the MS would be older, and conversely,
the MS would be younger if the MS tip is closer than 50 kpc \citep[as suggested
by the ram pressure models of][see their Fig.\ 7]{Mastro05}. 
A changing drift rate would also affect the MS age.
Even so, an MS age of $\sim$1.7 Gyr is fairly
close to the 1.5 Gyr age of the Stream found by \citet{MF80} at which time their
tidal simulations show an SMC-LMC-MW close encounter.
However, the orbit calculations using the new HST proper motions \citep{kalli06a,kalli06b,piatek07}
and an NFW MW potential by \citet{besla} give past MC encounters at roughly 0.2, 3 and 6 Gyr ago
(G. Belsa, private communication).  These numbers are highly dependent on the mass ratio of the MCs
($\sim$1:10) and the SMC's proper motion both of which are quite uncertain.
A close encounter at $\sim$2--2.5 Gyr would be consistent with the
bursts in the star formation rate seen in the star formation histories of {\em both} MCs
\citep{Smecker-Hane02,Harris04}.
Therefore, an age of $\sim$1.7 Gyr for the MS seems reasonable considering the uncertainties.

Even though the rotation hypothesis seems self-consistent there is one possible problem
-- the coherence of the sinusoidal pattern in the filaments.  If the rotation hypothesis
is correct then there should also be an oscillating pattern in the transverse velocity of the
filaments.  One would expect that the variation in velocities would pull apart or stretch the filament.
At the very least the different velocities should place the gas in different MW orbits because
of their differing energies, which would also destroy the coherence of the filament.
Since this is not seen, and the structure is coherent over at least $\sim$40$^{\circ}$,
there must be something else holding the structure together.
\citet{KBB02} show how the interaction of a cold cloud with its surrounding ambient medium can create
a magnetic barrier which helps keep the cloud intact.  In addition to explaining the survival times of
the MS filaments, the interaction of the clouds with their surrounding medium might help explain the
coherence of the Stream over such large distances.
More detailed N-body/hydrodynamic (and magneto-hydrodynamic) simulations are needed to investigate
how the MS filaments, with their oscillating velocity patterns, can remain coherent over such 
distances.

\begin{figure}
\includegraphics[scale=0.76]{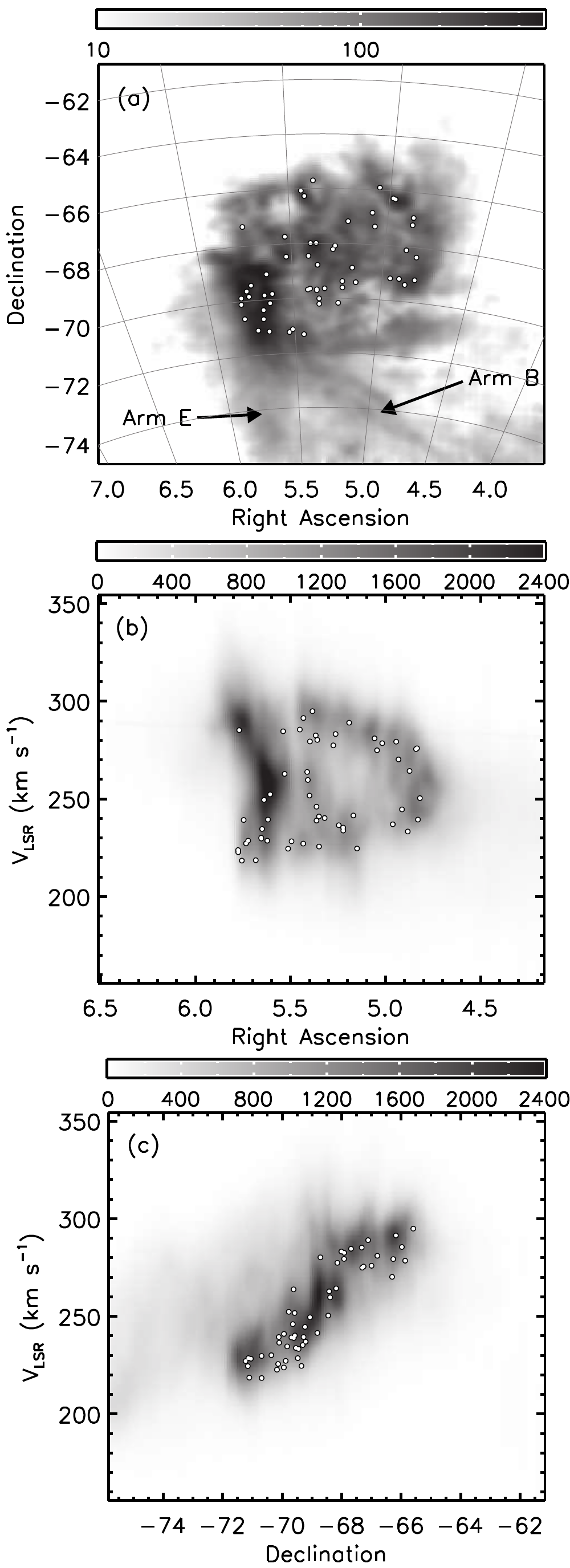}
\caption{Integrated intensity of the LMC \hi datacube (from S03) from three
perspectives: ({\em a}) Column density, \nhie, in units of $10^{19}$ atoms cm$^{-2}$,
({\em b}) \vlsr vs.~$\alpha$ (integrated in $\delta$; greyscale in units of K), and
({\em c}) \vlsr vs.~$\delta$ (integrated in $\alpha$; greyscale in units of K).  From
({\em c}) it is clear that there is gas coming off the LMC on the high-velocity
end and moving to the south.  The CO clouds from \citet{Mizuno01} (white dots in all panels) are almost always
associated with the LMC disk component (the regions with higher integrated intensity in panel ({\em  c})).}
\label{lmc_hires}
\end{figure}

For \lms$\lesssim -40$\dgr the sinsusoidal pattern diminishes substantially,
and the amplitude of any leftover variations is much smaller (Fig.\ \ref{msprofile}) and the behavior
of the MS is more nearly linear (\vlsr$ \approx 7.60$ \lms $ + 370$ \kmse; for \lms$< -40$\degr).
Something dramatic must happen at \lms$\sim-40$\dgr to cause this change.  One possible explanation is that this
is where drag from the MW halo gas becomes important, dampens the sinusoidal
pattern, and causes the MS to follow a quite linear curve.  Another possibility is that the
LMC is actually much larger than previously thought \citep[exceeding even the newly suggested limit in][]{Munoz06}
and the MS gas is escaping the tidal radius of the LMC at \lms$\approx -40$\degr.

If the LMC tidal radius is 40 kpc then from equation (45) in vdM02
(and a MW mass interior to the LMC of $4.9 \times 10^{11}$ \msune) this would imply
a minimum LMC mass of $1.68 \times 10^{11}$ \msun (or 34\% of $M_{\rm MW}$($<50$ kpc)).
The total intrinsic visible luminosity of the LMC is $L_V = 3.0 \times 10^9 L_{\odot}$ (vdM02)
which implies a mass-to-light ratio of $M/L_V = 56$.  This would mean that the LMC is very
dark matter dominated.

If we assume that ram pressure is the dominant force causing the MS to drift back behind the LMC
then, with some other assumptions, we can estimate the density of the hot MW halo gas.
The ram pressure on the MS is $P = \rho_{\rm MW} v_{\rm MS}^2$. If $L_{\rm MS}$ is the
approximate diameter of the MS, then the acceleration that the MS experiences is
$a_{\rm MS} \approx  \rho_{\rm MW} v_{\rm MS}^2 L_{\rm MS}^2/(\rho_{\rm MS} L_{\rm MS}^3) \approx
 \rho_{\rm MW} v_{\rm MS}^2/(\rho_{\rm MS} L_{\rm MS})$.  Solving for the ratio of
densities gives $\rho_{\rm MW}/\rho_{\rm MS} \approx a_{\rm MS} L_{\rm MS} / v_{\rm MS}^2$.
If we assume that the MS was undergoing a constant deceleration due to ram pressure
in the recent past then we can estimate this deceleration from the sinusoidal pattern
in the LMC filament,  $a_{\rm MS} \approx 2 \Delta{\rm x} / \Delta{\rm t}^2 \approx 2 \times 17.06
~{\rm kpc}/(340~{\rm Myr})^2 \approx 295.2~{\rm kpc/Gyr^2}$.  Approximating the diameter
of the LMC filament at the head of the Stream as $L_{\rm MS} \sim$2 kpc and $v_{\rm MS} \approx v_{\rm LMC} =
378~{\rm km/s}$ \citep{kalli06a,piatek07} we obtain $\rho_{\rm MW}/\rho_{\rm MS} \approx 0.004$.
The average column density of the LMC filament
is \nhi $\approx 1 \times 10^{20}~{\rm atoms/cm}^2$.  Assuming a distance of 50 kpc
and a width of $\sim$2 kpc gives a number density of $n_{\rm MS} \approx 0.016~{\rm atoms/cm}^2$.
Finally, we derive the number density of the hot MW halo as $n_{\rm MW} \approx 6.3 \times
10^{-5}~{\rm atoms/cm}^2$.  This rough estimate of the density of the hot MW halo
gas is consistent with most previous estimates \citep[e.g.][]{Stani02,Sembach03}.

\section{The Cause of the Outflow from the SE \hi Overdensity}
\label{sec:outflow}

\subsection{LMC High Velocity Gas Ejection}
\label{subsec:gasejection}

What is causing the Leading Arm and LMC filament
gas to flow out of the SEHO?  It is difficult to assess
this question with the LAB data due to its relatively low spatial resolution
(0.5\degr); however, the higher spatial and velocity resolution of the S03 \hi data of the LMC
from the Parkes telescope is well suited to a closer look at the SEHO.
In Figure \ref{lmc_hires} we show the integrated intensity of the LMC \hi
from three different perspectives: (a) $\delta$ vs.~$\alpha$, (b) \vlsr vs.~$\alpha$,
and (c) \vlsr vs.~$\delta$.  The CO clouds identified with NANTEN by \citet{Mizuno01}
are plotted as white dots in these figures.

\begin{figure}
\includegraphics[scale=0.46]{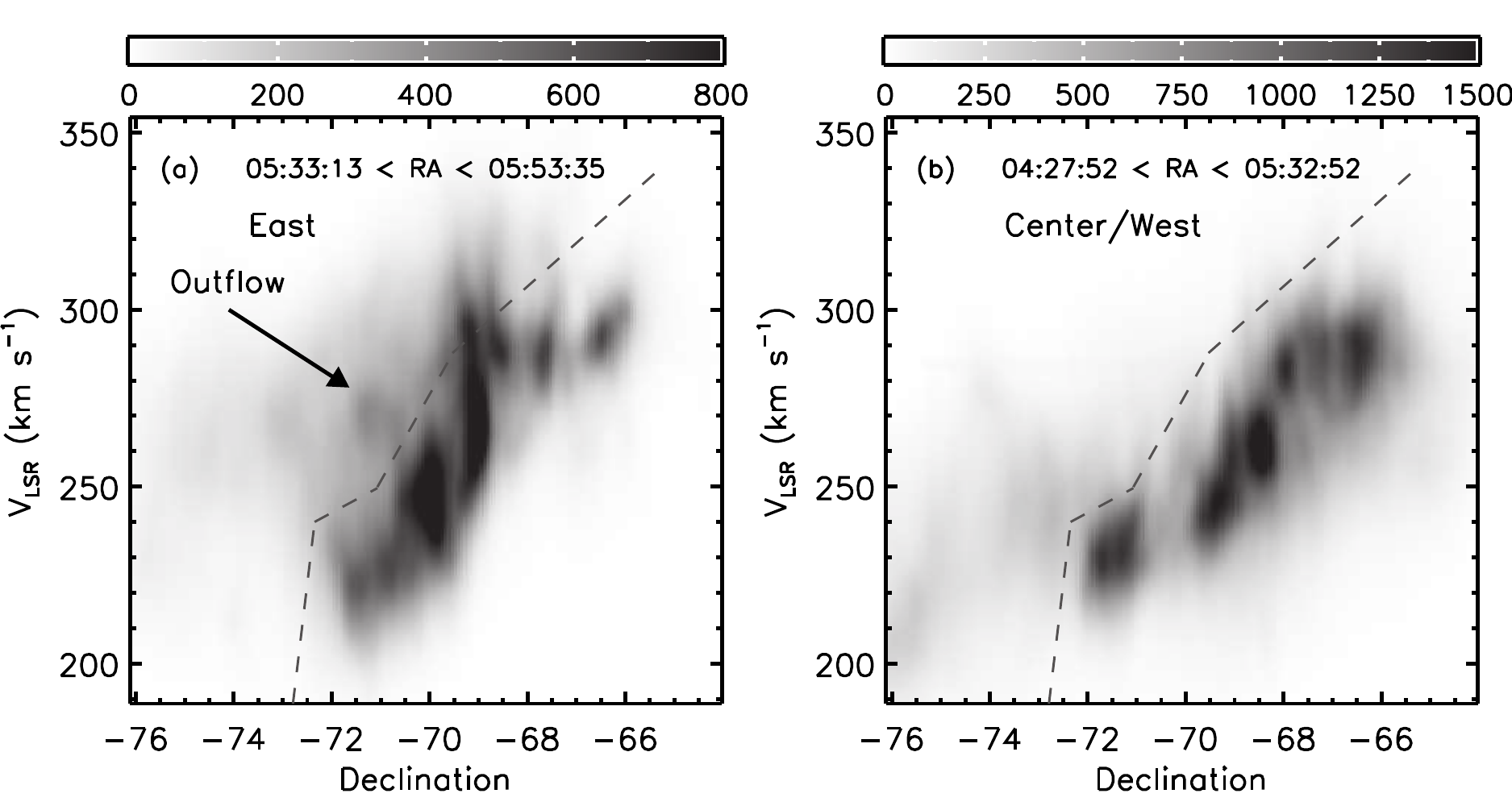}
\caption{Integrated intensity of the eastern and central/western regions of the
LMC from the S03 datacube. ({\em a}) The eastern region of the LMC showing the high and low velocity components
(for $-$72\degr$<$$\delta$$<-$69\degr).
({\em b}) The central and western regions of the LMC showing a velocity distribution with less outflow.
The high-velocity gas is coming mainly from the eastern part of the LMC.  The grescales are in units of K.
It is clear that the lower-velocity component in the south-east is the ``disk'' component by its high integrated
intensity (compared to the high-velocity gas) and its similarity to the velocity distribution to the west.  The
velocity criterion used to separate the high-velocity gas from the LMC disk gas at each declination
(see Fig.\ \ref{lmc_onsky_cut}) is shown by the dashed line.}
\label{lmc_vdec}
\end{figure}

The high density of the SEHO is readily apparent in the column density
plot (Fig.\ \ref{lmc_hires}a), as are great voids in the gas, which are evidence
of supergiant shells.\footnote{These supergiant shells are even more pronounced in the much higher
spatial resolution ATCA data used by \citet{Kim99} in their study of the supergiant shells
(as well as in the combined ATCA+Parkes \hi data by Kim et al.~2003).}
The Leading Arm can be seen stretching southward of the SEHO and the LMC
filament diagonally to the southwest (S03's arms E and B, respectively).
A position-velocity diagram of the LMC (Fig.\ \ref{lmc_hires}b; integrated along $\delta$ (which is
nearly parallel to the LMC \hi kinematical line-of-nodes,\footnote{Here we adopt the definition
(used by Luks \& Rohlfs 1992 and Kim et al.~1998) of the line-of-nodes as the axis of maximum
velocity gradient.  This differs from the definition of the line-of-nodes by vdM02 as the intersection
of the galaxy plane with the plane of the sky.} Kim et al.~1998)
shows that the LMC is not symmetrical in \hi but is lopsided and contorted.
Furthermore, there is a large build-up of neutral gas on its leading/eastern edge (where the
SEHO is located).  It has been suggested by \citet{deBoer98} that this is a result
of the LMC interacting with the diffuse MW halo gas (see further discussion of this
point in \S \ref{subsec:sinusoid}).

\begin{figure*}
\includegraphics[scale=0.96]{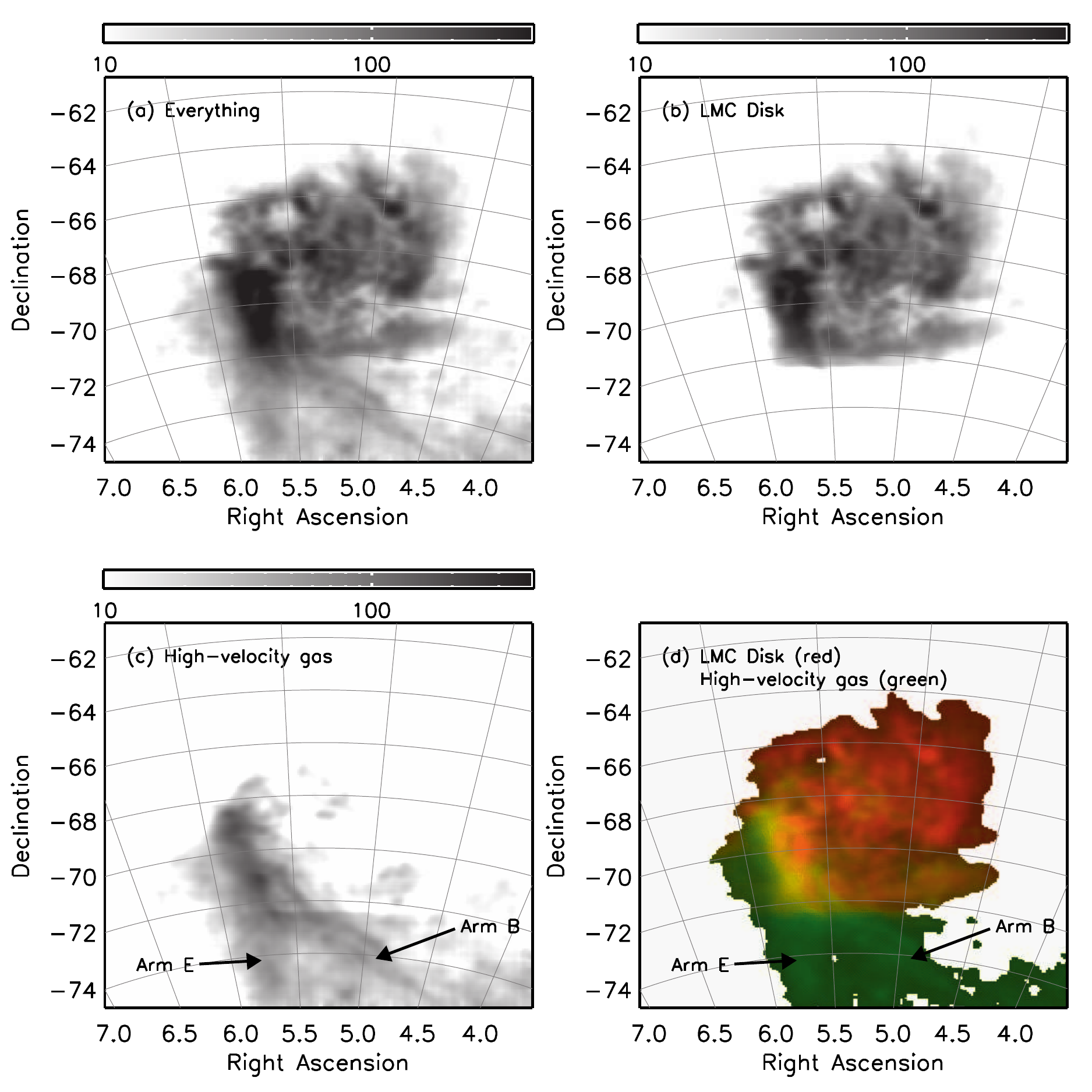}
\caption{The sky distribution (column density, \nhie, in units of $10^{19}$ atoms cm$^{-2}$)
of the different velocity components of the LMC:  ({\em a})  All Magellanic-associated gas, ({\em b}) LMC disk
gas only, and ({\em c}) high-velocity Magellanic gas only, using the velocity selection from Fig.\ \ref{lmc_vdec}.
Arm E and B are part of the high-velocity component. ({\em d}) The distribution of both the LMC disk (red)
and high-velocity (green) components of the LMC.  Near the LMC disk the high-velocity gas is predominantly on the
eastern/leading edge of the SEHO.}
\label{lmc_onsky_cut}
\end{figure*}

The velocity profile of the LMC in the third view (Fig.\ \ref{lmc_hires}c), a position-velocity diagram
integrated along $\alpha$ (i.e., nearly perpendicular to the Kim et al.~line-of-nodes)
appears, overall, to be much more symmetrical because it reflects very nearly the LMC rotation curve.
However, one can see in Fig.\ \ref{lmc_hires}c some 
gas at higher velocities than the LMC disk that has no corresponding, lower velocity counterpart.
This gas appears to come off of the
LMC disk (at the high-velocity side) and to stay at higher velocities than
the disk.  Both the high-velocity gas and the LMC disk gas follow a negative velocity
trend with $\delta$ (i.e. lower velocities toward the south), but the high-velocity gas
remains at higher velocities than the LMC disk at all positions.

In Fig.\ \ref{lmc_vdec} we compare the eastern versus the
central--western portions of the LMC (where we have integrated along $\alpha$).
The first panel reveals the ``high-velocity'' gas  clearly, with most of it coming from
the south--eastern portion of the LMC, especially from $\alpha$ $\gtrsim$ 05\h 30\m~and $\delta$ $\lesssim -$70\degr.
We isolate the high-velocity gas by making a rough velocity selection (the dashed line in Fig.\ \ref{lmc_vdec})
in the \vlsr vs.~$\delta$ plane, and show its sky distribution in Figure \ref{lmc_onsky_cut}c.
This high-velocity gas is indeed the Leading Arm and LMC filament (the same as S03's arm E and B), and
it is clearly coming from the south-eastern region of the LMC, or the SEHO.

Another obvious feature in Figure \ref{lmc_vdec}a is a strong
outflow of gas starting at $\delta$$\approx-$70\dgr and moving to higher velocities and southward; 
this feature 
might be coming from a supergiant shell (there are several in that region).
The high spatial resolution \hi data from \citet{Kim03} show this region to have many thin outflows
from the SEHO (see their Fig.\ 5, especially 05\h 37\m~$<$ $\alpha$ $<$ 05\h 48\m) that are not
resolved in the S03 data.  As Kim et al.~point out, these are probably
outflows from several smaller giant shells (they identified 16 giant shells
in this region).  In their figures it appears that there are many small fountains of gas coming out
of the SEHO which are contributing to the Leading Arm and trailing LMC filament.
These outflows are explored in more depth in the next subsection.

\citet{LR92} found that the \hi gas in the SEHO is 
disturbed and that there are two velocity components, which they called the high-velocity disk (``D'')
and low-velocity (``L'') components.  Figures \ref{lmc_hires}--\ref{lmc_onsky_cut} show, however,
that the high-velocity component in the south-eastern LMC is not the LMC disk, but an outflow that is the
beginning of the Leading Arm and the trailing LMC filament.  The low-velocity component has
a much higher column density than the high-velocity component (Figs.\ \ref{lmc_onsky_cut}a and c),
and is contiguous with the rest of the LMC disk in position-velocity slices (Fig.\ \ref{30dor_vdec}).
Also, the CO clouds in this region \citep{Mizuno01} are {\em all} associated with the low-velocity compononent and
none are associated with the high-velocity component (Figs.\ \ref{lmc_hires} and \ref{30dor_vdec}).
Therefore, we conclude (contrary to Luks \& Rholfs) that the
{\em low-velocity} gas is the disk component (albeit a disturbed disk) in the south-eastern part of the LMC.

\subsection{Source of the High-Velocity LMC Gas}
\label{subsec:hivelsource}

\begin{figure*}
\includegraphics[scale=0.96]{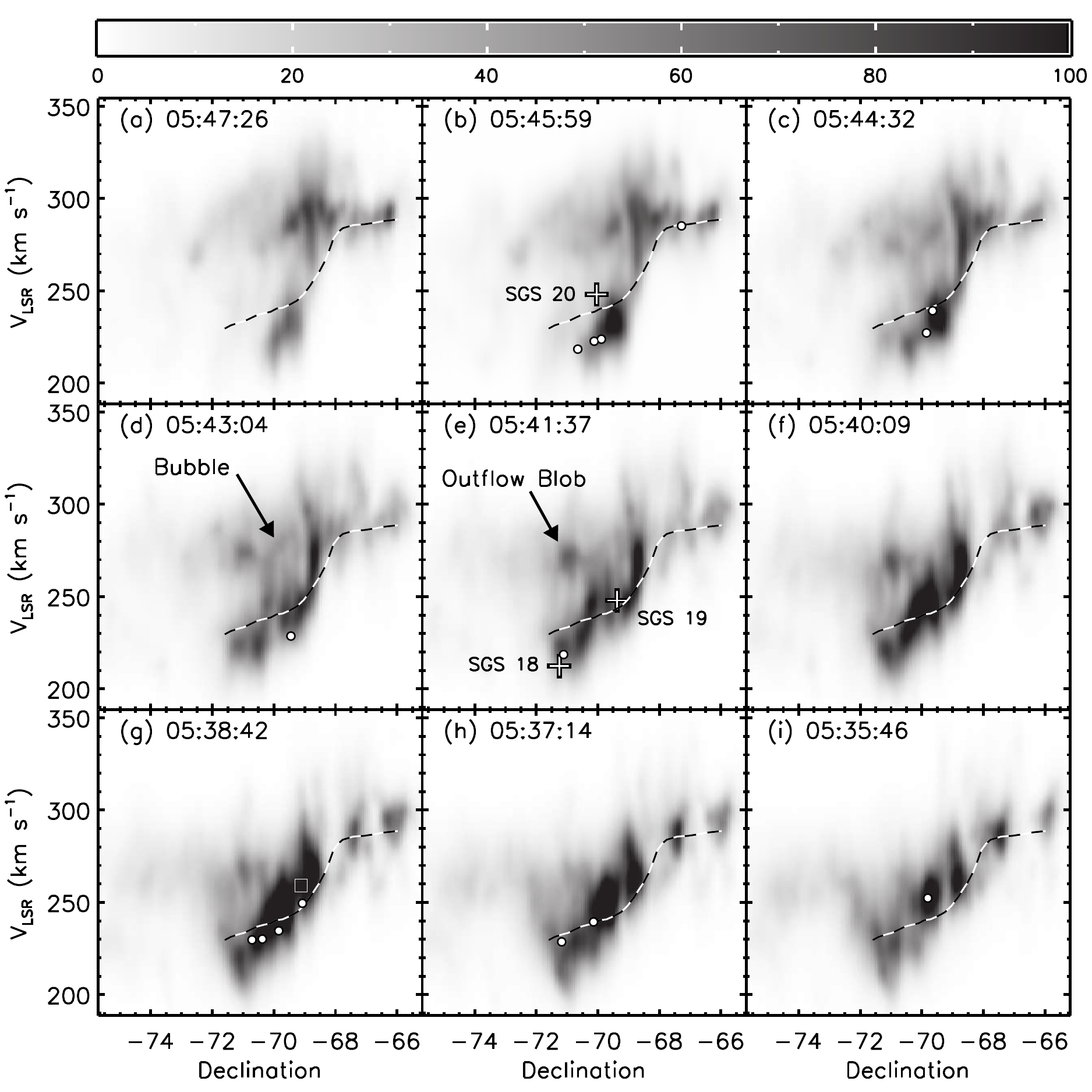}
\caption{Position-velocity cuts in $\alpha$ of the eastern part of the S03 LMC \hi datacube (going from east to west).
Two $\alpha$ bins in the datacube were combined to create each panel.  The central $\alpha$ is shown in the
upper-left hand corner of each panel and the greyscale is in units of K.
The dashed line shows the average rotation curve from the western part of the LMC, which helps separate
the high-velocity gas from the disk component.
White crosses indicate the central position and velocity of the three sugergiant shells from \cite{Kim99}
that are in this region (SGS 18, 19 and 20; the shells extend to the neighboring panels as well).
This figure shows the large outflow of gas from this region to high velocity.  A large
bubble of gas, most clearly seen in panels ({\em c})--({\em e}), (centered at $\delta$=$-69$\degr 28\arcmin 11\arcsec,
\vlsr=265 \kmse) is near the position of SGS 19 and 
an outflow blob of gas, most apparent in ({\em d})--({\em g}), seems to be coming off the bubble.
The CO clouds from \citet{Mizuno01} (white filled dots) are clearly not associated with the high-velocity outflow
gas but rather with the disk component.  The central position and velocity of the 30 Doradus star cluster R136 is
indicated by the white box in ({\em g}).  R136 is a little bit west of where most of the high-velocity gas is
being blown out.}
\label{30dor_vdec}
\end{figure*}

\begin{figure*}[!ht]
\includegraphics[scale=0.96]{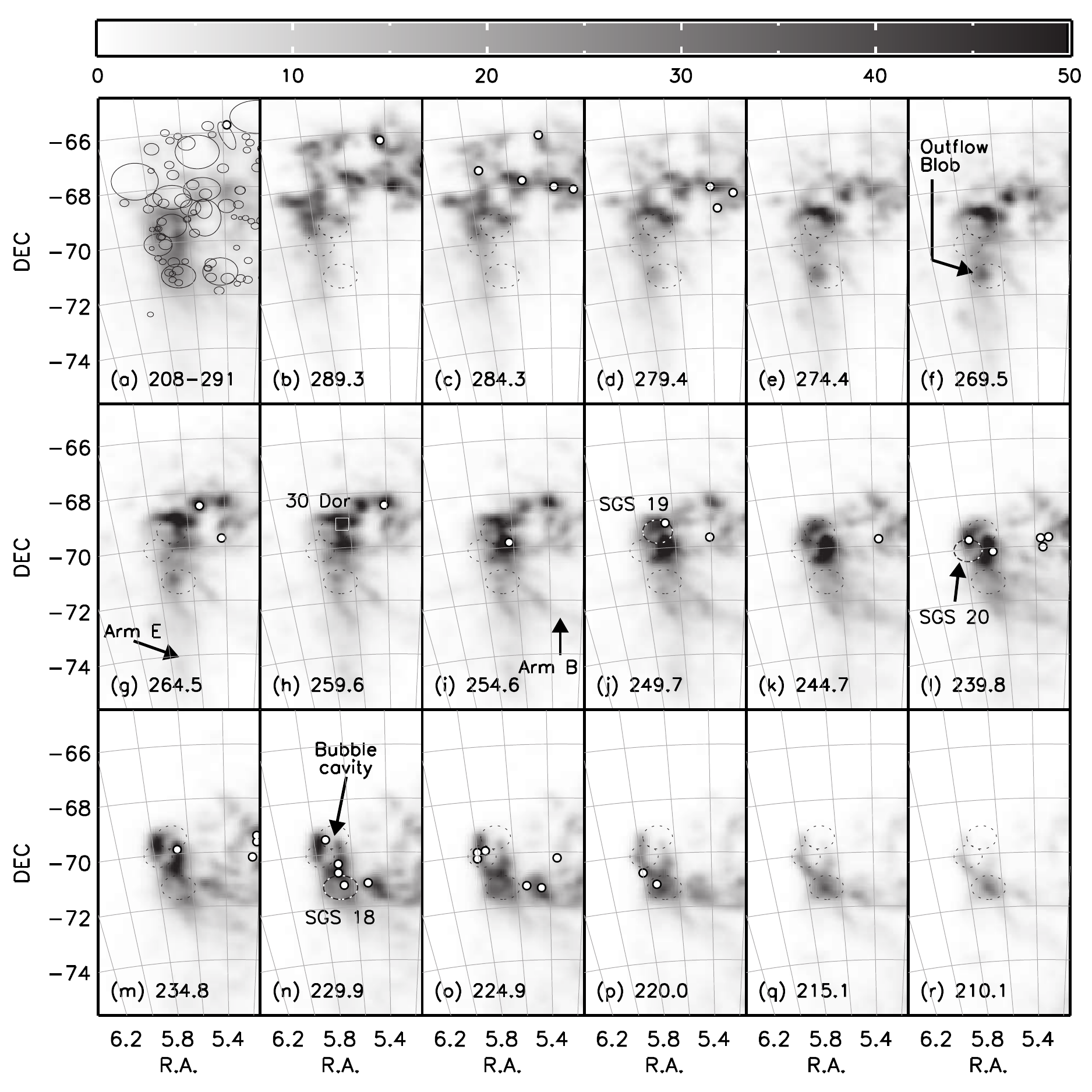}
\caption{Channel maps of the eastern part of the S03 LMC \hi datacube (from high to low velocity).  Six velocity
channels ($\sim$5 \kmse) were combined to form each panel.  The central \vlsr velocity (\kmse) is shown in the
lower-left hand corner of the panel.  Panel ({\em a}) shows the column density (for 208--291 \kmse) with the
supergiant shells and giant shells from \cite{Kim99} overplotted in black.
The scale at the top indicates the column density, \nhie, for panels ({\em b})--({\em r}) in units
of  $10^{19}$ atoms cm$^{-2}$.
White--black dashed lines indicate the three sugergiant shells SGS 18, 19 and 20 (at
their central velocity, but they extend to many of the neighboring panels in velocity) from \cite{Kim99}.
The outlines of these three supergiant shells are also shown as black dashed lines in panels ({\em b})--({\em r})
as reference positions.  
This figure shows the large outflow of gas from this region
to high velocity including arms E and B.  A large outflow blob of gas (also see Fig.\ \ref{30dor_vdec}),
most clearly seen in panels ({\em d})--({\em g}), is near the position where SGS 18 and 20 intersect, but
(Mizuno et al.\ 2001; white filled dots in all panels)
are located in regions of dense \hi and from which the outflows of gas originate (but at lower velocities
than the outflows).  The 30 Doradus star cluster R136 is indicated by the white box in {\em h} (at its
central velocity).}
\label{30dor_radec}
\end{figure*}

\begin{figure}
\begin{center}
\includegraphics[scale=0.70]{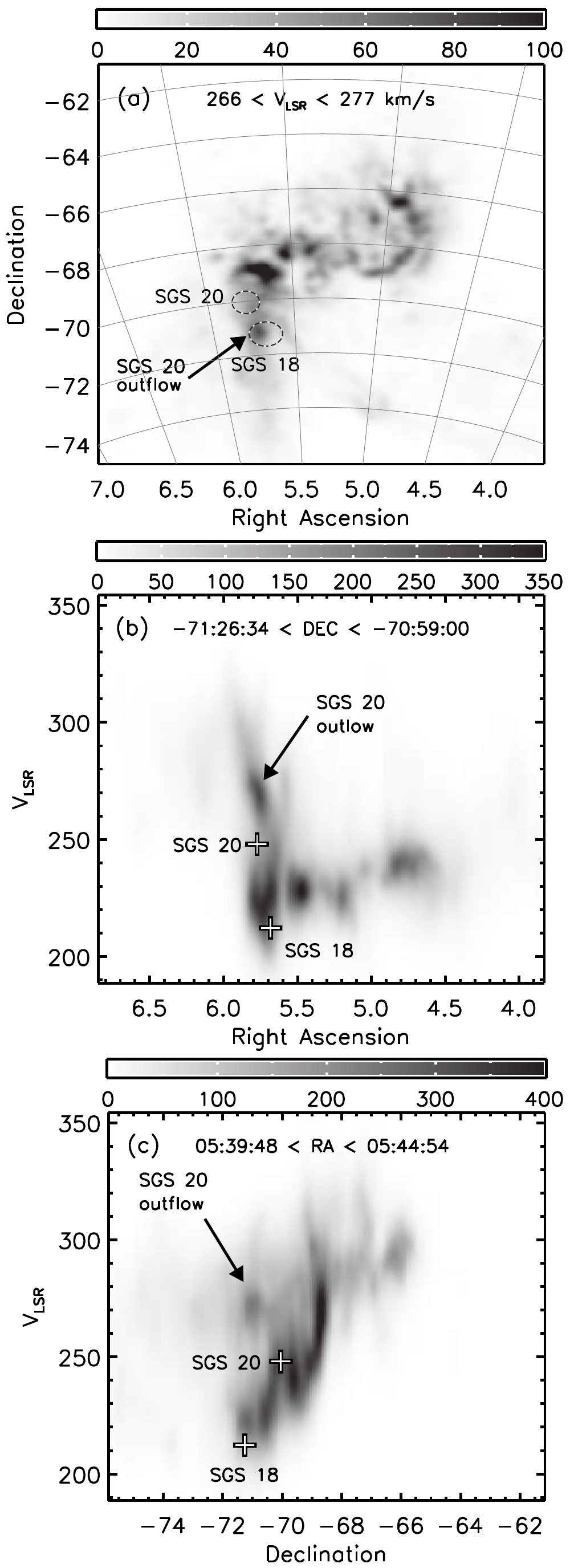}
\end{center}
\caption{Integrated intensity of the LMC \hi gas from S03 showing
the outflow of gas from SGS 20, centered at ($\alpha$, $\delta$, \vlsre)=
(05\h 43\m 09\se, $-$71\degr 08\arcmin 46\arcsec, 270 \kmse). ({\em a})
Column density, \nhie, in units of $10^{19}$ atoms cm$^{-2}$
(integrated from 266 $<$ \vlsr $<$ 277 \kmse), ({\em b})
\vlsr vs.~$\alpha$ (integrated from $-71$\degr 26\arcmin 34\arcsec $<$ $\delta$ $<$ $-70$\degr 59\arcmin 00\arcsec;
the greyscale is in units of K), and ({\em c}) \vlsr vs.~$\delta$ (integrated from
05\h 39\m 48\s $<$ $\alpha$ $<$ 05\h 44\m 54\se; the greyscale is in units of K).
The locations of SGS 20 and 18 are shown in all three panels.  Even though the outflow is spatially more
aligned with SGS 18 than SGS 20 (in panel ({\em a})) it is clear from panels ({\em b}) and ({\em c}) that this
is a projection effect.  The outflow of gas is connected in position and velocity to the bubble of gas
that is surrounding SGS 19 and 20 (see Figs.\ \ref{30dor_vdec} and \ref{30dor_radec}), but has moved farther to
the south and hence is spatially in front of SGS 18.
This figure shows that SGS 20 is blowing out a large amount of gas from the LMC.}
\label{sgs20}
\end{figure}

What might the source be of this high velocity outflow?
It has long been known that there are high-velocity gaseous flows emanating from star formation
regions in the LMC, particularly in the form of supergiant shells
(e.g., Meaburn \& Blades 1980; Meaburn 1984; Meaburn et al.\ 1987);
the most recent comprehensive study of these structures has been done by \citet{Kim99}.
Several of these supergiant shells are also visible in the Parkes \hi datacube and are, we believe, relevant to
the origin of the MS.
The $\alpha=$05\h 43\m 04\s slice in Figure \ref{30dor_vdec}d shows a bubble feature with a void of
gas in its center ($-$70.5\degr$\lesssim$ $\delta$ $\lesssim-$69.0\degr, 240 \kms $\lesssim$ \vlsr $\lesssim$ 280 \kmse).
This void of gas can also be seen in channel maps of the LMC (Fig.\ \ref{30dor_radec}j--r; e.g., the U-shaped
feature most obvious in Fig.\ \ref{30dor_radec}n) and near the position of the supergiant shells SGS 19 and
20 \citep{Kim99}.  In fact, the void is probably SGS 19 and 20 and the bubble their combined envelope.
There is a blob of gas that is ``attached'' near the edge of this bubble (Figs.\ \ref{30dor_vdec}d,e and
Fig.\ \ref{30dor_radec}e--g).  The blob and this linking gas has a ``tadpole''-like structure with the head
farther south than the tail; together these features have an upside-down L shape and end near the position
of the southern edge of SGS 20 (Figs.\ \ref{30dor_vdec}f,g).  This seems to be evidence of an outflow from SGS 19 and 20.
Figure \ref{sgs20} shows three perspectives of this outflow.
SGS 1 also has an attached blob of gas that exhibits the same tadpole features (see Fig.\ \ref{sgs1}c and
\S\ref{subsec:massaccounting}).  We conclude
that both of these are outflows being expelled by the supernovae (SNe) and high stellar winds coming from the
supergiant shells.  A possible explanation for the tadpole-like structure is that once the outflow
reaches a certain height above its origin site other forces, such as ram pressure and tidal forces, become
dominant and push/pull the outflow southward (in the same direction that arms B and E are being pushed/pulled).

We propose the dynamical kick of SNe-driven SGSs
as the mechanism that gave rise to the majority of the outflow in the SEHO.  
\citet{Mea84} previously found the 30 Doradus nebula to be a primary source of outflowing gas, but the
Parkes data cube reveals this phenomenon to be more widespread, encompassing the entire SEHO.

\subsection{Tracing Arms B and E}
\label{subsec:tracingarms}

The \hi arm B from the LMC
can be traced back to its origin in the region around SGS 18 and 20, as can be seen in Figures \ref{30dor_radec}f--r
and Figure \ref{lmc_vslices}.  There are two main strands of arm B, a northern and southern strand
(Figs.\ \ref{lmc_onsky_cut}a, c), each very thin (about $\sim$0.25\dgr wide) but with distinct velocities.
The northern strand appears to emanate around
($\alpha$, $\delta$, \vlsre) $\approx$ (05\h 38\m 32\se, $-70$\degr 39\arcmin 16\arcsec, 273 \kmse) = 
(5.6422\degr, $-70.6544$\degr, 273 \kmse)
 (Fig.\ \ref{30dor_radec}f and Fig.\ \ref{lmc_vslices}b, c) near the
intersection of SGS 20 and 18.  The southern strand of arm B is at lower velocity than the northern
strand (by $\sim$30 \kmse) and appears to emanate around
($\alpha$, $\delta$, \vlsre) $\approx$ (05\h 38\m 09\se, $-71$\degr 07\arcmin 08\arcsec, 200 \kmse)
= (5.6358\degr, $-71.1189$\degr, 200 \kmse)
(Fig.\ \ref{lmc_vslices}a) in SGS 18.  In Figure \ref{lmc_vslices}a there appears to be a
second parallel filament at the same velocity as the southern strand farther to the south.
Since we don't have distance information there are projection effects which make it difficult to pinpoint exact origins
for the arm B filaments.  Therefore, the locations given above should be considered rough estimates.
However, it is clear that arm B originates in the SEHO region.
Arm E is wider, more diffuse, and on average at higher velocity than arm B.  Arm E appears to emanate from many
outflows across the SEHO region.
There is a trend that, moving eastward, the outflow sites move to the north and to higher
velocity, which is also a general trend in the SEHO.


It is unclear why the two arms are moving in the directions they are
(arm E to the south, and arm B to the southwest) even though they originate
in the same place.  The two outflow blobs mentioned above show that
the gas must first reach a certain positive velocity offset ($\sim$30 \kms
for the SGS 1 outflow, and $\sim$20 \kms for the SGS 20 outflow) from the systematic
SGS or LMC disk velocity before it starts to move appreciably
from its place of origin.  This might tell us something about the forces
operating on the gas.  The motions of the two arms/filaments might be related to which
side of the disk they are on.  Since the LMC disk is inclined relative to
its direction of motion through the MW halo any gas blown out in ``front''
of the disk (towards the direction of LMC bulk motion)  will experience more ram
pressure than gas blown out ``behind'' the disk (away from the direction of motion).
All of the outflow gas is at higher positive velocities and it would seem
likely that the gas is on the opposite side of the LMC from our perspective
and in ``front'' of the LMC as it moves through the MW halo (this is discussed
further in \S \ref{subsec:reldistance}).  More investigation
is needed to explain the motions of the two filaments.

\subsection{Mass Accounting}
\label{subsec:massaccounting}

We can estimate the mass of \hi gas in the outflow blob (Fig.\ \ref{sgs20}) as well as the mass
evacuated from the SGSs and see how they compare.  The mass of the blob of gas
(within 05\h 39\m 32\s $<$ $\alpha$ $<$ 05\h 47\m 37\se, $-71$\degr 26\arcmin 02\arcsec 
$<$ $\delta$ $<-70$\degr 46\arcmin 53\arcsec, 267.4 \kms $<$ \vlsr $<$ 280.6 \kmse)
is $\sim$$1.5 \times 10^6$ \msune (using an LMC distance of 50 kpc).
We can estimate the mass of the \hi gas that has been evacuated from SGS 19 and 20
by calculating the difference in \nhi inside and outside the SGSs.  The
average \nhi in SGS 19 and 20 is $\sim$$3.54 \times 10^{21}$ \pcm and 
$\sim3.50 \times 10^{21}$ \pcm respectively, while for the surrounding region it is $\sim$$4.90 \times 10^{21}$ \pcme.
Using the area of SGS 19 and 20 from \citet{Kim99} (circles with diameters of 52.0\arcmin~and 49.4\arcmin~respectively)
we estimate the mass lost from SGS 19 to be $\sim$$1.6 \times 10^6$ \msun
and $\sim$$1.5 \times 10^6$ \msun lost from SGS 20.
These masses are remarkably close to the mass of the outflow blob.
However, this estimate ignores the possibility that a sizeable fraction of this lost gas has probably
been swept up by SGS winds and gone into compressing the surrounding ISM.  Nonetheless, even if only
half of the lost gas has been blown out of SGS 19 and 20 it is sufficient to explain the outflow blob.

\begin{figure}[!ht]
\begin{center}
\includegraphics[scale=0.71]{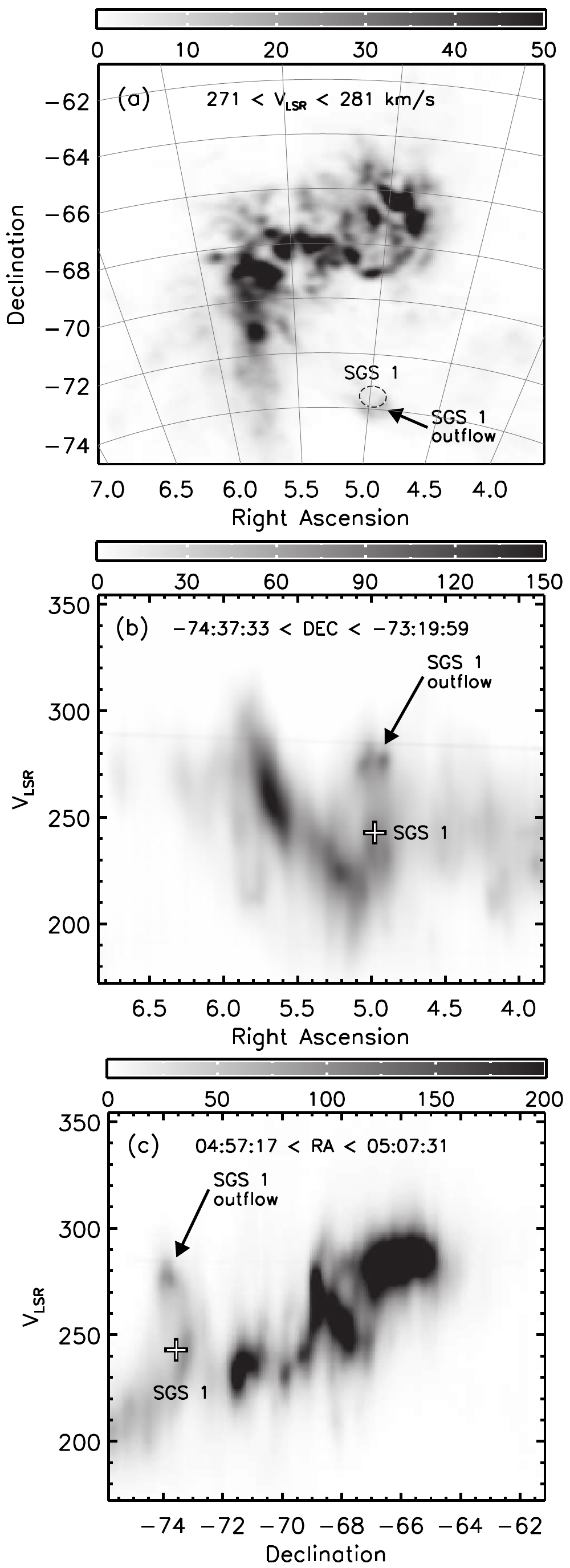}
\end{center}
\caption{Integrated intensity of the LMC \hi gas from S03 showing
the outflow of gas from SGS 1, centered at ($\alpha$, $\delta$, \vlsre) =
(04\h 58\m 36\se, $-73$\degr 33\arcmin 57\arcsec, 275 \kmse).
({\em a}) Column density, \nhie, in units of $10^{19}$ atoms cm$^{-2}$
(integrated along \vlsr for 271 $<$ \vlsr $<$ 281 \kmse),
({\em b}) \vlsr vs.~$\alpha$ (integrated along $\delta$ for $-74$\degr 37\arcmin 33\arcsec $<$ $\delta$ $<$ 
$-73$\degr 19\arcmin 59\arcsec; greyscale in units of K), and ({\em c}) \vlsr vs.~$\delta$
(integrated along $\alpha$ for
04\h 57\m 17\s $<$ $\alpha$ $<$ 05\h 07\m 31\se; greyscale in units of K).  This shows that SGS 1 is blowing
a substantial amount of gas out of the LMC filament.  This also suggests that there may be recent
star formation going on in the Magellanic Stream and there might be some young stars there.}
\label{sgs1}
\end{figure}

There is another outflow blob of gas to higher velocity from SGS 1 (Fig.\ \ref{sgs1}), which is a SGS
in arm B (the LMC filament) located at ($\alpha$, $\delta$)=(04\h 58\m 36\se, $-73$\degr 33\arcmin 57\arcsec)
with a mean velocity of \vlsr = 242 \kmse, size of (48.6\arcmin $\times$ 45.4\arcmin) and
age of 3.0 Myr \citep{Kim99}.  SGS 1 and the outflow can be seen in Figure
\ref{sgs1}.  The outflow first shoots to higher velocity by $\sim$27 \kms at nearly the
same position and then moves south $\sim$0.5\degr, where the densest portion is positioned (Fig.\ \ref{sgs1}c).
The blob appears to have two clumps that are separated in $\alpha$ by $\sim$0.5\dgr (Fig.\ \ref{sgs1}b)
(Fig.\ \ref{sgs1}a).
The outflow must have substantial tangential motion in order to have moved $\sim$0.5\dgr (corresponding
to $\sim$0.44 kpc) in $\sim$3 Myr, or it is older than \citet{Kim99} determined.  The mass of the blob
(within 04\h 49\m 30\s $<$ $\alpha$ $<$ 05\h 05\m 48\se, $-74$\degr 20\arcmin 04\arcsec 
$<$ $\delta$ $<-73$\degr 31\arcmin 10\arcsec, 267.4 \kms $<$ \vlsr $<$ 288.0 \kmse)
is $\sim$$7.1  \times 10^5$ \msune, which is about half the mass of the outflow blob
coming from SGS 19 and 20.  These examples of SGSs blowing out massive amounts of gas are 
strong evidence that the SGSs in the SEHO are capable of blowing out
substantial amounts of gas, and, if chained in a series of propagated star formation events, creating
the Leading Arm and the trailing LMC filament of the MS.

SGS 1 lies well outside (by $\sim$2\degr) the LMC \hi disk as defined in Figure \ref{lmc_onsky_cut},
and matches arm B perfectly in position and velocity (Fig.\ \ref{sgs1}).  We have argued in \S\ref{subsec:twofil}
that arm B is the beginning of the LMC filament of the MS.  If SGS 1 is indeed a part of arm B, as
all evidence points, and our explanation of arm B is correct, then this implies that there is ongoing
star formation at the head of the MS and that there are probably some young stars in the Stream.

The \hi mass of the entire LMC is $\sim$4.8$\times 10^8$ \msun (S03), while
$\sim$1.3 $\times 10^8$ \msun of that is in the high-velocity gas.
The mass of arm B is $\sim$4.9 $\times 10^7$ \msune, that of arm E is $\sim$6.9 $\times 10^7$
\msune, and $\sim$1.2 $\times 10^7$ \msun is to the west of arm B.
This high-velocity gas constitutes 27\% of the measured \hi mass of the LMC
and $\sim$1/3 the combined mass of the MS and LAF (which have $\sim$3.8 $\times 10^8$ \msun
and $\sim$3 $\times 10^7$ \msun respectively; B03).  Thus 
the LMC has just recently blown out and lost a substantial amount of mass from the SEHO.
It therefore seems plausible that over an even more extended period of time
the SEHO might have produced the Magellanic Stream and LAF.


What is the mass ledger like for this hypothesis?
The total mass within the volume of a SGS in the SEHO of the LMC
with an average diameter of 50\arcmin~(or 0.72 kpc) is $\sim1 \times 10^7$ \msune.
If 10\% of the mass within a SGS is blown out of the LMC disk, then
each SGS contributes $\sim1 \times 10^6$ \msun of gas to the MS and/or
LAF.  If the Magellanic Stream originated in the SEHO
and was blown out by SGSs, then some $\sim$410 shells
would have been required to have created the Stream.  If the Stream is $\sim$1.74 Gyr old
(from \S \ref{subsec:periodic}), then the
average SGS creation rate would have to be 1 per $\sim$4.2 Myr.
If the shells are coherent and observable for $\sim$10 Myr, which appears to be the
maximum age of shells in the LMC \citep{Kim99}, then approximately 2--3 should be seen in
the SEHO at any time.  This is the number currently observed (SGS 18, 19, and 20).

Of course, one issue with the MS and LAF originating in the SEHO is that the current
\hi mass of the SEHO ($\sim$1.3$\times 10^8$ \msune) is now smaller than the \hi mass of
the entire Stream.  There are several possible explanations for this discrepancy:
(1) the SEHO was more massive in the past, (2) the SEHO is continually
being replenished by gas from the LMC disk, or
(3) the SEHO is continually
growing by accreting from the hot, coronal MW it is plowing through.  Since the current
\hi masses of the LMC and SMC are roughly
the same, moving the origin of the MS and LAF from the LMC to the SMC does not solve the mass problem.
The system from which the MS and LAF originate has lost a substantial amount of gas recently
and must have contained at least double the \hi in the past.

On the other hand, is it possible for the SGS creation rate and gas outflow to be sustained
for $\sim$1.74 Gyr?  According to \citet{Kim99}, the continual creation of SGSs
can be achieved by self-propagating star formation, which is observed in the 30 Doradus complex.
Therefore, it appears plausible for there to have been continuous star
formation and SGSs blowing out gas sufficient to form the MS+LAF for an extended amount of time
in the SEHO.  This is discussed further in section \ref{subsec:originseho}

\subsection{Energetics of the Supergiant Shell Blowout}
\label{subsec:sgsenergetics}

Is it energetically realistic to suppose that SGSs have blown out the MS and LAF?
The energy that would be required to blow out the entire mass of the MS and LAF
($\sim$4.1$\times 10^8$ \msune) to a velocity of $\sim$50 \kms is
$E = (1/2) M_{\rm MS} v^2 \approx 1.0 \times 10^{55}$ ergs.  Since the average kinetic energy released
by a SNe is $\sim$$10^{51}$ ergs \citep{WW86,Cho07} this would require $\sim$10,000
SNe.  If half of the SNe kinetic energy in an SGS goes into sweeping
up gas and compressing the ISM in the disk and the other half blows out gas perpendicular
to the disk, then the combined energy of $\sim$20,000 SNe are required.
Spread out over $\sim$1.74 Gyr gives an average SNe rate of 1 every $\sim$87,000 years.
This is fairly low compared to the MW SNe rate of 1 every $\sim$100 years
and the LMC rate of 1 every $\sim$500 years \citep{vdB91}.  If the SNe are divided
into $\sim$410 SGSs then each SGS would contain $\sim$49 SNe.  The 30 Doradus nebula
contains 39 O stars, 12 Wolf-Rayet stars and 8 B supergiants that will
all eventually go supernova \citep{Melnick85} a tally that validates the plausibility of our estimates.
Therefore, the energetics appear to be favorable to the SGS blowout hypothesis.

\subsection{Magnetic Fields in the SE \hi Overdensity}
\label{subsec:magfields}

\citet{Haynes91} performed a radio continuum survey of the Magellanic
Clouds at 2.45, 4.75, and 8.55 GHz.  In the linearly polarized maps of the LMC
Haynes et al.~discovered that two thin, long ($\sim$3--4 kpc) filaments (stretching
south of the 30 Doradus nebula) dominated the emission (most obvious at 2.45 GHz or 12 cm;
see Fig.\ \ref{magbar}a here or their Fig.\ 7).  \citet{Xu92} separated the thermal (free-free) from the
nonthermal (synchrotron) emission and found that, while there is relatively
uniform, nonthermal radiation present across most of the LMC (indicating
pervasive magnetic fields), the region extending south of the 30 Doradus nebula
(where the filaments are located) contains almost entirely thermal emission (see their
Figs.\ 10a and b).  

\begin{figure}[!ht]
\begin{center}
\includegraphics[scale=0.70]{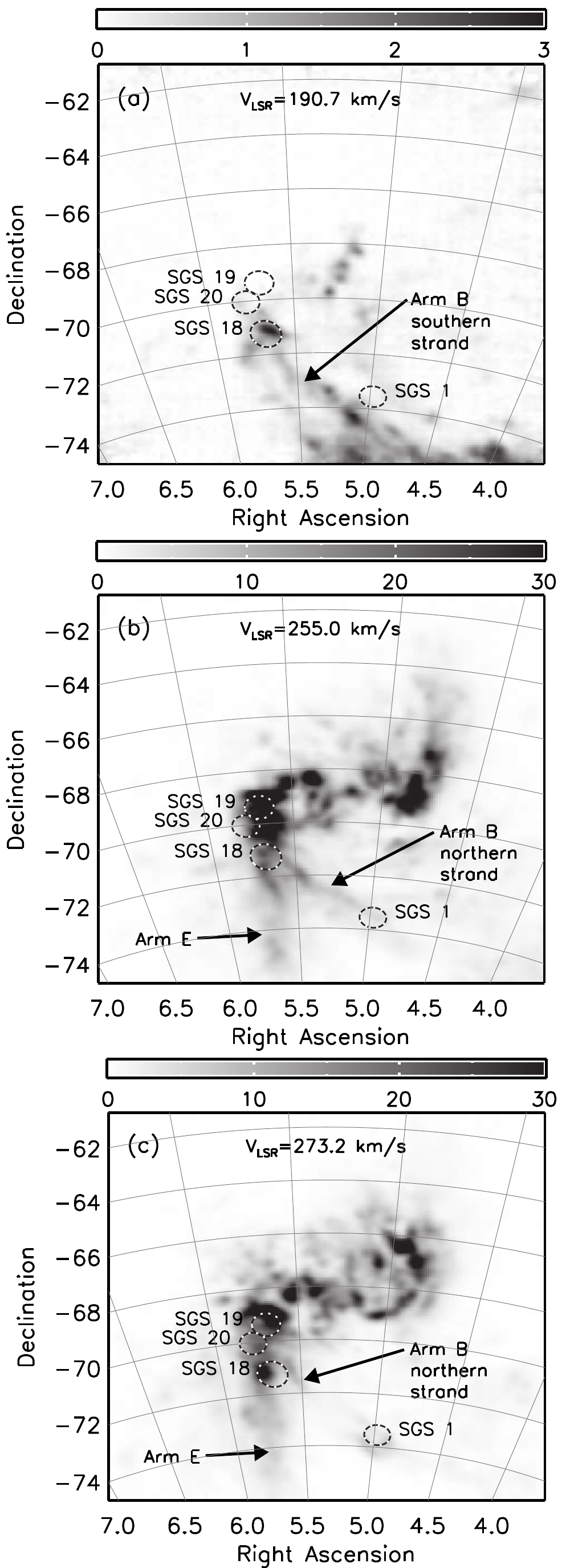}
\end{center}
\caption{Velocity slices of the \hi datacube from S03 showing where the filaments of arm B
(the LMC filament) and arm E (the LAF) originate (the greyscales are in units of K).  The outlines
of SGS 1, 18, 19 and 20 from \citet{Kim99} are also shown.  In panel ({\em a}) at \vlsr$=190.7$ \kmse, the two
southern filaments of arm B (with fairly faint $T_{\rm B}$) appear to originate in SGS 18.  In panel ({\em b})
at \vlsr$=255.0$ \kmse, the northern filament of arm B and a part of arm E seem to originate in SGS 18.
Panel ({\em c}) at \vlsr$=273.2$ \kmse, the northern filament of arm B emanates from the north-western
part of SGS 18.  Dense knots of \hi gas around SGS 19 and 20 seem to be the source of some of the arm E gas.}
\label{lmc_vslices}
\end{figure}

\begin{figure*}
\includegraphics[scale=0.95]{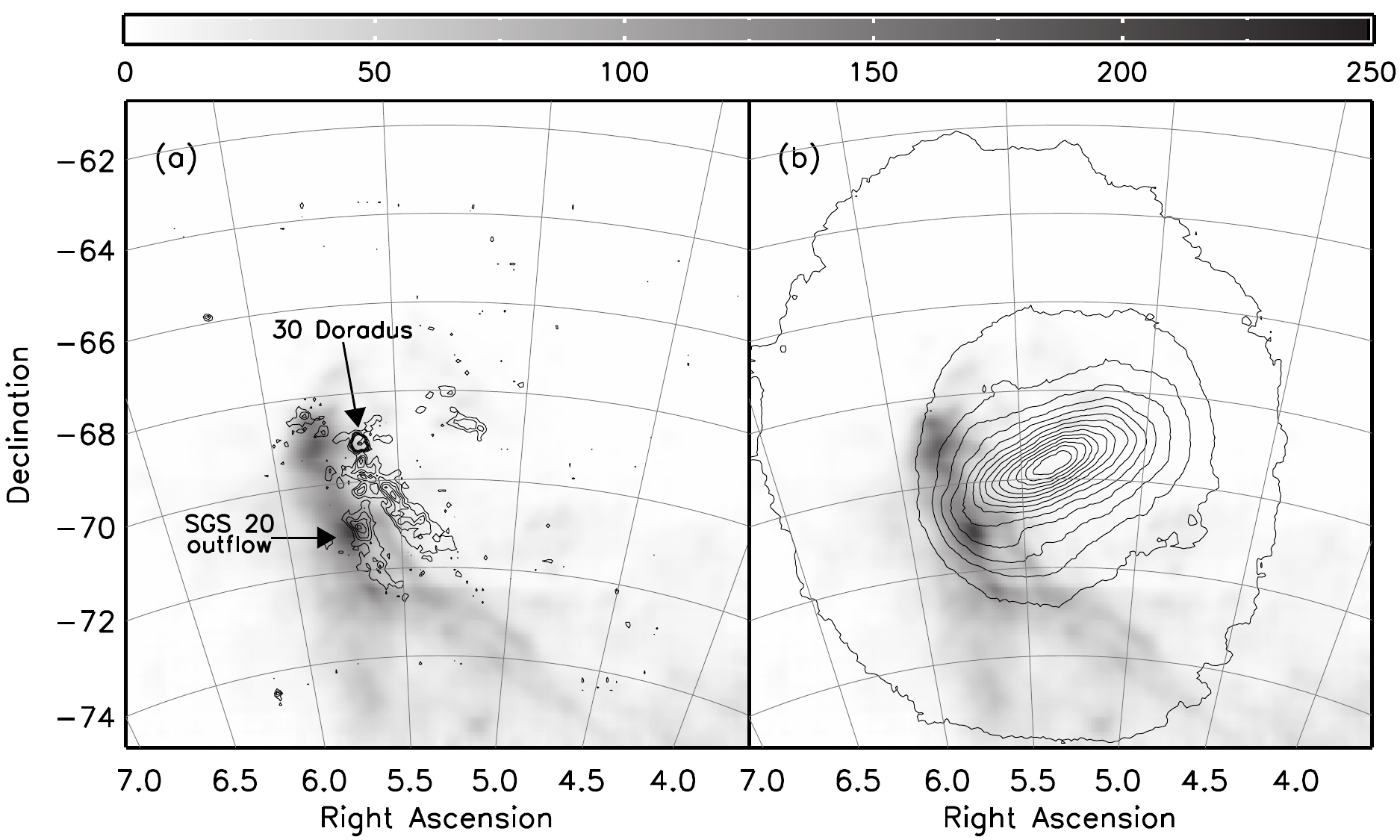}
\caption{The sky distribution (column density, \nhie, in units of $10^{19}$ atoms cm$^{-2}$)
of the high-velocity component of the LMC \hi gas from the S03 datacube (same as Fig.\ \ref{lmc_onsky_cut}c).
({\em a}) The 12 cm linearly polarized radio emission from \citet{Klein93} is overplotted as contour lines
and shows the two magnetic ``filaments''.  There is strong emission from 30 Doradus, the
SGS 20 outflow blow (also see Fig.\ \ref{sgs20}), as well as two filaments which are parallel to the
high-velocity gas, especially arm B.  The eastern polarized filament appears to be associated with
the SGS 20 outflow blob.  The polarized and \hi sets of filaments seem to be anti-correlated and a
small correlation coefficient of 0.30 bears this out.
({\em b}) The density of 2MASS LMC red giant branch stars \citep{Skrutskie06} is overplotted with contour lines
showing the central stellar LMC bar.  The eastern end of the bar is  near the region where
most of the high-velocity gas is originating.  The proximity of the end of the bar might be partly
responsible for the high-density of \hi gas and perturbed dynamics in this region of the LMC.}
\label{magbar}
\end{figure*}

With the addition of observations at 1.4 GHz, \citet{Klein93} used
the radio continuum data to study the 3D magnetic field structure of the LMC, and
found that the magnetic field in the region of the filaments was significantly aligned
(i.e.~not random), and oriented more or less parallel to the filaments' direction.
In addition, the most extreme rotation measures (RM) --- an indication of
the magnetic field parallel to the line-of-sight --- were in this region.  Klein et al.~conclude
that there is a giant magnetic loop bending out of the plane of the LMC
towards us (see their Fig.\ 6).
We compare the high-velocity gas with the linearly polarized filaments from Klein et al.~in Fig.\ \ref{magbar}a.
The two groups of filaments (high-velocity \hi and ``polarized'') are
offset from each other.  In fact, they appear to be anti-correlated with a low
correlation coefficient of 0.30.
One exception is a small region near SGS 18 
(and the SGS 20 outflow blob) that is bright in the polarized emission (and in CO as well:
Cohen et al.~1988, Fukui et al.~1999), probably indicating that this region is very active.
It appears that the eastern polarized filament might originate from there. The
western polarized filament might be related to the 30 Doradus nebula.
The proximity to each other, filamentary structure, and abnormality (compared to
the rest of the LMC) of the two sets of filaments indicates a probable physical relationship
between them;  the exact nature of that relationship and the
origin of the spatial offset remain unclear at this time.

These polarization and magnetic-field studies show that there is something violent
happening in the region south of 30 Doradus that is likely related to the high-velocity
gas and the SGSs.  This corroborates circumstantially
our suggestion that the SGSs may be blowing out the \hi gas and creating the
LMC filament of the Magellanic Stream and the Leading Arm.  It is possible that
the relativistic ions are being accelerated by the supernovae shocks from the
SGSs, and that the aligned magnetic field is being caused by the
stellar winds and supernovae shocks (also from the SGSs).  It is not
entirely clear how the magnetic loop fits into this picture.  However, if there
are magnetic fields in the Magellanic Stream it might explain how the filaments
can remain so coherent over large distances.

\subsection{Relative Distance of the High-Velocity Gas}
\label{subsec:reldistance}

There is some diagreement about whether the anomalous (``high-velocity'')
gas is in front of or behind the LMC disk.  \citet{LR92} conclude that
the low-velocity component is in front of the high-velocity component since they
find no absorption of the 30 Doradus emission (associated with the low-velocity
component) by the high-velocity component.
\citet{Points99} used ROSAT data to study the supergiant shell LMC 2 (SGS 19).
They detected no X-ray absorption features that correlated spatially with the distribution
of the high-velocity gas.  Based on this evidence they concluded that the high-velocity
gas is behind the plasma in LMC 2 (which presumably is in the disk of the LMC).
However, this study only looked at a small subsection of the LMC.

\citet{Dickey94} performed an absorption-line study
of several sight-lines in the LMC.  They used a kinematical argument to conclude
that the high-velocity component is in front of the low-velocity component. Since
the LMC disk is inclined, a cloud at a height above the LMC disk plane (and following
the same rotation curve) will have an actual LMC-centric distance that is different
than the projected one and therefore will have a different rotational velocity than
the rest of the gas along that same sight-line (that is in the disk).  Whether the
rotation velocity will increase or decrease depends on the exact sight-line, and this
information is embodied in a ``velocity gradient''($dv/dr$).  Dickey et
al.~calculated the velocity gradient for each position at which they detected
absorption lines using the smoothed velocity field by Luks \& Rohlfs (1992; their Fig.\ 9) and
assuming that the north-east is the near side of the LMC.  Based on the velocity gradient
at each position and the radial velocity offset from the disk radial velocity at that point,  they
conclude that the low-velocity gas is on the far side of the disk and the high-velocity
gas is in front.

The kinematical argument by Dickey et al.~works fairly well for arm B, which moves across
a large part of the LMC.  Based on the velocity gradient its velocity offset should
at first be positive, then zero, and then negative, and this is actually what is observed.
However, there is a problem with the gas ``blobs'' that are being blown out of the SGS
mentioned above (\S \ref{subsec:hivelsource}).  The gas moves to higher velocity at nearly the same position
which would be interpreted to mean, without considering the rotation curve, that the
gas is moving away from us.  Furthermore, since it seems clear now that most, if not
all, of the high-velocity gas in the south-eastern portion of the LMC is coming from
the SGSs, the high-velocity gas should be on the same side of the disk as the ``blob'' gas.
Even though the kinematical argument of Dickey et al.~seems realistic, it assumes
that the gas out of the disk plane is also in circular motion.  However, we question if this
assumption is a valid one.  It is clear that the LMC's gravity
is not the only force acting on the gas.  As already stated, it seems clear that the
SGSs are blowing gas out of the disk (probably away from us),
and since arm B becomes the $\sim$100\dgr trailing LMC filament of the MS, 
tidal or ram pressure forces must be acting on it.  Therefore, it seems unlikely that a
simple model with the gas in relative equilibrium, such as proposed by Dickey et al., will
work correctly in this highly dynamical situation with various contributing forces.

Finally, \citet{Klein93} claim that two filaments visible in the linearly polarized emission
are on the near-side of the disk.  They conclude that the filaments are outside the disk
since their rotation measures are much lower than would be expected if they were inside the
LMC disk.  The rotation measures would also be higher than expected if the filaments were on
the far side of the disk.
Due to the anti-correlation of the high-velocity \hi and ``polarized'' filaments (mentioned above)
it is doubtful that the claims by Klein et al.~about the polarized filaments can be used to ascertain
the position of the high-velocity \hi gas relative to the disk.  However, the possible association of the
\hi filaments with the polarized filaments, and the ``magnetized loop'' hypothesized by Klein et al.,
deserve further study.

We conclude that the high-velocity gas of both arms E and B is on the far side of the LMC, mainly
because we are attributing the dynamics to the force of the SGSs blowing out gas to
higher velocity, and this must put higher-velocity gas away from us.
This is consistent with the conclusions of both Luks \& Rohlfs and Points et al., but inconsistent with
Dickey et al.~as well as with the Klein et al.~assessment of the polarized filaments.

\section{Discussion and Summary}
\label{sec:discussion}

Our exploration of the Magellanic System using a Gaussian decomposition of the LAB data in
combination with other radio data has led us to several conclusions regarding the nature
and origin of the Magellanic Stream.

\subsection{The Large Magellanic Cloud as Progenitor of the Magellanic Stream and Leading Arm Feature}

We have found evidence that one of the filaments of the Magellanic Stream and the Leading Arm can
be traced back to the LMC and that both of these \hi structures originate there.  In our database
of \hi Gaussians the MS appears as two filaments as has been previously reported by \citet{Put03}.
But, capitalizing on the coherence of the filaments in
ppv (position--position--velocity) space,
we were able to track one of them back to the LMC in both velocity and position.  After isolating the \hi of
this filament in one of the position-velocity diagrams
(\vlsr vs.~\lmse), we were able to show that it originates in the SEHO of the LMC.  Likewise
we showed that the first complex of the LAF (LA I) begins (in position and velocity)  near the end of
S03's arm E from the LMC.  The spatial and velocity progression of the three concentrations of LA I and arm E,
as well as their continuity in the deeper HIPASS data, strongly suggest that they are physically connected.

An LMC origin of the MS is contrary to much of the current
literature, which has largely supported an SMC/Bridge origin for the MS
(e.g., P03 and B05).  Even most modeling papers
(especially those based on a tidal origin of the MS) have started with the SMC/Bridge assumption and adopt an
N-body representation of the SMC influenced by only a static LMC (and MW) potential
(e.g.,~Gardiner \& Noguchi 1996; Yoshizawa \& Noguchi 2003;
Connors et al.~2004, 2006).  Therefore, these models {\em by design} rely on the MS
forming from the SMC and not the LMC.
Models that represent {\em both} the LMC and SMC as N-bodies (Murai \& Fujimoto 1980;
R\accent23u\v{z}i\v{c}ka et al.~2006) have also concluded that most of the Magellanic Stream came from
the SMC; however, these simulations do show at least some debris coming off the LMC and
this debris follows the position and velocity distribution of the main body of the Stream.
The main reason that an SMC origin was preferred is because {\em more} material came off the SMC
than the LMC in these models.  The reasons cited for this observed mass-loss inequity is that the SMC
potential is weaker than the LMC's and that the LMC has a much larger tidal influence on the SMC than
vice versa.  Even though these statements may be correct in the context of these models,
it is likely that these models are missing important physical processes: none
of the simulations include the potentially important dynamics of
LMC star formation and SGSs that may {\em greatly} enhance gas outflow from the LMC.
If SGSs propel gas faster than the escape velocity, this gas
can then be swept away from the MCs by tidal and/or ram pressure forces.
Future simulations including these effects {\it subsequent} to blowout may come closer to
matching the \hi structural features that we have observed.
First steps towards this goal have already been taken by Olano (2004) who has already demonstrated
that blowout from the MCs can create the large scale features of the MS, as well as the HVC system of the MW.  Olano's
model explored blowout from the MCs that results from an interaction between the two Clouds 570 Myr ago.
Our analysis here identifies a more specific site (the SEHO of the LMC) and timescale ($\sim$1.7 Gyr)
for the creation of the MS that we believe ought to be accommodated by such models.

With the discovery of the two filaments of the MS (P03) it was suggested
that one of the filaments originates from the Bridge, and that therefore the Bridge might
be older than the $\sim$200 Myr suggested by the models of \citet{GSF94} and the young blue
stars discovered between the Magellanic Clouds by \citet{IDK90}.
However, now that this same \hi filament
can be traced to the LMC there is no reason to believe that the Bridge is old.
In fact, the recent study of stellar populations in the Bridge by \citet{Harris06}
shows that there are only young stars there and that star formation
started in the Bridge only some 200--300 Myr ago.

According to Russell \& Dopita (1992), the current mean metallicities of gas in the LMC and SMC are
[Fe/H]$=-$0.2$\pm$0.2 and $-$0.6$\pm$0.2, respectively, while the average metallicity of the MS is 
[Fe/H]$=-$0.6$\pm$0.2 (Wakker 2001).  At first glance it appears that the metallicity information
points towards an SMC origin of the MS.  However, it is not the current metallicity
of the MCs that is important, but the metallicity at the time that each part of the
MS left the Clouds, starting around 1.74 Gyr ago (so that one might expect a metallicity gradient
along the Stream).  According to \citet{PT98}, the
metallicities of the Clouds were $\sim$0.3 dex lower $\sim$1.7 Gyr ago, which would put them
at [Fe/H]$\approx-0.5$ (LMC) and [Fe/H]$\approx-0.9$ (SMC).  Thus, even though metallicity
provides only a weak discriminant of the MS origins, it appears to slightly favor an LMC origin
over an SMC origin.

Moreover, there continues to be a global
problem reconciling large observed variations in the metallicities of different gaseous Magellanic features.  
For example, the low metallicity of the Bridge, [Fe/H]$=-1.1$ \citep{Lehner02}, compared to
[Fe/H]$=-0.6$ for the SMC (where it is thought to have originated), has been used to argue that
the Bridge is old.  However, because other evidence supports a young age ($\sim$200 Myr) for
the Bridge (see above), the enrichment level of the gas must be more complex than mere mass loss from
a simple closed box model of the MCs.  For example, \citet{Lehner02} suggests that the Bridge gas from
the SMC may have mixed with other, less enriched gas.  One possible source of less enriched gas could
be earlier gas ejections from the SMC itself which could be accreted.
Another possible explanation for the low metallicity of the Bridge gas is that it could have originated
from gas in the outskirts of the SMC that might be more metal-poor than gas in the central part of the
SMC.  Whatever the explanation is, it seems clear that it is difficult at present to interpret the relative
metallicities of Magellanic gas and the MCs.

\subsection{Origins in the SE \hi Overdensity}
\label{subsec:originseho}

We have been able to track the MS back to the LMC, and argue, in \S \ref{sec:outflow}, for its origin in the 
SEHO.  What remains unresolved is whether the physics creating this intense star-forming region is tied
to a particular, fixed dynamical hot spot that is the product of the bulk motion of the
LMC, its rotation, and interaction with MW halo gas, or whether the SEHO participates in the
general LMC rotation.  The answer bears on how to evaluate an accounting of relative ages, masses, and
energetics between the SEHO site and what is needed to produce the MS and LAF and their apparent
oscillations.

For example, \citet{deBoer98} suggest that the leading edge of
the LMC (the SE, where the SEHO is presently located) is being compressed as the LMC moves through
the diffuse halo of the MW (creating a bow-shock), which produces a steep \hi gradient (as observed),
and triggers star formation.  In this scenario, the southeast quadrant of the LMC will always
remain an active star-forming site, with previous generations of stars formed there rotating
off of the hot-spot due to LMC rotation, but with the evolution of massive stars constantly regenerating
SGSs near their birthplace.  This argues for considering a broader perspective in evaluating
the age and energetics of MS production.  Unfortunately, while it is obvious that there has been much recent
star formation occuring in the 30 Doradus (Tarantula) nebula and its {\it immediate} surroundings,
the star formation history of the SEHO and the annulus of the LMC it occupies have been less
well established.  But from our assessment of the mass, age, and required energy needed to produce
the MS and LAF (\S\S \ref{subsec:massaccounting} and \ref{subsec:sgsenergetics}) we can hypothesize some general expectations.

We have given evidence in \S \ref{sec:outflow} that SGSs in the SEHO are currently
blowing out gas and creating the MS and LAF.  This scenario is supported by the high-speed
outflows from the 30 Doradus starburst \citep{redman03} and by the energetic outflows from the LMC \citep{lehner07}.
From the continuity of the MS we can infer that this
process must have been going on continuously at a rate steady enough to not create gaps in the MS.
In the de Boer et al.~scenario one might expect a continuous cycle of star formation,
SGSs, and gas blowout as long as the compression and bow-shock are present.  As discussed in \S6.3, 
to eject the $\sim$4.1$\times 10^8$ \msun 
of the MS and LAF out of the SEHO in $\sim$1.7 Gyr would require $\sim$400
SGSs (each blowing out $\sim$10$^6$ \msune). This exceeds the current \hi mass of the SEHO
($\sim$8$\times 10^7$ \msune, excluding the high-velocity gas), but this comparison belies a complex
calculus of mass exchange.
On the one hand, some \hi from the SEHO is converted into stars and stellar remnants, while some
is lost to SGS blowout to produce the MS and LAF, and additional \hi gas can be generated from the
destruction of H$_2$, although this is likely to be a small amount.\footnote{According to \citet{YS91}
the mass in molecular hydrogen is typically an
order of magnitude less than that in atomic form for very late-type spirals.}
On the other hand some gas may be inflowing from the MW halo (and indeed may be the source
of less enriched material complicating the metallicity accounting discussed above).

If the MS and LAF were formed as part of the de Boer et al.~scenario, then a $\gtrsim$1.7 Gyr age for 
these gaseous structures is viable, but the star formation that produced these features would
have created stellar populations that would have rotated around the LMC about five times.
Therefore, we would expect that the stellar populations at any position in a 2 kpc radius to
have five distinct peaks in age corresponding to the times when this position was in the SE
corner of the LMC where the vigorous star formation is occurring.  These peaks should be
separated by $\sim$340 Myr (i.e.~the LMC rotation period at this radius) but with approximately
the same star formation rate.  The exact ages of these peaks should be shifted with position
in the 2 kpc annulus -- older as you move clockwise (with the LMC rotation) and younger
in the opposite direction.  The total number of stars younger than $\sim$1.7 Gyr should be
approximately constant with position around the 2 kpc annulus.
Thus, a detailed analysis of the star formation history of the 2 kpc annulus could be used
to check the validity of the de Boer et al.~scenario.  Unfortunately, radial and azimuthal
mixing of stars in the LMC disk might smear out some of these patterns which could make this
analysis difficult in practice.

It is not clear what caused the vigorous star formation in the SEHO region to start $\sim$1.7 Gyr
ago and commence the SGS blowout and formation of the MS and LAF.  One possibility is that
the LMC and SMC had a close encounter that could have triggered star formation.  The integrated
star formation histories of the LMC \citep{Harris04} and SMC \citep{Smecker-Hane02} both show a
``burst'' of star formation $\sim$2--2.5 Gyr ago which suggests that an interaction between the MCs might
indeed have occurred around that time.  The  LMC globular cluster age--gap between 3 and 13 Gyr
\citep[e.g.,][]{DaCosta91,Geisler97,Rich01,Piatti02} indicates an onset of star formation in the LMC
around $\sim$3 Gyr ago, however, such an age--gap is not seen in the SMC clusters \citep{DaCosta91}.
As previously mentioned in \S \ref{subsec:periodic}, the orbits
by \citet{besla} indicate possible encounters of the MCs with each other around 0.2, 3 and 6 Gyr ago.  Therefore,
it does not seem unrealistic to suppose that a close encounter of the MCs $\sim$2--3 Gyr ago triggered
star formation in the LMC and SEHO region.  Another
possibility might be that $\sim$1.7 Gyr ago the LMC was close enough to the MW that ram pressure due to the
hot MW halo created a bow shock and started star formation in the SEHO
(jump starting the de Boer et al.~scenario mentioned above).
However, according to the hyperbolic orbits of \citet{besla} the LMC was $\sim$300--500 kpc from the MW
at this time (depending on the MW mass); even though the density of the gaseous MW halo is not well known
it is doubtful that the density at those distances could be large enough to create the bow shock
necessary to start star formation in the LMC.  The trigger of the recent, vigorous star formation
in the SEHO region of the LMC therefore remains an open, and interesting question.

It should be acknowledged that the SMC also has many giant shells and SGSs \citep{ss97}, and
it might be expected that the SMC is also blowing out gas that might be contributing to the MS.  
For example, it may be that the second filament of the MS is coming from the SMC.  Unfortunately,
we have no evidence to support or disprove this based on our analysis.
The lack of an obvious major source of frenetic star formation activity in the SMC may
substantially lower the efficiency of and amount of SMC blowout relative to that in the LMC.
Whether the SMC contributes to the MS and by how much requires further work to determine.

\subsection{The Sinusoidal Velocity Pattern of the LMC Filament}
\label{subsec:sinusoid}

Among the most striking features of the MS, as seen in Figures \ref{msprofile}b and \ref{ms_vmlon_zoom},
are the velocity oscillations of the two filaments.  These sinusoidal patterns
have not been previously observed or predicted.
We have hypothesized that this pattern for the LMC filaments may be an imprint
of the motion of the gas ejection site according to the LMC rotation curve.
Using this hypothesis, we estimated that the drift rate of the MS
gas away from the LMC is $\sim$49 \kmse, and, based on the length of the entire MS, we surmise
that the MS is $\sim$1.7 Gyr old.  As we mentioned in the previous subsection, this
estimate of the age of the MS is fairly consistent with other studies.

However this hypothesis contradicts the de Boer et al.~(1998) scenario by invoking the SEHO as a
rotating star-formation site.
In this case, the multiple generations of stars formed in this process should still be relatively
near the ejection site.
But without replenishment of gas, this scenario may require
inordinately large amounts of mass to be processed through one particular star-formation site with no 
apparent driver and for a relatively long time.

A hybrid hypothesis that could incorporate the driving physics of the de Boer et al.
scenario and still produce oscillating patterns in the MS and LAF might include gas streaming
along the LMC bar, creating a ``perfect storm" of compressional activity when combined
with bulk LMC motion and LMC rotation.
\citet{Kim98} show that the LMC \hi velocity field deviates from circular rotation,
especially at the north-western
end of the stellar bar, and take this as evidence for large-scale streaming motions in the \hie.
In Figure \ref{magbar}b we show the high-velocity \hi gas together with the density of 2MASS LMC 
red giant branch stars \citep{Skrutskie06} indicating the stellar bar.
The eastern end of the bar is close to the SEHO where the \hi filaments
originate, suggesting that the bar may play a role in (1) the accumulation of the
dense gas in SEHO and (2) the origin of the high-velocity filaments there.
However, whereas the end of 
the bar on the {\it leading} side of the LMC coincides with intense star formation in
the SEHO, no equivalent feature is found on the western side of the bar.  
If bar streaming provides the critical, extra dynamical contribution necessary for 
star formation activity at the level needed for persistent blowout, then oscillating
motion of the ejection site would naturally correspond to the bar pattern rotation (which is not
well known).
But this would also imply periodic jumps in location every half period when there is a transition from
one end of the bar being on the leading side of the LMC to the other end leading. 
This motion would be imprinted
in the shape and velocity of the MS.  Spatial jumps similar to
that suggested may be visible (e.g., Fig.\ \ref{sep_fils}), but they are not as evident in the velocity 
distribution.

Clearly hydrodynamical modeling is needed to resolve the dilemma of how the 
ejection site is generated and to understand the expected effects of its motion, or 
lack thereof, on the MS and LAF.  We consider this one of the most important
challenges remaining to complete this new picture of how the MS and LAF formed.

\subsection{Relevance to the Tidal vs.~Ram Pressure Models}

The MS origins debate has focused on the tidal versus ram pressure
models.  We have added a new mechanism, SGS blowout, to explain
how the MS gas is removed from the MCs.  Once the gas has escaped its host galaxy, ram
pressure and/or tidal
forces are still required to disperse the unbound gas and move part of it forward
(to create the LAF) and other parts backward (to create the MS).  Therefore, the blowout
model moves the tidal versus ram pressure debate from the mechanism for {\it removal} of
gas from the MCs, to the mechanism for {\it dispersal} of the unbound gas.

It seems likely that {\em both} ram pressure and tidal forces are
present and needed to explain the characteristics of the MS and LAF:
The LAF {\em cannot} be satisfactorily explained without a
tidal force, and ram pressure forces are evidently at work in building
the steep density gradient in the leading edge of the LMC.
The {\em combination} of ram pressure and tidal forces can also help explain the
column density gradient along the Magellanic Stream \citep{MD94,Mastro05}
{\em and} the imbalance of mass in the MS and LAF, since it will be more difficult
to move material ahead of the MCs due to the extra force pushing it backwards.

Our proposed scheme for the origin of the MS through blowout resolves
a problem that has plagued tidal models, namely the
lack of observed stars in the Stream.  If most of the MS and LAF gas was blown
out of the LMC from SGSs in the SEHO, as suggested here,
then no stars would be expected in the Stream.  The forces operating in the
SGSs that blow out the gas (superwind and supernovae shocks) do
not affect the stars, and therefore the mystery of the lack of stars in the Stream,
even in the presence of tidal mechanisms, is easily explained.

Our model also resolves the paradox recently posed by \citet{besla} that if the
MCs are on hyperbolic orbits, as now indicated by HST proper motions \citep{kalli06a,kalli06b,piatek07},
then the distance of the MCs from the MW at the time that the MS originated
is too large ($\sim$300-500 kpc) for the ram pressure and tidal forces to strip the gas out of the LMC or
the SMC.  SGS blowout can do the work of moving the gas to large enough LMC radii to where even
weak ram pressure and tidal forces can take over for dispersal in the hyperbolic orbit scenario.

\subsection{New Constraints For Modeling of the Magellanic Stream}

In this paper we have used the detailed spatial and velocity distributions of 
Magellanic \hi gas to lead us to a new paradigm
for how the MS and LAF formed, namely through blowout from intense star formation
in the SEHO of the LMC.
But there are intriguing patterns in the MS and LAF that can provide additional clues at a 
more detailed level to the
processes that shaped them.  The most obvious clues are the periodic patterns.
Among them:
\begin{enumerate}
\item The two MS filaments exhibit large velocity and spatial oscillations. The
oscillation of the LMC filament has a velocity amplitude of 26.4 \kms with an angular period of
19.5\dgr (17.1 kpc at a distance of 50 kpc) in \lms (Fig.\ \ref{lmcfil_fit}).  The amplitude of the spatial
oscillation in \bms is $\sim$2\dgr (Fig.\ \ref{sep_fils}).
\item There are three concentrations in LA I, elongated along \bms ($\sim$$2 \times 7$\degr)
that look very similar to one another and that are each offset by $\sim$$12$\dgr in \lms and $\sim$$9.5$\dgr in \bms
from the previous one (Fig.\ \ref{msskyplot}).
\item LA II and LA III are very similar in appearance (Fig.\ \ref{msskyplot}).  They are both elongated
along \lms with sizes of $\sim$$21 \times 5$\dgr and parallel to each other (offset by $\sim$20\dgr in \bmse).
\item The two filaments of the Magellanic Stream exhibit strong periodic patterns in position,
and are composed of clumps elongated along \lms with sizes of $\sim$6.0$ \times 1.5$\degr, surprisingly similar
in size to the LA I clumps and highly suggestive that they may have formed by the same process
(Fig.\ \ref{msskyplot}b).
\item The two MS filaments are quite similar in appearance and mirror each other in their shape
(from \lms $\approx -15$\dgr to $-45$\degr), only shifted by $\sim$1\dgr in \lms and $\sim$4\dgr in \bms
(Figs.\ \ref{msskyplot}b and \ref{sep_fils}).
\end{enumerate}
Future modeling efforts of the MS should not only incorporate an LMC/SEHO origin
for both the MS and LAF, but strive to reproduce these other newly found distinctive observational
characteristics of the MS.  We propose that such models account for the energetics of supergiant shell
blowout in the creation of the MS.

\acknowledgements

We appreciate useful discussions with Bill Kunkel, Gurtina Besla, Pavel Kroupa, Jon Hibbard,
Remy Indebetouw, Ed Murphy, Zhi-Yun Li, Jake Simon, John Hawley and Amy Reines.  Our work has
made use of the IDL program MPFIT by Craig Markwardt, and we would like to thank him for making
this and other useful programs available to the public.  We would also like to thank Lister Staveley-Smith for
providing us with the Parkes \hi datacube of the LMC as well as Bryan Gaensler and Uli Klein for providing us with 
the 12 cm polarization data of the LMC.  D.L.N. is supprted by the ARCS Foundation,
the Green Bank Telescope Student Support Program, a University of Virginia President's Fellowship,
and the Virginia Space Grant Consortium.  S.R.M. acknowledges funding from NSF grant AST-0307851,
NASA/JPL contract 1228235, and the generous support of Frank
Levinson and through the Celerity Foundation.
This publication makes use of data products from the Two Micron All Sky Survey, which is a joint project
of the University of Massachusetts and the Infrared Processing and Analysis Center/California Institute
of Technology, funded by the National Aeronautics and Space Administration and the National Science Foundation.
We would like to thank the referee Jacco van Loon for numerous useful comments and suggestions.


\end{document}